\documentclass[english]{article}
\usepackage[OT1]{fontenc}
\usepackage[latin9]{inputenc}
\usepackage{float}
\usepackage{setspace}
\RequirePackage{amsthm}
\RequirePackage[cmex10]{amsmath}
\usepackage{pgf,tikz}
\usetikzlibrary{shapes,decorations,arrows,calc,arrows.meta,fit,positioning}
\tikzset{
	-{Latex[length=4mm, width=1.1mm]},auto,node distance =1 cm and 1 cm,semithick,
	state/.style ={ellipse, draw, minimum width = 0.7 cm},
	point/.style = {circle, draw, inner sep=0.04cm,fill,node contents={}},
	bidirected/.style={dashed,arrows={Latex[length=6mm, width=1.2mm]-Latex[length=4mm, width=1.4mm]}},
	el/.style = {inner sep=2pt, align=left, sloped}
}

\pgfdeclarelayer{bg}    % declare background layer
\pgfdeclarelayer{mid}    % declare background layer
\pgfsetlayers{bg,mid,main}

\usepackage{amssymb}
\usepackage{ctable}
\usepackage{colortbl}
\usepackage{footnote}
\usepackage{graphicx}
\usepackage[authoryear]{natbib}
\RequirePackage[colorlinks,citecolor=blue,urlcolor=blue]{hyperref}
\PassOptionsToPackage{normalem}{ulem}
\usepackage{ulem}
\usepackage[position=bottom]{subfig}
\numberwithin{equation}{section}
\theoremstyle{plain}

\numberwithin{equation}{section}
\makeatletter

\providecommand{\tabularnewline}{\\}

\@ifundefined{showcaptionsetup}{}{%
	\PassOptionsToPackage{caption=false}{subfig}}
\usepackage{subfig}
\makeatother
\usepackage{xcolor}

\makeatletter

\newcommand{\indep}{\perp \!\!\! \perp}
\newcommand{\globalcolor}[1]{%
	\color{#1}\global\let\default@color\current@color
}
\makeatother

\AtBeginDocument{\globalcolor{black}}

\begin{document}
	
	\title{Proxy Controls and Panel Data}
	\author{Ben Deaner \thanks{University College London. Email at bendeaner@gmail.com. This version resubmitted 20th July 2023. Previous editions of this work can be found at https://arxiv.org/abs/1810.00283 (first version September 30th 2018). Many thanks to my advisers Whitney Newey, Anna Mikusheva, and Jerry Hausman for invaluable advice. Thanks to those who attended the Econometric society session C1 at the ASSA 2020 =Annual Meeting in San Diego for useful feedback and likewise to those who attended via Zoom the Treatment Effects 2 session of the 2020 Econometric Society World Congress. Thanks as well to those who attended the many internal presentations of this project at MIT.}}
	\maketitle
	
\begin{abstract}
	
	We provide new results for nonparametric identification, estimation, and inference of causal effects using `proxy controls': observables that are noisy but informative proxies for unobserved confounding factors. Our analysis applies to cross-sectional settings but is particularly well-suited to
	panel models. Our identification results motivate a simple and `well-posed' nonparametric 
	estimator. We derive convergence rates for the estimator and construct uniform confidence bands with asymptotically correct size. In panel settings, our methods provide a novel approach to the difficult problem of identification with non-separable, general
	heterogeneity and fixed $T$. In panels, observations
	from different periods serve as proxies for unobserved
	heterogeneity and our key identifying assumptions follow from restrictions
	on the serial dependence structure. We apply our methods to two empirical settings. 
	We  estimate consumer demand counterfactuals using panel data and we estimate causal effects of grade retention on cognitive performance.
\end{abstract}

A sizable portion of the empirical economist's working life is dedicated to diagnosing and accounting for the presence of confounding factors. Ideally, a researcher would control for these factors, but some important confounders may not be measured in the data. We suppose that the data contain some observables that can act as proxies (broadly defined) for these latent factors. 

We refer to observables that provide noisy and possibly biased signals for unobserved confounders as `proxy controls' because they can act as proxies for factors one would like to control for. Test scores may be  informative proxies for academic ability and years of experience for human capital. With panel data, past observations can provide a wealth of information about the latent characteristics of an individual. For example, past consumption habits are likely informative about the individual's consumption preferences.

The problem of mismeasured controls has long been acknowledged in the labor economics literature, particularly in the context of returns to schooling (see \cite{Griliches1977} for an early empirical example). Classical analyses assume additive linear specifications for the potential outcomes and the measurement error. Linearity may be implausible in some settings and precludes the study of nonlinearities and heterogeneity in treatment effects. Following the seminal work of \cite{Miao2018}, an emerging literature considers proxy controls in non-linear, semiparametric,  and nonparametric settings.

We develop new nonparametric identification results in the context of proxy controls. We identify the conditional (on observed treatments and possibly other covariates) average potential outcomes. Causal objects can be identified using either of two conditional moment restrictions. Using these dual characterizations, we show that estimation of causal effects in this setting is `well-posed' under our identifying assumptions. Well-posedness is crucial for deriving simple and transparent convergence rates for estimation methods based on our identification results. We show that the problem of proxy controls can be adapted to causal analysis using panel data and develop new nonparametric identification results for panel models with a fixed number of time periods.

In panels, observations from other time periods can be informative proxies
for latent confounding factors. By definition, confounders are associated with treatments and potential outcomes, and so, if the confounding factors are persistent, past treatments and past outcomes are informative about the confounding factors. We provide conditions on the  serial dependence structure of the data and
latent variables so that one can form proxy controls from past observations
that satisfy our identifying assumptions.

We suggest estimation and inference
procedures based on our identification results. Our methods are fully non-parametric, they allow for continuous treatments and identification of causal effects conditional on continuous observable covariates. The procedures can be applied in both cross-sectional and panel settings. The estimation method is based on series regression, and therefore it also suggests a flexible parametric method if the number of series terms is simply held fixed rather than allowed to grow with the sample size.

We establish consistency and a convergence rate for our estimator under our identifying assumptions and primitive conditions of the kind employed in the literature on standard nonparametric regression. We give conditions under which our estimator can be asymptotically approximated by a Gaussian process. We develop a method for constructing uniform confidence bands that is based on the multiplier bootstrap and a related specification test.

To demonstrate the usefulness of our methodology we apply it to two
very different real-world data problems. We use data from the Panel Survey of Income Dynamics
(PSID) to estimate a structural Engel curve for food.  We revisit the empirical setting of \citet{Fruehwirth2016} who use
data from the Early Childhood Longitudinal Study of Kindergartners
(ECLS-K) to estimate the causal impact of grade retention on the performance
of US students in cognitive tests.

\subsection*{Related Literature}

This paper contributes to an expanding body of recent research on the use of proxy controls. \citet{Miao2018} first established nonparametric identification of average potential outcomes and the marginal distribution of potential outcomes in cross-sectional settings when controls are mismeasured.

\citet{Miao2018} follows earlier work that detects and sometimes adjusts for the presence of measurement error or confounding using `negative controls': variables which lack a direct causal relationship to either the treatments or to outcomes (for a review see \cite{Shi2020a}). The conditions \citet{Miao2018} and others place on proxy controls means that they are in fact, negative controls. Some of this earlier work utilizes serial dependence in the data to select valid negative controls, similar to our use of serial dependence restrictions to motivate choices of proxy controls (for example, \cite{Flanders2009}). 

We extend the identification results of \citet{Miao2018} to identify  conditional average potential outcomes and we provide a dual identification result based on a different conditional moment restriction. This dual result plays a crucial role in our analysis of estimation using proxy controls. Our identifying assumptions also differ from \citet{Miao2018} in that we do not require one vector of proxies to be statistically complete for another. Instead, we use a condition of the type employed by \cite{Shi2020} in the context of categorical confounding and proxies.

Recent work adapts the proximal approach to a broad range of causal inference problems. For example, mediation analysis \cite{Dukes},   synthetic control \cite{Qiu2022}, and complex longditudinal studies \cite{Ying2021}. The latter of these uses proxy controls to estimate causal effects in dynamic panel models as in this work, but has somewhat different aims. We use the panel structure to motivate choices of valid proxies formed from past treatments and/or outcomes. By contrast \cite{Ying2021} uses time-varying proxies to identify rich dynamic causal effects.

A number of papers that consider estimation using proxy controls utilize  parametric specifications (for example, \citet{Tchetgen} and \cite{Miao2018a}). Much of the work in this literature considers semiparametric estimation, for example \citet{Cui2020} who derive semiparametric efficiency bounds. A flurry of recent work apply a kernel minimax estimation approach to this semiparametric problem, including \cite{Ghassami2021}, \cite{Kallus2021}, and \cite{Bennett2022}. One innovation in these recent works is the use of Neyman-orthogonal learning, which can reduce bias and may have other advantages (\cite{Chernozhukov}). Some of these later works utilize a dual characterization of causal objects using a `treatment bridge' akin to our result in Theorem 1.1.i. \cite{Nagasawa2018} considers identification and estimation with proxy controls using a control function approach. 

We consider the fully nonparametric estimation and inference problems. Other work on these problems includes \cite{Singh2020} and \cite{Mastouri2021}, who employ RKHS-based approaches.

One important point of distinction between our analysis of estimation using proxy controls and that elsewhere in the literature, is that we avoid assuming smoothness of solutions to the identifying conditional moment restrictions. These functions, sometimes referred to in the literature as `bridge functions', lack a clear structural interpretation, and under our identifying assumptions they may not be unique. As such, it is difficult to motivate a priori conditions that these functions have say, bounded derivatives or are well approximated by polynomials. Fortunately, we are able to establish asymptotic guarantees for our estimates without the need for such restrictions on the bridge functions as we discuss in Section 4.  

Our work contributes to the broader literature on nonparametric and non-separable measurement error models and their applications. \citet{Hu2008} provide identification results for nonparametric and non-separable models with measurement error and present a related estimator. The results of \citet{Hu2008} have been used for estimation in panel models using a factor analytic approach. Notably in \citet{Hu2012} and \citet{Arellano2016}. \citet{Wilhelmb} applies their results to a panel model with noisy measurements of the covariates of interest. The  results of \citet{Hu2008} is adapted to causal inference with mismeasured confounders in the context of regression discontinuity by \cite{Rokkanen2015} and may be adapted to the problem of proxy controls more generally, as discussed in \cite{Deaner2022}. As shown in \cite{Deaner2022}, the exclusion restrictions of \citet{Hu2008} differ from those in this paper and other closely related works. Moreover, the corresponding characterization of causal objects differs markedly and motivates a distinct estimation strategy.

Our panel analysis follows a long line of work in which observations
from other periods are used to account for unobserved heterogeneity.
This approach is the basis of classic methods like those of \citet{Hausman1981b}, \citet{Holtz-Eakin1988b}, and \citet{Arellano1991} and some more recent work that allows for non-linearity like \citet{Freyberger2018}
and \citet{Evdokimov2009}.

An extensive literature examines the effects of grade
retention on cognitive and social success.
For a meta-analysis see \citet{Jimerson2001}.
We build on the work
of \citet{Fruehwirth2016}. We use the cleaned data available with their paper
and we estimate some of the same causal effects. Recent work to estimate consumer demand counterfactuals (in particular, structural Engel curves) in nonparametric/semi-parametric models includes the instrumental
variables approach of \citet{Blundell2007} and the panel approach
of \citet{Chernozhukov2015a}. For a short survey see \citet{Lewbel}.
\section{Identification with Cross-Sectional Data}

Consider an outcome of interest $Y$ and a column vector of treatments $X$. We are interested in the causal effect of $X$ on $Y$ which we define in terms of potential outcomes. We denote by $Y(x)$ the potential outcome from a counterfactual level $x$ of the treatment.

The identification of causal effects of $X$ on $Y$ is complicated by the presence of confounders: factors which jointly determine $X$ and $Y$. In order to adjust for the confounding, researchers may control for a set of pre-treatment covariates $D$. A problem arises when there are additional confounding factors that are not captured in $D$. Let $W$ be vector of latent factors that includes all the remaining confounders. If $W$ were observed then we could control for $W$ and $D$ in the usual way to recover the effect of $X$ on $Y$.

For example, suppose $X$ is some feature of a student's education and $Y$ measures the student's adult wages. We may have access to covariates $D$ that represent demographic characteristics of the student. Controlling for $D$ is likely insufficient to recover the causal effect of $X$ on $Y$. This is because $D$ does not fully account for a persistent component of academic aptitude $W$ which may confound $X$ and $Y$.

While $W$ is not observed, we suppose that the researcher has access to two vectors of covariates $V$ and $Z$ which provide some noisy information about $W$. For example, $V$ and $Z$ may include sets of test scores which are informative about the student's underlying academic ability $W$. We refer to $V$ and $Z$ as `proxy controls' because they act as  stand-ins (or `proxies') for latent factors which we would like to control for.\footnote{In the econometrics literature one sometimes says that a variable $\tilde{X}$ is a `proxy' for $X$ if $X=\tilde{X}+\epsilon$ with $\epsilon$ independent of $\tilde{X}$. We use the term more loosely to simply refer to a variable that is informative about another.} 

The conditions we place on the two vectors of proxies are not symmetric and in particular they differ in terms of the restrictions on their relationships with treatments and outcomes. We place relatively weak restrictions on the relationship between $V$ and the outcome $Y$ and so we refer to $V$ as `outcome-aligned' proxies. We place only weak restrictions on the association between $Z$ and $X$, and thus we refer to $Z$ as `treatment-aligned' proxies. 

\begin{table}[h]
	\caption{List of Variables}
	\centering
	\begin{tabular}{c||c}
		\multicolumn{1}{c}{Variable} & \multicolumn{1}{c}{Description}
		\tabularnewline
		\specialrule{.2em}{.1em}{.1em} 
		$Y$  & Outcome of interest. \tabularnewline
		$X$  & Vector of treatments. \tabularnewline
		$W$ & Vector of latent factors. \tabularnewline
		$V$ & Vector of outcome-aligned proxies for $W$. \tabularnewline
		$Z$ & Vector of treatment-aligned proxies for $W$. \tabularnewline
		$D$ & Vector of additional controls. \tabularnewline
	\end{tabular}	
\end{table}

An informal summary of the key identifying conditions is provided below. This may act as a check-list for applied researchers interested in applying proxy control methods. We formalize and examine these conditions later in this section.

\begin{enumerate}
	\item After controlling for the latent factors $W$ and observed confounders $D$, the potential outcomes $Y(x)$ and outcome-aligned proxies $V$ must each be jointly independent of the treatments $X$ and treatment-aligned proxies $Z$. 
	\item The proxies $V$ and $Z$ must each be sufficiently informative about $W$. In particular,  $V$ and $Z$ must each satisfy an instrumental relevance condition in an IV model in which $W$ is a vector of endogenous regressors, with $X$ and $D$ acting as exogenous regressors.
	\item The confounders $W$ must satisfy a full support condition given the treatments and additional controls $D$.
\end{enumerate}

We suggest researchers check these conditions in turn, first attempting to write down a minimal set of latent factors $W$ so that the first condition is credible before proceeding to assess the next.

\theoremstyle{definition}
\newtheorem*{GRS}{Example 1: Returns to Schooling}
\begin{GRS}
	\cite{Griliches1977} provides a classical analysis using proxy controls. In that work the outcome $Y$ is the logarithm of an individual's adult wages and the treatment $X$ is years in education. Griliches observes a vector of family background characteristics $D$.
	
	Griliches suggests that even after controlling for family characteristics, the treatments and outcomes may be confounded due to the influence of a persistent component of academic aptitude $W$. However, Griliches observes two sets of test scores $V$ and $Z$ which may be informative but noisy proxies for unobserved aptitude. The $V$ tests are taken early on in school and the $Z$ tests later in school.
	
	Griliches makes a number of assumptions regarding the structural relationships between the observables and latent variables. He captures these assumptions in the linear structural model below. For simplicity we omit family characteristics $D$ and intercepts: one can imagine that the variables have all been residualised with respect to $D$ and a constant.
	\begin{align}
		Y&=\beta' X + G_{1}W + U_Y\nonumber\\
		X&=G_{2} W + G_3 S+ U_X\nonumber\\
		V&=G_{4} W + U_V\nonumber\\
		Z&=G_{5} S +G_{6} W  +U_Z\nonumber\\
		S&=G_7 W +U_S \label{linearmodel}
	\end{align}
	
	The coefficients in the model are taken to represent direct causal effects, thus the exclusion of a variable from an equation indicates a lack of any direct impact. Griliches  assumes there are no omitted joint causes of any of the variables, and so $W$ and the noise terms $U_Y$, $U_X$, etc. are uncorrelated. Interest is in $\beta$.
	
	Griliches supposes that the test scores have no impact on length of schooling and wages. This is credible if the scores are measured privately by researchers or are otherwise given negligible weight when making schooling and employment decisions. In addition, he assumes that family background $D$ and the persistent component of academic ability $W$ are the only factors that jointly determine a) late test scores or length of schooling, and b) early test scores or adult wages.
	
	However, length in schooling and the late-tests may have other shared sources of variation. In particular, Griliches suggests that expected length in schooling at the time of the late tests, denoted by $S$, could impact both length in schooling and performance on the late tests. This does not preclude identification because the late test scores are treatment-aligned proxies, and so their association with treatment need not be explained by $W$ and $D$.
	
%	The structural model of Griliches can be relaxed. Let us assume only that the linear models for $Y$ and $V$ in (\ref{linearmodel}) hold. We allow $U_Y$, $U_V$, and $W$ to be correlated. Let $U_{Y(x)}=G_1 W+U_Y$, which represents variation in potential outcomes of $Y$. It is enough that after residualizing $U_{Y(x)}$ and $U_V$ with respect to $W$, these variables are each uncorrelated with both $X$ and $Z$. In other words, after controlling for $W$ and $D$, both $X$ and $Z$ must be uncorrelated with the residual variation in  both potential outcomes and $V$.
	
	Identification requires that the proxies are sufficiently informative about the latent factors $W$. Let $\tilde{V}$ and $\tilde{Z}$ denote $V$ and $Z$ residualised with respect to $X$. Suppose $E[W \tilde{V}']$ and $E[W \tilde{Z}']$ each have full row rank and no components of $X$ are perfectly multicolinear. Then given the model, $\beta$ is identified from the moment condition below.\footnote{See \cite{Deaner2021} for further detail regarding  identification in linear models with proxy controls.}
	\begin{equation}
		E[(Y-\beta' X-\xi V)(Z',X')']=0 \label{linan}
	\end{equation}
	
	The rank conditions ensure that $V$ and $Z$ are sufficiently informative about $W$. Full row rank of $E[W \tilde{V}']$ is precisely the rank condition for identification when $V$ is used as an instrument for $W$ with $X$ acting as exogenous regressors, and similarly for $E[W \tilde{Z}']$. Thus the proxies are sufficiently informative  if they are relevant instruments for $W$. 
	
	The strength of the rank condition depends on the richness of the latent factors $W$ and the proxies. In Griliches' model, $W$ is a scalar that represents aptitude, thus it suffices that (after controlling for treatment) the test scores $V$ and $Z$ are correlated with aptitude.
	
	However, suppose test scores also depend upon a second latent factor that captures a student's test-taking skill and this is not perfectly correlated with aptitude. Then Assumption 1 requires that $W$ consist of two latent factors $W_1$ and $W_2$. In this case the rank condition requires both $V$ and $Z$ are at least two-dimensional i.e., they each consist of at least two test scores. Suppose $V$ contains two scores $V_1$ and $V_2$. Again, let tildes indicate residualisation with respect to $X$.  The rank condition for $V$ holds if and only if the ratio of $E[W_1 \tilde{V}_{1}]$ to $E[W_2 \tilde{V}_{1}]$ differs from the ratio of $E[W_1 \tilde{V}_{2}]$ to $E[W_2 \tilde{V}_{2 }]$. In other words, one of $V_1$ and $V_2$ must be strictly more sensitive to test-taking skill relative to aptitude than the other.
	
	Given the discussion above, we suggest that in order to justify the rank condition researchers should first list a set of factors $W$ so that the required conditional independence restrictions are plausible. The rank condition is then credible if, for each factor in $W$, there is at least one proxy in each of $V$ and $Z$ that is likely to be particularly strongly associated with that factor compared to the remaining factors.
\end{GRS}

Assumptions 1-3 below strengthen the identifying assumptions in linear case so that we can achieve identification in nonparametric, non-separable models.

\theoremstyle{definition}
\newtheorem*{A1}{Assumption 1 (Independence)}
\begin{A1}
	 i. $Y(x) \perp\!\!\!\perp (X,Z)|(W,D)$, and  ii.  $V \perp\!\!\!\perp (X,Z)|(W,D)$
\end{A1}

\newtheorem*{A2}{Assumption 2 (Informative Proxies)}
\begin{A2}
	For every function $\delta$ with the property that $0<E[\delta(W)^2|X=x,D=d]<\infty $:	
	\[
	\text{i.}\, E\big[E[\delta(W)|X,D,Z]^2|X=x,D=d\big]>0 
	\]
	\[
	\text{ii.}\,E\big[E[\delta(W)|X,D,V]^2|X=x,D=d\big]>0
	\]
\end{A2}

\theoremstyle{definition}
\newtheorem*{A3}{Assumption 3 (Full Support)}
\begin{A3}
	i. $W$ has a conditional probability density function $f_{W|XD}$ (with respect to some dominating measure) that is strictly positive on the support of $W$. ii. $E[Y^2]<\infty$ and
	$E[Y(x)^2]<\infty$.
\end{A3}

Assumption 1 effectively requires $W$ contains not only all the latent confounders, but any unobserved factor that directly impacts both a) $V$ or $Y$, and b) $Z$ or $X$. If there were additional factors that determined say, $Z$ and $Y$, then $Z$ and $Y(x)$ would likely be correlated even after controlling for $X$ and $W$.

In practice, we suggest that researchers list a minimal set of factors $W$ that satisfy the condition in the previous paragraph before proceeding to justify the remaining assumptions. For example, in the Griliches setting, one must consider whether univariate ability and family characteristics are the only variables that impact one of adult wages or early test scores, and which also affect either time is schooling or late test scores. If the researcher believes that test-taking ability may also be an important determinant of both early and late test scores then this should be included in $W$.

Assumption 1 rules out certain causal relationships between the proxies, treatments and outcomes. The condition proscribes any direct  causal relationship between outcome-aligned proxies $V$ and treatments, as well as any direct causation between treatment-aligned proxies $Z$ and the outcome. These exclusion restrictions hold in the Griliches model, for example the early test scores $V$ do not impact time in schooling $X$.

Assumption 1 is implied by a nonparametric analogue of the Griliches model. The model on the right of Figure 1.a has the same exclusion restrictions as (\ref{linearmodel}) and these can be justified by the same arguments. The structural functions $g_Y$, $g_X$ etc. are possibly non-linear and the noise terms are taken to be jointly independent rather than uncorrelated. In this model $Y(x)$ is equal to $g_Y(x,W,U_Y)$. The directed graph on the left of Figure 1.a encodes the exclusion restrictions of the model.

The model in Figure 1.a is a nonparametric structural equations model (NSEM) of the type in \cite{Pearl2009}. NSEMs are structural (i.e., causal) models that can act as intuitive primitive conditions for conditional independence restrictions. As in Figure 1.a, the restrictions of any NSEM can be neatly encoded in a directed graph. The NSEMs associated with the other graphs in Figure 1 also imply Assumption 1.  This would remain true if we include $D$ as a cause of all other variables. These models are not exhaustive. The two graphs generalize the model in Figure 1.a. They allow $V$ to impact $Y$, include unobservables $Q$ which may affect both $Y$ and $V$, and allow direct causation between $X$ and $Z$.

Any NSEM that implies Assumption 1 must have the following properties. There must be no direct causation between treatment-aligned proxies $Z$ and the outcome $Y$, nor between outcome-aligned proxies $V$ and treatment $X$. Neither vector of proxies $Z$ nor $V$ must directly impact the other, and $Y$ must not impact $V$.

\begin{figure}[h]
	\centering
	\caption{Causal Structure of Proxy Controls}
	
	\centering
	
	\subfloat[Nonparametric Griliches Model]{
		\resizebox{!}{87pt}{%
			
			\begin{tikzpicture}
				% nodes %
				
				\node (y) at (1.1,-1.22) [label=below right:$Y$,point];
				\node (x) at (-1.1,-1.22) [label=below left:$X$,point];
				\node (w) at (0,1.5) [label=above:$W$,point];
				\node (s) at (-1.8,-0.8) [label=left:$S$,point];
				\node (v) at (1.6,0.46) [label=above right:$V$,point];
				\node (z) at (-1.6,0.46) [label=above left:$Z$,point];
				%\node (d) at (0,-0.1) [label=below:$D$,point];			
				\node (y1) at (3.5,1.6) [label=right:{$Y {=}g_Y (X,W,U_{Y})$},point];
				\node (y2) at (3.5,1.1) [label=right:{$X {=}g_X(S,W,U_{X})$},point];
				\node (y3) at (3.5,0.6) [label=right:{$V {=}g_V(W,U_{V})$},point];
				\node (y4) at (3.5,0.1) [label=right:{$Z {=}g_Z(W,S,U_{Z})$},point];		
				\node (y5) at (3.5,-0.4) [label=right:{$S {=}g_{S}(W,U_{S})$},point];	
				\node (y6) at (3.5,-0.9) [label=right:{$W$, $U_{Y}$, $U_{X}$, $U_{V}$, $U_{Z}$, $U_{S}$ jointly independent.},point];

				%\path (d) edge (x);
				%\path (d) edge (y);
				%\path (d) edge (z);
				%\path (d) edge (v);
				%\path (d) edge (w);
				
				\path (x) edge (y);
				\path (w) edge (x);
				\path (w) edge (y);
				\path (w) edge (z);
				\path (w) edge (v);
				\path (s) edge (z);
				\path (s) edge (x);
				\path (w) edge (s);
				%\path (v) edge (y);
				
			\end{tikzpicture}

		}		
	}
	\centering
	
	\subfloat[Additional Models]{
		\resizebox{!}{67pt}{%
			
			\begin{tikzpicture}
				% nodes %
				
				\node (y) at (1.1,-1.22) [label=below right:$Y$,point];
				\node (x) at (-1.1,-1.22) [label=below left:$X$,point];
				\node (w) at (0,1.5) [label=above:${W}$,point];
				\node (v) at (1.6,0.46) [label=above right:$V$,point];
				\node (z) at (-1.6,0.46) [label=above left:$Z$,point];
				%	\node (d) at (0,-0.1) [label=below:$D$,point];			
				\node (s) at (-1.8,-0.8) [label=left:$S$,point];
				\node (q) at (1.8,-0.8) [label=right:$Q$,point];	
				
				%	\path (d) edge (x);
				%	\path (d) edge (y);
				%	\path (d) edge (z);
				%	\path (d) edge (v);
				%	\path (d) edge (w);
				%	\path (d) edge (s);
				%	\path (d) edge (q);
				
				\path (x) edge (y);
				\path (w) edge (x);
				\path (w) edge (y);
				\path (w) edge (z);
				\path (w) edge (v);
				\path (x) edge (z);
				\path (v) edge (y);
				\path (w) edge (s);
				\path (w) edge (q);
				
				\path(s) edge (x);
				\path(s) edge (z);
				\path(q) edge (v);
				\path(q) edge (y);	
			\end{tikzpicture}
			
			\begin{tikzpicture}
				% nodes %
				
				\node (y) at (1.1,-1.22) [label=below right:$Y$,point];
				\node (x) at (-1.1,-1.22) [label=below left:$X$,point];
				\node (w) at (0,1.5) [label=above:${W}$,point];
				\node (v) at (1.6,0.46) [label=above right:$V$,point];
				\node (z) at (-1.6,0.46) [label=above left:$Z$,point];
				%	\node (d) at (0,-0.1) [label=below:$D$,point];			
				\node (s) at (-1.8,-0.8) [label=left:$S$,point];
				\node (q) at (1.8,-0.8) [label=right:$Q$,point];	
				
				%	\path (d) edge (x);
				%	\path (d) edge (y);
				%	\path (d) edge (z);
				%	\path (d) edge (v);
				%	\path (d) edge (w);
				%	\path (d) edge (s);
				%	\path (d) edge (q);
				
				\path (x) edge (y);
				\path (w) edge (x);
				\path (w) edge (y);
				\path (w) edge (z);
				\path (w) edge (v);
				\path (z) edge (x);
				\path (v) edge (y);
				\path (w) edge (s);
				\path (w) edge (q);
				
				\path(s) edge (x);
				\path(s) edge (z);
				\path(q) edge (v);
				\path(q) edge (y);	
			\end{tikzpicture}	

%			\begin{tikzpicture}
				% nodes %
				
%				\node (y) at (1.1,-1.22) [label=below right:$Y$,point];
%				\node (x) at (-1.1,-1.22) [label=below left:$X$,point];
%				\node (w) at (0,1.5) [label=above:$W$,point];
%				\node (v) at (1.6,0.46) [label=above right:$V$,point];
%				\node (z) at (-1.6,0.46) [label=above left:$Z$,point];
				%	\node (d) at (0,-0.1) [label=below:$D$,point];			

				%	\path (d) edge (x);
				%	\path (d) edge (y);
				%	\path (d) edge (z);
				%	\path (d) edge (v);
				%	\path (d) edge (w);
				
%				\path (x) edge (y);
%				\path (w) edge (x);
%				\path (w) edge (y);
%				\path (w) edge (z);
%				\path (v) edge (w);
%				\path (z) edge (x);
%				\path (v) edge (y);
				
%			\end{tikzpicture}	
		
%			\begin{tikzpicture}
				% nodes %
				
%				\node (y) at (1.1,-1.22) [label=below right:$Y$,point];
%				\node (x) at (-1.1,-1.22) [label=below left:$X$,point];
%				\node (w) at (0,1.5) [label=above:$W$,point];
%				\node (v) at (1.6,0.46) [label=above right:$V$,point];
%				\node (z) at (-1.6,0.46) [label=above left:$Z$,point];
				%	\node (d) at (0,-0.1) [label=below:$D$,point];
				
				%	\path (d) edge (x);
				%	\path (d) edge (y);
				%	\path (d) edge (z);
				%	\path (d) edge (v);
				%	\path (d) edge (w);
				
%				\path (x) edge (y);
%				\path (w) edge (x);
%				\path (w) edge (y);
%				\path (z) edge (w);
%				\path (w) edge (v);
%				\path (z) edge (x);
%				\path (v) edge (y);
				
%			\end{tikzpicture}	
		}		
	}
	
\end{figure}

Assumption 2 ensures that the proxies $V$ and $Z$ are each  sufficiently informative about the unobserved confounders. Similar conditions are used for identification in nonparametric factor models (see \cite{Hu2008}). Assumption 2 is a nonparametric analogue of the rank conditions for identification in the linear model in Example 1. In fact, the two coincide when the variables are jointly normal (see \cite{Newey2003}).

For this reason we suggest that researchers assess the credibility of Assumption 2 by considering whether the corresponding rank conditions are credible, noting that the two coincide in Gaussian models but are otherwise non-nested. At the end of our discussion of Example 1 we note that researchers might justify the rank conditions by arguing that, for each factor in $W$, there is a proxy in both $V$ and $Z$ that is  particularly strongly associated with this factor compared to the other factors.

The rank conditions require that the dimensions of $V$ and $Z$ are each weakly greater than that of $W$. That is, we have at least as many proxies in each vector as there are confounding factors. Strictly speaking, this order condition is not necessary for Assumption 2. However, because Assumption 2 coincides with the rank condition in the Gaussian case, it seems prudent that researchers consider the order condition when determining whether Assumption 2 is credible.

Recall that the rank conditions in Example 1 ensure that the proxies $V$ and $Z$ are relevant instruments for $W$ in the sense of linear instrumental variables. Similarly,  Assumption 2 ensures $V$ and $Z$ are relevant instruments for $W$ as required in Nonparametric Instrumental Variables (NPIV) models. To be more precise, Assumption 2.i is a statistical completeness condition which identifies an NPIV model of the kind in \cite{Newey2003} and \cite{Ai2003} where $Z$ is an instrument for $W$, and $X$ and $D$ are exogenous regressors. 2.ii is identical but with $V$ in place of $Z$. \cite{Newey2003} state that completeness is analogous to the rank condition in linear instrumental variables. Strictly speaking, Assumption 2 imposes $L_2$-completeness, which is slightly weaker than the standard completeness condition (see \citet{Andrews2017b}).

Note that Assumption 2 differs from the completeness conditions in \cite{Miao2018}. In particular, \cite{Miao2018} require that one vector of proxies is statistically complete for the other. In Gaussian models this rules out the possibility that both vectors of proxies are of higher dimension than $W$. By contrast, if Assumption 2 holds for some $V$ and $Z$ then it also hold for any ${V}^*$ and ${Z}^*$ of which $V$ and $Z$ are subvectors.\footnote{Strictly speaking neither Assumption 2 nor the completeness conditions of \cite{Miao2018} are stronger. However, as we note in the proof of Theorem 1.1, Assumption 2 can be weakened slightly and identification still holds. If Assumption 1 holds then the weakened version of Assumption 2 is in fact weaker than the conditions in \cite{Miao2018}.} Conditions of essentially the same form to Assumption 2 are employed by \cite{Shi2020} in the context of categorical confounding and proxies.

The full support condition in Assumption 3.i is required for identification even in the case with $W$ observed. In the case of a binary treatment, the condition ensures that the propensity score (from covariates $W$) is non-zero. 

We state our main identification result Theorem 1.1 below. In the condition $\mathcal{V}_{x,d}$ and $\mathcal{Z}_{x,d}$ respectively denote the supports of $V$ and $Z$ conditional on $X=x$ and $D=d$.

\theoremstyle{plain}
\newtheorem*{Th11}{Theorem 1.1 (Identification)}
\begin{Th11}
	Suppose Assumptions 1-3 hold for $x=x_1,x_2$.
	 i. For any function $\varphi$ with $E[\varphi(Z)^{2}|X=x_{1},D=d]<\infty$
	and:
	\begin{equation}
		E[\varphi(Z)|X=x_{1},D=d,V=v]=\frac{f_{V|XD}(v|x_2,d)}{f_{V|XD}(v|x_1,d)},\,\forall{v \in \mathcal{V}_{x_1,d}}\label{eq1}
	\end{equation}
	the conditional average potential outcome satisfies:
	\begin{equation}
	E[Y(x_{1})|X=x_{2},D=d]=E[\varphi(Z)Y|X=x_{1},D=d]\label{char1}
	\end{equation}
	ii.
	For any function $\gamma$ with  $E[\gamma(V)^{2}|X=x_1,D=d]<\infty$ and:
	\begin{equation}
		E[Y-\gamma(V)|X=x_1,D=d,Z=z]=0,\,\forall z\in \mathcal{Z}_{x_1,d}\label{eq2}
	\end{equation}
	the conditional average potential outcome satisfies:
	\begin{equation}
	E[Y(x_{1})|X=x_{2},D=d]=E[\gamma(V)|X=x_{2},D=d]\label{char2}
	\end{equation}
\end{Th11}

So long as either (\ref{eq1}) or (\ref{eq2}) has a solution, Theorem 1.1 identifies the conditional average potential outcome. Note $\varphi$ that satisfies (\ref{eq1}) may depend on $x_1$,  $x_2$, and $d$. Similarly $\gamma$ in (\ref{eq2}) may depend on $x_1$ and $d$.  $\gamma$ is sometimes referred to elsewhere in the literature as a `confounding bridge function' or `outcome bridge' and $\varphi$ (or a closely related object) as a `treatment bridge'.

The existence of solutions to (\ref{eq1}) and (\ref{eq2}) follows from Assumption 2 and a set of regularity conditions. The regularity conditions, which apply Picard's Criterion to this setting, are somewhat technical and so we relegate them to  Appendix A along with further discussion. We provide semiparametric examples in which (\ref{eq1}) and (\ref{eq2}) admit solutions below. Note that  our conditions do not imply that the solutions to these equations are unique.

Theorem 1.ii is similar to the result of \cite{Miao2018} but identifies the conditional average potential outcome rather than just an  unconditional average. Thus it allows for identification of say, the average effect of treatment on the treated and other conditional average treatment effects.

Theorem 1.i is fully original. It provides an alternative path to identification using a moment condition that does not depend on the outcome $Y$. This result plays a crucial role in our analysis of nonparametric estimation using proxy controls, as we discuss in Section 4.

Note the similarity between (\ref{eq2}) and the moment condition that identifies $\beta$ in the linear model of Example 1. Assuming no additional controls $D$, and making the dependence of $\gamma$ on $x_1$ explicit, (\ref{eq2}) can be written as: 
\[
E[Y-\gamma(V,x_1)|X=x_1,Z=z]=0
\]
If the above holds for all $x_1$ and $z$  in the joint support of $X$ and $Z$, then we have:
\[
E\big[\big(Y-\gamma(V,X)\big)(X',Z')\big]=0
\]
If we restrict $\gamma$ to be linear then the above is precisely the moment condition that identifies $\beta$ in the linear model.

\theoremstyle{definition}
\newtheorem*{EE1}{Semiparametric Example 1: Multivariate Normal}
\begin{EE1}
	Suppose for simplicity that there are no additional confounders $D$ and that $W$ and $Z$ are jointly normal conditional on each value $x$ of $X$, with a strictly positive definite conditional variance-covariance matrix.
	\[
	\begin{pmatrix}W\\
		Z
	\end{pmatrix}|X=x\sim\mathcal{N}\bigg(\begin{pmatrix}\mu_{W|x}\\
		\mu_{Z|x}
	\end{pmatrix},\begin{pmatrix}\Sigma_{WW|x} & \Sigma_{WZ|x}\\
		\Sigma_{WZ|x}' & \Sigma_{ZZ|x}
	\end{pmatrix}\bigg)
	\]
	
	Then Assumption 2.i holds if and only if $\Sigma_{WZ|x}$ has rank equal to the dimension of $W$.  Note that this implies $Z$ has weakly greater dimension than $W$. We can replace $Z$ with $V$ to get an equivalent characterization of Assumption 2.ii. 
	
	Suppose Assumptions 1-3 hold and the conditional variance of $W$ does not vary with $X$ (i.e., $\Sigma_{WW|x_{2}}=\Sigma_{WW|x_{1}}$). This holds for example if $W$, $Z$, and $X$ are jointly Gaussian. Then (\ref{eq1}) holds with $\varphi$ of the form below, where $\alpha$ is a scalar and $M$ a vector:
	\begin{align*}
		\varphi(z) & =\alpha  exp(Mz)
	\end{align*}
	More generally, if $\Sigma_{WW|x_{2}}\neq \Sigma_{WW|x_{1}}$ but the two matrices are sufficiently close, (\ref{eq1}) holds with:
	\begin{align*}
		\varphi(z) & =\alpha  exp(z'Pz+Mz)
	\end{align*}
	Where $P$ is a matrix which may depend on $x_1$ and $x_2$. The precise condition on $\Sigma_{WW|x_{2}}$ and $\Sigma_{WW|x_{1}}$ is given in Appendix A along with a proof of this result.

\end{EE1}

\theoremstyle{definition}
\newtheorem*{SE2}{Semiparametric Example 2: Partially Linear Model}
\begin{SE2}
	
		Suppose for simplicity that there are no additional confounders $D$, and that conditional on $X=x_1$, $V$ and $W$ are jointly normally distributed with strictly positive definite variance-covariance matrix. In addition suppose that $E[Y^2|X=x_{1}]$ is finite and the conditional mean of $Y$ is linear in $W$ (but not necessarily $X$):
	\[
	E[Y|W=w,X=x_{1}]=A(x_{1})+B(x_{1})w
	\]
	If Assumption 2 holds there exist a scalar $a$ and vector $b$ (possibly dependent on $x_1$) so that (\ref{eq2}) is satisfied by:
	\[
	\gamma(V)=a+b'V
	\]
	
\end{SE2}
\section{Identification with Panel Data}

In the previous section we assume the existence of two vectors of proxies $V$ and $Z$ for the unobserved confounding factors. In panel settings, past and future treatments and outcomes are a natural source of relevant proxies. By definition, unobserved confounders affect both outcomes and treatments. If there is confounding in every period, and if the confounders are persistent, then treatments and outcomes from other periods are likely to be correlated with today's confounders.

We leverage serial independence restrictions on the treatments and outcomes to ensure that proxies $V$ and $Z$, and controls $D$, each composed of treatments and outcomes, are valid in the sense that Assumption 1 holds.

Our interest is in panel data with a fixed number of time periods. Observations are available from periods $t=1,...,T$, where $T$ is fixed. When $T$ is large there are more potential proxies and thus we can allow for richer panel dynamics while still finding proxies that satisfy Assumption $1$. Moreover, if the proxies $V$ and $Z$ contain more observables, then the informativeness conditions in Assumption 2 are more plausible.

In the discussion below we let $X_t$ denote the treatment at period $t$,  and $Y_t$ the outcome at $t$. $W_t$ denotes the unobserved  confounding factors at time $t$. In addition, we subscript a variable with $[s:r]$ with $s\leq r$ to denote the observations of that variable from periods $s$ to $r$ inclusive, for example $X_{[s:r]}=\{X_s,X_{s+1},...,X_{r-1},X_r\}$. For convenience, if $s> r$ we take $X_{[s:r]}=\emptyset$.

\subsection{Nonparametric Fixed Effects}

We begin by considering models in which the unobserved confounders are constant over time. That is, $W_t=W$ for all $t$. These can be understood as non-parametric fixed effects models in which a set of time-invariant individual-specific factors (or `fixed effects') $W$, can determine both treatments and outcomes.

Crucial to our analysis is that some of the observables exhibit  Markov dependence. By this we mean that given the unobserved individual-characteristics $W$, the association between observables in the past and future is explained by observables in the intervening periods.

To see why Markov dependence is useful, suppose that treatments (and not necessarily outcomes) are first-order Markov dependent. That is, for each $t=3,4,..,T$:
\begin{equation}
X_{[1:t-2]}\indep X_{[t:T]}|\{X_{t-1},W\}\label{fo1}
\end{equation}
Let $Y_t(x_t)$ denote the potential value of $Y_t$ under a counterfactual intervention that sets $X_t$ equal to the (non-random) value $x_t$. In addition to the above, suppose that the fixed-effects $W$ explain all confounding between the outcome in some period and the treatments up to that point so that the following holds for each $t=1,2,...,T$:
\begin{equation}
Y_t(x_t)\indep X_{[1,t]}|W\label{fo2}
\end{equation}
The condition above is a nonparametric analogue of the standard `weak exogeneity' or `predetermination' condition in linear panel data and time-series models. This condition allows for feedback from past outcomes to future treatments. However, it rules out any effect of past treatments on today's outcome unless that effect is mediated by today's treatment.

The Markov dependence condition (\ref{fo1}) and weak exogeneity (\ref{fo2}) imply vectors of proxies $V$ and $Z$ formed from past treatments are valid in the sense of Assumption $1.1$. In particular, if $T>3$ then  Assumption $1$ holds with $X=X_T$, $Y=Y_T$, $D=X_{t-1}$, $V=X_{[1:t-2]}$, and $Z=X_{[t:T-1]}$, where $t$ can be any number between $3$ and $T-1$ inclusive. This follows from the more general results later in this section.

The Markovian dependence in (\ref{fo1}) is crucial because it ensures that after conditioning on $W$ and on treatment in period $t-1$, the treatments prior to this period are independent of those after. This allows us to compose $V$ and $Z$ from the history of treatments without violating  Assumption 1.ii. In fact, as we discuss below, we can choose valid $V$ and $Z$ using treatments and/or outcomes from the past and/or future under a range of different modeling assumptions.

To better understand the kinds of modeling assumptions that justify Markovian dynamics, we frame the remainder of this subsection in the dynamic, nonparametric structural (i.e., causal) model given below which holds for $t=1,...,T$.
\begin{align}
	&Y_t=g_{Y,t}(Y_{[t-\kappa_{YY}:t-1]},X_{[t-\kappa_{YX}:t]}, W,U_{Y,t})\nonumber\\
	&X_t=g_{X,t}(Y_{[t-\kappa_{XY}:t-1]},X_{[t-\kappa_{XX}:t-1]}, W,U_{X,t})\nonumber\\
	&\{U_{Y,t}\}_{t=1}^T\text{, }\{U_{X,t}\}_{t=1}^T\text{, and } W\text{ are jointly independent.}\label{markov1}
\end{align}

In the model above $g_{Y,t}$ and $g_{X,t}$ are non-random structural functions. The time subscripts indicate that these functions may vary over time. $U_{Y,t}$ and $U_{X,t}$ are period-specific and individual-specific noise terms. The model allows the dynamics of treatments and outcomes to vary both between individuals and over time due to the presence of $W$ and the time-dependence of the structural functions. In the model all observables are directly impacted by the confounders $W$, suggesting they constitute informative proxies.

The constants $\kappa_{YY}$, $\kappa_{YX}$, $\kappa_{XY}$, and $\kappa_{XX}$ determine how far back in the past an observable can be and still directly impact treatments and outcomes today. For example, if $\kappa_{YY}=1$ then $Y_{t-1}$ may directly affect $Y_t$ but earlier outcomes only impact $Y_t$ indirectly.

Note that treatments or outcomes in earlier periods may be impacted by treatments and outcomes from periods prior to period $1$, the first period for which there is data. Because the model need only hold for $t=1,...,T$ we avoid any initial conditions assumption.

In a structural model like (\ref{markov1}) potential outcomes can be  defined by replacing a variable with a fixed value in every equation in which it appears. We focus on the potential value of $Y_t$ under a counterfactual value $x_t$ of $X_t$ which is given by:
\[Y_t(x_t)= g_{Y,t}(Y_{[t-\kappa_{YY}:t-1]},X_{[t-\kappa_{YX}:t-1]},x_t,W,U_t)
\]

\begin{figure}
	\caption{Nonparametric Fixed Effects Models}
	\centering
	\subfloat[$\kappa_{XX}=\kappa_{XY}=\kappa_{YY}=1$, $\kappa_{YX}=0$]{
		
		\resizebox{!}{53pt}{%
			
			\begin{tikzpicture}
				% nodes %
				
				\node (y1) at (1,1.4) [label=above:$Y_1$,point];			
				\node (y2) at (3,1.4) [label=above:$Y_2$,point];
				\node (y3) at (5,1.4) [label=above:$Y_3$,point];
				\node (y4) at (7,1.4) [label=above:$Y_4$,point];

				\node (x1) at (0,0) [label=below:$X_1$,point];			
				\node (x2) at (2,0) [label=below:$X_2$,point];
				\node (x3) at (4,0) [label=below:$X_3$,point];
				\node (x4) at (6,0) [label=below:$X_4$,point];
				
				\path (x1) edge (x2);
				\path (x2) edge (x3);
				\path (x3) edge (x4);
				
				\path (y1) edge (x2);
				\path (y2) edge (x3);
				\path (y3) edge (x4);
				
				\path (y1) edge (y2);
				\path (y2) edge (y3);
				\path (y3) edge (y4);

				\path (x1) edge (y1);
				\path (x2) edge (y2);
				\path (x3) edge (y3);
				\path (x4) edge (y4);

				\node (wfull)[rounded corners=10,fill=red,opacity=0.3,fit=(x3), inner sep=0.3cm]{} ; 
				\node (wfull)[rounded corners=10,fill=teal,opacity=0.3,fit=(y4), inner sep=0.3cm]{} ; 
				\node (wfull)[rounded corners=10,fill=violet,opacity=0.3,fit=(x1), inner sep=0.3cm]{} ;    
				\node (wfull)[rounded corners=10,fill=blue,opacity=0.3,fit=(x4), inner sep=0.3cm]{} ;
				\node (wfull)[rounded corners=10,fill=yellow,opacity=0.3,fit=(x2), inner sep=0.3cm]{} ;
				
				\node (wfull)[rounded corners=10,fill=yellow,opacity=0.3,fit=(y3), inner sep=0.3cm]{} ; 
				\node (wfull)[rounded corners=10,fill=violet,opacity=0.3,fit=(y1), inner sep=0.3cm]{} ;    
				\node (wfull)[rounded corners=10,fill=yellow,opacity=0.3,fit=(y2), inner sep=0.3cm]{} ;
				
				\matrix [draw,below right,ampersand replacement=\&] at (8.5,1.4) {
					\node [shape=circle, fill=violet,opacity=0.3,label=right:$V$] {}; \&
					\node [shape=circle, fill=red,opacity=0.3,label=right:$Z$] {}; \&
					\node [shape=circle, fill=yellow,opacity=0.3,label=right:$D$] {}; \\
					\node [shape=circle, fill=blue,opacity=0.3,label=right:$X$] {}; \&
					\node [shape=circle, fill=teal,opacity=0.3,label=right:$Y$] {}; \&\\
				};

			\end{tikzpicture}
			
	}}\\
	\subfloat[$\kappa_{XX}=1$, $\kappa_{XY}=0$, $\kappa_{YX}=2$, $\kappa_{YY}=2$]{
		
		\resizebox{!}{60pt}{%		
			
			\begin{tikzpicture}
				% nodes %
				
				\node (y1) at (1,1.4) [label=above:$Y_1$,point];			
				\node (y2) at (3,1.4) [label=above:$Y_2$,point];
				\node (y3) at (5,1.4) [label=above:$Y_3$,point];

				\node (x1) at (0,0) [label=below:$X_1$,point];			
				\node (x2) at (2,0) [label=below:$X_2$,point];
				\node (x3) at (4,0) [label=below:$X_3$,point];
				
				\path (x1) edge (x2);
				\path (x2) edge (x3);
				
				\path (x1) edge[teal,opacity=0.6] (y2);
				\path (x2) edge[teal,opacity=0.6] (y3);
				\path (x1) edge[teal,opacity=0.6] (y3);

				\path (y1) edge (y2);
				\path (y2) edge (y3);	
				\path (y1) edge[red,opacity=0.6,bend left=30] (y3);

				\path (x1) edge (y1);
				\path (x2) edge (y2);
				\path (x3) edge (y3);

				\node (wfull)[rounded corners=10,fill=blue,opacity=0.3,fit=(x2), inner sep=0.3cm]{} ; 
				\node (wfull)[rounded corners=10,fill=teal,opacity=0.3,fit=(y2), inner sep=0.3cm]{} ; 
				\node (wfull)[rounded corners=10,fill=violet,opacity=0.3,fit=(y1), inner sep=0.3cm]{} ;   
				\node (wfull)[rounded corners=10,fill=red,opacity=0.3,fit=(x3), inner sep=0.3cm]{} ;
				\node (wfull)[rounded corners=10,fill=yellow,opacity=0.3,fit=(x1), inner sep=0.3cm]{} ;
				
			\end{tikzpicture}
			
	}}\,\,\,\,\,\,\,\,\,
	\subfloat[$\kappa_{XX}=1$, $\kappa_{YX}=0$, $\kappa_{XY}=\kappa_{YY}=2$]{
		
		\resizebox{!}{60pt}{%		
			\begin{tikzpicture}
				% nodes %
				
				\node (y1) at (1,1.4) [label=above:$Y_1$,point];			
				\node (y2) at (3,1.4) [label=above:$Y_2$,point];
				\node (y3) at (5,1.4) [label=above:$Y_3$,point];
				\node (y4) at (7,1.4) [label=above:$Y_4$,point];

				\node (x1) at (0,0) [label=below:$X_1$,point];			
				\node (x2) at (2,0) [label=below:$X_2$,point];
				\node (x3) at (4,0) [label=below:$X_3$,point];
				\node (x4) at (6,0) [label=below:$X_4$,point];
				
				\path (x1) edge (x2);
				\path (x2) edge (x3);
				\path (x3) edge (x4);
				
				\path (y1) edge (x2);
				\path (y2) edge (x3);
				\path (y3) edge (x4);	
				\path (y1) edge[blue,opacity=0.6] (x3);
				\path (y2) edge[blue,opacity=0.6] (x4);
			%	\path (y1) edge[blue,opacity=0.6] (x4);
				
				\path (y1) edge (y2);
				\path (y2) edge (y3);
				\path (y3) edge (y4);	
				\path (y1) edge[red,opacity=0.6,bend left=30] (y3);
				\path (y2) edge[red,opacity=0.6,bend left=30] (y4);
			%	\path (y1) edge[red,opacity=0.6,bend left=40] (y4);

				\path (x1) edge (y1);
				\path (x2) edge (y2);
				\path (x3) edge (y3);
				\path (x4) edge (y4);

				\node (wfull)[rounded corners=10,fill=red,opacity=0.3,fit=(x3), inner sep=0.3cm]{} ; 
				\node (wfull)[rounded corners=10,fill=teal,opacity=0.3,fit=(y4), inner sep=0.3cm]{} ; 
				\node (wfull)[rounded corners=10,fill=violet,opacity=0.3,fit=(x1), inner sep=0.3cm]{} ;    
				\node (wfull)[rounded corners=10,fill=blue,opacity=0.3,fit=(x4), inner sep=0.3cm]{} ;
				\node (wfull)[rounded corners=10,fill=yellow,opacity=0.3,fit=(x2), inner sep=0.3cm]{} ;
				\node (wfull)[rounded corners=10,fill=yellow,opacity=0.3,fit=(y1), inner sep=0.3cm]{} ;	
				\node (wfull)[rounded corners=10,fill=yellow,opacity=0.3,fit=(y2), inner sep=0.3cm]{} ;	
				\node (wfull)[rounded corners=10,fill=yellow,opacity=0.3,fit=(y3), inner sep=0.3cm]{} ;

			\end{tikzpicture}	
			
	}}	\\
	{\scriptsize{}}%
	\noindent\begin{minipage}[t]{1\columnwidth}%
		{\scriptsize{}}%
	\end{minipage}{\scriptsize\par}
\end{figure}

Figure 2.a depicts the exclusion restrictions of the model (\ref{markov1}) in the special case of $\kappa_{YX}=0$ and $\kappa_{YY}=\kappa_{XY}=\kappa_{XX}=1$. We omit $W$, which directly affects all variables, for the purpose of legibility.

The model (\ref{markov1}) implies conditional independence restrictions. These conditions imply Assumption 1 holds for appropriate choices of proxies $V$ and $Z$, and conditioning variables $D$. In particular, in the special case depicted in Figure 2.a, one can confirm that Assumption 1 holds for $V=\{X_1,Y_1\}$, $Z=\{X_3\}$, $D=\{X_2,Y_2,Y_3\}$, $X=X_4$, and $Y=Y_5$.

These choices are indicated on the graph by color-coding the nodes. The nodes for variables in $V$ are colored purple, those in $Z$ are red, $D$ is yellow, $X$ is blue, and $Y$ is green. Note that the choice of $D$ `blocks' (or `$d$-separates') all paths from variables in $V$ to variables in $Z$, ensuring that $V$ and $Z$ are independent conditional on $D$ and $W$ (see Section 1.2.3 of \cite{Pearl2009}).

We first consider the special case of (\ref{markov1}) in which $\kappa_{XY}=0$. Under this restriction $Y_t$ has no direct affect on future treatments. This has important consequences for the choice of proxies because our assumptions generally preclude the possibility that $V$, $Z$, or $D$ are caused by the outcome of interest (directly or indirectly). Thus when $\kappa_{XY}\neq0$ we can only form proxies using past observables whereas when $\kappa_{XY}=0$ we may use future treatments or outcomes as proxies. 

\subsubsection{Future and Past Observables as Proxies}

We begin by considering the special case of $\kappa_{XY}=0$. Under this restriction we may use future treatments as treatment-aligned proxies. If, in addition, $\kappa_{YY}=0$ then we can use future outcomes as treatment-aligned proxies as well. The outcome-aligned proxies consist of lagged outcomes.

\theoremstyle{plain}
\newtheorem*{P333}{Proposition 2.1}
\begin{P333}
	Suppose $\kappa_{XY}=0$. Then for any $t$ the model (\ref{markov1}) implies the following:
	\begin{align}
		Y_{[1:t-1]}&\indep X_{[t:T]}|(X_{[t-\kappa_{XX}:t-1]},W)\label{WD2}\\
		Y_{t}(x_t)&\indep X_{[t:T]}|(X_{[t-\kappa_{XX}:t-1]},W)\label{WD3}
	\end{align}
	If we also have $\kappa_{YY}=0$ then letting $k_X=\max\{\kappa_{XX},\kappa_{YX}-1\}$ we have:
	\begin{align}
		Y_{[1:t-1]}&\indep (X_{[t:T]},Y_{[t+1:T]})|(X_{[t-k_X:t-1]},W)\label{WD4}\\
		Y_{t}(x_t)&\indep (X_{[t:T]},Y_{[t+1:T]})|(X_{[t-k_X:t-1]},W)\label{WD5}
	\end{align}
\end{P333}

Under the conditional independence results from Proposition 2.1 we can compose the treatment aligned proxies $Z$ from future treatments and outcomes, and the outcome-aligned proxies $V$ using past outcomes. Theorem 2.2 does not require that the model (\ref{markov1}) holds. Rather, it requires the conclusions of Propositions 2.1 which are implied by, and are thus weaker than, the model  (\ref{markov1}) with the restriction $\kappa_{XY}=0$.

\theoremstyle{plain}
\newtheorem*{T222}{Theorem 2.1}
\begin{T222}
	Suppose that for each $1\leq t\leq T$, (\ref{WD2}) and (\ref{WD3}) hold. Then for each $1 < t < T$, Assumption $1$ holds with $Y=Y_{t}$, $X=X_{t}$, $V=Y_{[1:t-1]}$, $Z=X_{[t+1:T]}$, and $D=X_{[t-\kappa_{XX}:t-1]}$. If (\ref{WD4}) and (\ref{WD5}) hold and we instead set $D=X_{[t-k_X:t-1]}$, then we can also include $Y_{[t+1:T]}$ in $Z$.
\end{T222}

Theorem 2.1 suggests choices of proxies $V$ and $Z$, and additional controls $D$, so that Assumption 1 holds. The theorem defines these as sets of random variables but we can simply stack the elements to form vectors. The treatments and outcome of interest are those from period $t$. Note that $1<t<T$ and so we require at least three periods of data and we cannot use Theorem 2.1 to identify causal effects in the first and final period.

The controls $D$ are composed of lagged treatments. Consider the choice  $D=X_{[t-\kappa_{XX}:t-1]}$. For all components of $D$ to be included in the observed data, we require that $t> \kappa_{XX}$. To ensure this to holds for at least one choice of $t$, the number of periods for which data is available must weakly exceed $\kappa_{XX}+2$.

Note that Proposition 2.1 and Theorem 2.1 together allow one to form valid proxies without any restrictions on $\kappa_{YX}$. That is, we can allow treatments to directly impact outcomes indefinitely far in the future and still form proxies so that Assumption 1 holds.

In the case of $\kappa_{XY}=0$ only, $V$ can be composed ot $t-1$ lagged outcomes and $Z$ composed of $T-t-1$ future treatments. If many periods of data are available then this may amount to a large number of proxies in $V$ and $Z$. While a larger set of proxies increases the credibility of the completeness conditions (Assumptions 2.i and 2.ii), nonparametric estimation with many proxies may result in noisy estimates. Therefore it may be  preferable to include only a subset of these proxies. This is not problematic for Assumption 1 because if the condition holds for some choice of $V$ and $Z$, then it also holds for subvectors of these variables.

We apply Theorem 2.1 to the model in Figure 2.b. We let $D=X_1$, $V=Y_1$, $Z=X_3$, $X=X_2$, and $Y=Y_2$. As in Figure 2.a, the nodes are color-coded to indicate inclusion of a variable in either $D$, $V$, $Z$, $X$, or $Y$ and we omit $W$ which is a cause of all other variables. Some of the arrows are colored to aid legibility.

\subsubsection{Past Observables as Proxies}

We now consider the case in which lagged outcomes may impact future treatments (that is, $\kappa_{XY}\neq 0$). This generally precludes the use of future observables as proxies. However, lagged observables may still form valid proxies under sufficient restrictions on the model (\ref{markov1}).

\theoremstyle{plain}
\newtheorem*{P331}{Proposition 2.2}
\begin{P331}
	For any $t$, the model (\ref{markov1}) implies the following conditional independence restriction where we define  $k_X=\max\{\kappa_{XX},\kappa_{YX}\}$ and $k_Y=\max\{\kappa_{XY},\kappa_{YY}\}$.
	\begin{equation}
		(X_{[1:t-k_X-1]},Y_{[1:t-k_Y-1]})\indep (X_{[t:T]},Y_{[t:T]})|(X_{[t-k_X:t-1]},Y_{[t-k_Y:t-1]},W)\label{Markov}
	\end{equation}
Moreover, we have:
	\begin{equation}
		Y_{t}(x_t)\indep (X_t,X_{[1:t-\kappa_{YX}-1]},Y_{[1:t-\kappa_{YY}-1]})|(X_{[t-\kappa_{YX}:t-1]},Y_{[t-\kappa_{YY}:t-1]},W)\label{WD1}
	\end{equation}
	
\end{P331}

Given the independence conditions in Proposition 2.2, Theorem 2.2 below states that Assumption $1$ holds for proxies $V$ and $Z$, and conditioning variables $D$, formed from appropriate choices of lagged observables. We focus on the case in which the causal effect of interest is of $X_T$ on $Y_T$, however for earlier periods causal effects can be identified in the same way by ignoring data from subsequent periods.

\theoremstyle{plain}
\newtheorem*{T221}{Theorem 2.2}
\begin{T221}
Suppose for every $1\leq t\leq T$, (\ref{Markov}) and (\ref{WD1}) hold. Then for every $t$ that is strictly greater than $\min\{k_X,k_Y\}+1$ and strictly less than $T-\max\{\kappa_{YX},\kappa_{YY}\}$, Assumption $1$ holds with $Y=Y_T$ and $X=X_T$ so that $Y(x)=Y_T(x_T)$, and with $V$, $Z$, and $D$ defined as follows.
 
  $V=X_{[1:t-k_X-1]}\cup Y_{[1:t-k_Y-1]}$, $Z=X_{[t:T-\kappa_{YX}-1]}\cup Y_{[t:T-\kappa_{YY}-1]}$, and $D=D_1\cup D_2$ where $D_1=X_{[t-k_X:t-1]}\cup Y_{[t-k_Y:t-1]}$ and $D_2=X_{[T-\kappa_{YX}:T-1]}\cup Y_{[T-\kappa_{YY}:T-1]}$.
	
\end{T221}

Theorem 2.2 shows that, under the conditional independence restrictions from Proposition 2.2, we can form valid proxies using past observables. $V$ consists of observables from some periods prior to a period $t$, $Z$ contains  observables from periods after $t$. For there to exist a $1\leq t\leq T$ that satisfies the conditions of the theorem, the number of periods $T$ must be sufficiently large. In particular, we need  $\min\{k_X,k_Y\}+\max\{\kappa_{YX},\kappa_{YY}\}+2<T$.

Note that the vector of conditioning variables $D$ is composed of two sub-vectors. Conditioning on the first of these, $D_1$ (along with $W$), renders $V$ independent of $Z$ and $X$. Conditioning on $D_2$ ensures potential outcomes are jointly independent of $X$ and $Z$.

We apply Theorem 2.1 to the model in Figure 1.c. As in Figures 1.a and 1.b, we omit $W$ from the graph.  Applying Theorem 2.2 with $t=3$, we have $V=X_1$, $Z=X_3$, $D=X_2\cup Y_{[1:3]}$. These choices are color-coded as in Figures 1.a and 1.b. Applying Theorem 2.1 to the setting in Figure 1.a yields the choices for $V$, $Z$, $D$, $X$, and $Y$ described earlier and indicated by the color-coding on that graph. The stronger restrictions in Figure 1.a allow us to use $Y_1$ as an outcome-aligned proxy, whereas in Figure 1.c it must be included as an additional control. 

\subsection{Time-Varying Confounding}

We now consider models in which the confounding factors $W_t$ may vary over time. We generalize (\ref{markov1}) to the following model: 

\begin{align}
	&Y_t=g_{Y,t}(Y_{[t-\kappa_{YY}:t-1]},X_{[t-\kappa_{YX}:t]}, W_{[t-\kappa_{YW}:t]},U_{Y,t})\nonumber\\
	&X_t=g_{X,t}(Y_{[t-\kappa_{XY}:t-1]},X_{[t-\kappa_{XX}:t-1]}, W_{[t-\kappa_{XW}:t]},U_{X,t})\nonumber\\
	&W_t=g_{W,t}(Y_{[t-\kappa_{WY}:t-1]},X_{[t-\kappa_{WX}:t-1]}, W_{[t-\kappa_{WW}:t-1]},U_{W,t})\nonumber\\
	&\{U_{Y,t}\}_{t=1}^T\text{, }\{U_{X,t}\}_{t=1}^T\text{, and } \{U_{W,t}\}_{t=1}^T\text{ are jointly independent.}\label{markov2}
\end{align}

In this model past values of the confounders $W_t$ from multiple times periods may directly affect outcomes and treatments in each period. Moreover past confounding may directly determine future confounders. 

We show that the results of Proposition 2.1 and Theorem 2.1 generalize straight-forwardly to this case. As before we must restrict the model so that past outcomes have no direct affect on future treatments. Similarly, we assume that past values of the outcomes do not directly impact future confounders. That is, we assume both $\kappa_{XY}=0$ and $\kappa_{XW}=0$.

\theoremstyle{plain}
\newtheorem*{P24}{Proposition 2.3}
\begin{P24}
		Suppose $\kappa_{XY}=\kappa_{WY}=0$. Letting ${k}_X=\max\{\kappa_{XX},\kappa_{WX}-1\}$ and defining  ${k}_W=\max\{\kappa_{XW},\kappa_{WW}-1\}$,Then for any $t$ the model (\ref{markov1}) implies the following:
	\begin{align}
		Y_{[1:t-1]}&\indep X_{[t:T]}|(X_{[t-k_{X}:t-1]},W_{[t-k_{W}:t]})\label{WD6}\\
		Y_{t}(x_t)&\indep X_{[t:T]}|(X_{[t-k_{X}:t-1]},W_{[t-k_{W}:t]})\label{WD7}
	\end{align}
	If we also have $\kappa_{YY}=0$ then letting $\bar{k}_X=\max\{\kappa_{XX},\kappa_{WX}-1,\kappa_{YX}-1\}$ and defining  $\bar{k}_W=\max\{\kappa_{XW},\kappa_{WW}-1,\kappa_{YW}-1\}$, we have:
	\begin{align}
		Y_{[1:t-1]}&\indep (X_{[t:T]},Y_{[t+1:T]})|(X_{[t-\bar{k}_X:t-1]},W_{[t-\bar{k}_W:t]})\label{WD8}\\
		Y_{t}(x_t)&\indep (X_{[t:T]},Y_{[t+1:T]})|(X_{[t-\bar{k}_X:t-1]},W_{[t-\bar{k}_W:t]})\label{WD9}
	\end{align}
\end{P24}

\theoremstyle{plain}
\newtheorem*{T223}{Theorem 2.3}
\begin{T223}
	Suppose (\ref{WD6}), and (\ref{WD7}) hold for each $1\leq t\leq T$. Then for any $1<t<T$, Assumption $1$ holds with $Y=Y_{t}$, $X=X_{t}$, $V=Y_{[1:t-1]}$, $Z=X_{[t+1:T]}$,  $D=X_{[t-k_{X}:t-1]}$, and $W=W_{[t-k_{W}:t]}$. If (\ref{WD9}) and (\ref{WD9}) hold and we instead set $D=X_{[t-\bar{k}_X:t-1]}$ and  $W=W_{[t-\bar{k}_W:t]}$, then we can also include $Y_{[t+1:T]}$ in $Z$.
\end{T223}

We apply Theorem 2.3 to the model in Figure 3.a. The choices of $V$, $Z$, $D$, $X$, and $Y$ are the same as in Figure 2.b. The nodes are color coded as before with the addition that the node for $W_2$ is colored orange to indicate that this variable plays the role of $W$ in Assumption 1.

In the case of time-varying confounders, the informativeness of the proxies is less evident. $W$ in Theorem 2.3 is composed of confounders from various periods, and these may not all directly cause or be caused by all elements of $V$ and $Z$. Of particular concern is that the choices of $D$, $Z$, $V$, and $W$ in Theorem 2.3 may lead to a necessary violation of Assumption 2.  Consider an element $W_s$ of $W$ and let $W_{-s}$ denote the other elements of $W$. Assumption 2.i cannot hold if $W_s\indep Z|(X_t,D,W_{-s})$, and similarly for Assumption 2.ii.\footnote{Strictly speaking Assumption 2 can hold only in the trivial case in which $W_s$ is non-random given $X_t$, $D$, and $W_{-s}$.}

However, we can rule out this possibility so long as we assume that the confounders are sufficiently persistent, that is, $\kappa_{WW}$ is sufficiently large.  Consider the case in Theorem 2.3 in which $Z$ is composed of future treatments. If $\kappa_{WW}>\kappa_{XW}$, then one can show there is a path from each component of $W$ in the directed graph corresponding to (\ref{markov2}) to each component of $Z$, that is `unblocked' by  $D$, $X_t$, and $W_{-s}$. This means that generically, each element of $Z$ is statistically associated with each element of $W$ even controlling for $D$, $X_t$, and the other elements of $W$. The same holds for each element of  $V$. The same is true when $Z$ includes future treatments if we also assume that $\kappa_{WW}>\kappa_{YW}$. Note that these conditions on are sufficient to rule out necessary violations of Assumption 2 but they may not be necessary.

In Figure 2.a, we see that $V=Y_1$ is generically associated with $W_2$, even controlling for $D$ and $X_2$, because both are directly impacted by $W_1$. $Z=X_3$ is generically associated with $W_2$ even after controlling for $D$ and $X_2$ because $W_2$ impacts $W_3$ which is a direct cause of $X_3$.

Unfortunately, the violation of Assumption 2 described above precludes an extension of the case in which proxies are composed only of lagged observables to the setting with time-varying confounding.

%The analysis in Section 2.1.2 can also be generalized to the case of time-varying confounding. However, this setting presents problems for the proxy informativeness conditions in Assumption 2. In order to select valid proxies $V$ and $Z$ using only lagged observables, we must condition on some sub-history of the confounders $W_{[a,b]}$ so that $V$ is rendered independent of $Z$, and also a sub-history $W_{[c,d]}$ so that $Z$ is independent of $Y_T(x_T)$. Apart from in some special cases, $V$ will be independent of $W_{[c,d]}$ conditional on $W_{[a,b]}$, $D$, and $X_T$. This implies that either $W_{[c,d]}$ is a deterministic function of $W_{[a,b]}$ (in which case we are back to the nonparametric fixed effects world) or else the completeness condition Assumption 2.ii fails.

\begin{figure}[h]
	\caption{Time-Varying Confounding}
	\centering
	\subfloat[$\kappa_{XX}=\kappa_{WX}=\kappa_{WW}=1$, $\kappa_{XY}=\kappa_{XW}=\kappa_{WY}=0$, $\kappa_{YX}=\kappa_{YY}=\kappa_{YW}=2$]{
		
		\resizebox{!}{75pt}{%
			
			\begin{tikzpicture}
				% nodes %

				\node (y1) at (1,2) [label=above:$Y_1$,point];			
				\node (y2) at (3,2) [label=above:$Y_2$,point];
				\node (y3) at (5,2) [label=above:$Y_3$,point];
				
				\node (x1) at (0,0) [label=below:$X_1$,point];			
				\node (x2) at (2,0) [label=below:$X_2$,point];
				\node (x3) at (4,0) [label=below:$X_3$,point];
				
				\node (w1) at (-0.5,1) [label=above:$W_1$,point];			
				\node (w2) at (1.5,1) [label=above:$W_2$,point];
				\node (w3) at (3.5,1) [label=above:$W_3$,point];

				\path (w1) edge (w2);
				\path (w2) edge (w3);

				\path (w1) edge (x1);
				\path (w2) edge (x2);
				\path (w3) edge (x3);

				\path (w1) edge (y1);
				\path (w2) edge (y2);
				\path (w3) edge (y3);

				\path (x1) edge (x2);
				\path (x2) edge (x3);

				\path (x1) edge (w2);
				\path (x2) edge (w3);

				\path (x1) edge[teal,opacity=0.6,bend right=10] (y2);
				\path (x2) edge[teal,opacity=0.6,bend right=10] (y3);
				
				\path (x1) edge[teal,opacity=0.6] (y3);
				
				\path (w1) edge[blue,opacity=0.5] (y3);
				\path (w1) edge[blue,opacity=0.5] (y2);
				\path (w2) edge[blue,opacity=0.5] (y3);	
				
				\path (y1) edge (y2);
				\path (y2) edge (y3);	
				\path (y1) edge[red,opacity=0.6,bend left=30] (y3);
				
				\path (x1) edge (y1);
				\path (x2) edge (y2);
				\path (x3) edge (y3);

				\matrix [draw,below right,ampersand replacement=\&] at (6.5,2) {
					\node[shape=circle, fill=violet,opacity=0.3,label=right:$V$] {};\&
					\node[shape=circle,fill=red,opacity=0.3,label=right:$Z$] {};\&
					\node[shape=circle,fill=yellow,opacity=0.3,label=right:$D$] {}; \\
					\node[shape=circle, fill=blue,opacity=0.3,label=right:$X$] {};\&
					\node[shape=circle, fill=teal,opacity=0.3,label=right:$Y$] {};\&
					\node[shape=circle, fill=orange,opacity=0.3,label=right:$W$] {};\\
				};
				
				\begin{pgfonlayer}{bg}  	
					
					\node (wfull)[rounded corners=10,fill=blue,opacity=0.3,fit=(x2), inner sep=0.3cm]{} ; 
					\node (wfull)[rounded corners=10,fill=teal,opacity=0.3,fit=(y2), inner sep=0.3cm]{} ; 
					\node (wfull)[rounded
					corners=10,fill=violet,opacity=0.3,fit=(y1), inner sep=0.3cm]{} ;    
					\node (wfull)[rounded corners=10,fill=red,opacity=0.3,fit=(x3), inner sep=0.3cm]{} ;
					\node (wfull)[rounded corners=10,fill=yellow,opacity=0.4,fit=(x1), inner sep=0.3cm]{} ;
					\node (wfull)[rounded corners=10,fill=orange,opacity=0.4,fit=(w2), inner sep=0.3cm]{} ;	
					
				\end{pgfonlayer}
				
			\end{tikzpicture}

	}}\\
	{\scriptsize{}}%
	\noindent\begin{minipage}[t]{1\columnwidth}%
		{\scriptsize{}}%
	\end{minipage}{\scriptsize\par}
\end{figure}

\section{Estimation and Inference}

In this section we describe our estimation and inference procedures. The key step in estimation corresponds to penalized
sieve minimum distance (PSMD) estimation (see \cite{Cheng} and \cite{Chen2015}). Inference is based on the multiplier bootstrap (see for example \citet{Belloni2015}).
Our methods can be applied in panel settings or to cross-sectional
data. To emphasize this generality we return to the notation in Section
1 in which we suppress time subscripts.

Let $\{(Y_{i},X_{i},Z_{i},V_{i},D_{i})\}_{i=1}^{n}$ be a random iid sample of $n$
observations of the variables $Y$, $X$, $Z$, $V$, and $D$. In the panel case, $Y_i$ and $X_i$ should be understood to come from one fixed period $t$. Our estimation method is of the sieve-type, and so we need to specify some vectors of basis functions. For each
$n$ let  $\phi_{n}(V,X,D)$ be a vector of transformations of $V$, $X$, and $D$. The practitioner estimates conditional means $g_i$,  $\pi_{n,i}$, and $\alpha_n$ which are defined by:
\begin{align*}
	g_i & =E[Y_i|Z_i,X_i,D_i]\\
	\pi_{n,i} & =E[\phi_{n}(V_i,X_i,D_i)|Z_i,X_i,D_i]\\
	{\alpha}_{n}(x_1,x_2,d)&=E[\phi_n(V_i,x_1,d)|X_i=x_2,D_i=d]
\end{align*}

The objects above can be estimated by fitted values from  nonparametric regression. Many methods are available, for example local-linear
regression, or series least-squares. We describe a particular procedure later in this sub-section.

Note that $\pi_{n,i}$ is a vector of the same length as $\phi_{n}(v,x,d)$ and in general we must perform a separate regression in order to estimate each of its components. However, we can reduce the number of regressions if $\phi_n$ has a multiplicative structure. Suppose that $\phi_{n}(v,x,d)$ is of the form $\rho_{n}(v)\otimes\chi_n(x,d)$ where `$\otimes$' is the Kronecker product, then we have:
\[
\pi_{n,i} =E[\rho_{n}(V_i)|Z_i,X_i,D_i]\otimes\chi_{n}(X_i,D_i)
\]
 Thus we need only perform one regression for each element of $\rho_n(v)$ rather than for each element of $\phi_n(v,x,d)$. The same applies for estimation of ${\alpha}_{n}(x_1,x_2,d)$.

Having obtained estimates $\hat{g}_i$, $\hat{\pi}_{n,i}$, and $\hat{\alpha}_n$ of $g_i$, $\pi_{n,i}$, and $\alpha_n$, the researcher evaluates
a vector of coefficients $\hat{\theta}$. These coefficients minimize the penalized least-squares objective below:  
\begin{equation}
	\hat{\theta}=\arg\min_{\theta}\frac{1}{n}\sum_{i=1}^{n}\big(\hat{g}_{i}-\hat{\pi}_{n,i}'\theta\big)^{2}+\lambda_{0}\|\theta\|^2\label{eq:penalreg}
\end{equation}
$\lambda_{0}$ is a positive
scalar penalty parameter that may change with the same size and controls the degree of regularization in the second stage. In our empirical applications we set $\lambda_0=\sqrt{n}$. The ridge regression problem has the closed-form solution $(\hat{\Sigma}_{\hat{\pi}}+\lambda_0 I)^{-1}\frac{1}{n}\sum_{i=1}^n \hat{\pi}_{n,i} \hat{g}$, where $I$ is the identity matrix and  $\hat{\Sigma}_{\hat{\pi}}$ is equal to $\frac{1}{n}\sum_{i=1}^n \hat{\pi}_{n,i}\hat{\pi}_{n,i}'$. An estimate of the conditional average potential outcome  is then given by:
\begin{equation}
	E[Y(x_1)|X=x_2,D=d]\approx\hat{\alpha}_{n}(x_1,x_2,d)'\hat{\theta}\label{eq:resulti-1}
\end{equation}

A researcher may be interested in average potential outcomes conditional only on a function of $X_i$ and $D_i$. In order to achieve this one can replace $\alpha_n(x_1,x_2,d)$ with the mean of $\phi_n(V_i,x_1,D_i)$ conditional on that function of $X_i$ and $D_i$. For example, in order to estimate $E[Y(x_1)|X=x_2]$ we can replace $\hat{\alpha}_{n}(x_1,x_2,d)$ with $\hat{\alpha}_{n}(x_1,x_2)$ which estimates the regression function below:
\[{\alpha}_{n}(x_1,x_2)=E[\phi_n(V_i,x_1,D_i)|X_i=x_2]\]

Suppose we are interested in the unconditional mean of potential outcomes $E[Y(x)]$. In this case we can replace $\hat{\alpha}_n(x_1,x_2,d)$ with $\hat{\alpha}_{n}(x)$ an estimate of $\alpha_n(x)=E[\phi_n(V,x,D)]$.  We can simply use the sample mean:
\[
\hat{\alpha}_n(x)=\frac{1}{n}\sum_{i=1}^n \phi_n(V_i,x,D_i)
\]
Then $\hat{\alpha}_{n}(x)'\hat{\theta}$ is an estimate of the average potential outcome $E[Y(x)]$.

\subsection{First-Stage Regressions}

In our empirical applications we use series ridge regression to estimate $g_i$, $\pi_{n,i}$, and $\alpha_n$. Some of our asymptotic results pertain to this particular choice of regression method. We assume that $\phi_n$ has the multiplicative structure defined earlier in this section. We must specify additional basis functions. Let $\zeta_{n}(Z,X,D)$ be a  vector of transformations of $Z$, $X$, and $D$ and likewise for $\psi_n(Z,X,D)$. We let $\zeta_{n,i}=\zeta_{n}(Z_i,X_i,D_i)$ and similarly for $\psi_{n,i}$, $\rho_{n,i}$ and $\chi_{n,i}$. In addition, take $\rho_{n,i,k}$ to be the $k$-th component of the vector $\rho_{n,i}$. Define regression estimates $\hat{\beta}_g$, $\hat{\beta}_{\pi,k}$, and $\hat{\beta}_{\alpha,k}$ as follows:

\begin{align}
	\hat{\beta}_g & =\arg\min_{\beta} \frac{1}{n}\sum_{i=1}^n  (y_i-\psi_{n,i}'\beta)^2 +\lambda_{g}\|\beta\|\label{greg}\\
	\hat{\beta}_{\pi,k} & =\arg\min_{\beta} \frac{1}{n}\sum_{i=1}^n (\rho_{n,i,k}- \zeta_{n,i}'\beta)^2+\lambda_{\pi}\|\beta\|\label{pireg}\\
	\hat{\beta}_{\alpha,k} & =\arg\min_{\beta} \frac{1}{n}\sum_{i=1}^n (\rho_{n,i,k}- \chi_{n,i}'\beta)^2+\lambda_{\alpha}\|\beta\|\label{alphareg}
\end{align}
$\lambda_g$, $\lambda_{\pi}$, and $\lambda_{\alpha}$ are penalty parameters. In our empirical applications we set these to zero. Stack the vectors $\hat{\beta}_{\pi,k}$ into a matrix $\hat{\beta}_{\pi}$ and similarly for $\hat{\beta}_{\alpha,k}$. Our estimates of $g_i$, $\pi_{n,i}$, and $\alpha_n$ are then:
\begin{align*}
	\hat{g}_i&=\pi_{n,i}'\hat{\beta}_g\\
	\hat{\pi}_{n,i}&=(\hat{\beta}_{\pi}'\zeta_{n,i})\otimes\chi_{n,i}\\
	\hat{\alpha}_{n}(x_1,x_2,d)&=\big(\hat{\beta}_{\alpha}'\chi_{n}(x_2,d)\big)\otimes\chi_{n}(x_1,d)
\end{align*}

\subsection{Inference}

In order to perform inference we specify a multiplier bootstrap procedure. Let $Q_{b,i}$ for each $i=1,...,n$ and $b\in\{1,...,B\}$
be iid standard exponential random variables that are independent of the data.\footnote{We follow (\citet{Belloni2015}) and use the standard exponential, but  $Q_{b,i}$ may have any distribution with mean and variance both equal to $1$ and $\max_{1\leq i\leq n }|Q_{b,i}|\precsim_{p}ln(n)$.}

For each $b$ we re-estimate $g_i$, $\pi_{n,i}$, and $\alpha_i$ using the method in the previous subsection, but weighting the contribution of the $i^{th}$ observation by $Q_{b,i}$ in each summation.
 
For example, consider $\hat{g}_i$, our estimate of $g_i$ defined in the previous subsection. We define the $b^{th}$ bootstrap estimate of $g_{i}$ by $\hat{g}_{b,i}=\psi_{n,i}'\hat{\beta}_{g,b}$ where $\hat{\beta}_{g,b}$ is given below:
\[
\hat{\beta}_{g,b}=\big(\frac{1}{n}\sum_{i=1}^n Q_{b,i}\psi_{n,i}\psi_{n,i}'+\lambda_g  I\big)^{-1}\frac{1}{n}\sum_{i=1}^n Q_{b,i}\psi_{n,i}y_{i}
\]
We define the $b^{th}$ bootstrap estimates of $\pi_{n,i}$ and ${\alpha}_{n}$, denoted by $\hat{\pi}_{b,n,i}$ and $\hat{\alpha}_{b,n}$, similarly, again replacing averages in the formulas with weighted averages.

Let $\hat{\theta}_{b}$ solve the weighted, penalized, least squares objective below:
\[
	\hat{\theta}_b=\arg\min_{\theta}\frac{1}{n}\sum_{i=1}^{n}Q_{b,i}\big(\hat{g}_{b,i}-\hat{\pi}_{b,n,i}'\theta\big)^{2}+\lambda_{0}\|\theta\|^2
\]
 Thus we obtain a bootstrap sample $\{\hat{\alpha}_{b,n}(x_{1},x_{2},d)'\hat{\theta}_{b}\}_{b=1}^B$. We can form a bootstrap sample that corresponds to estimates of say, the average potential outcome  $E[Y(x)]$, similarly. We obtain bootstrap analogues of the estimator of  $E[\phi_n(V,x_1,D)]$, denoted $\hat{\alpha}_{n}(x)$ earlier in this section. Again we achieve this simply by replacing averages with weighted averages in the formula. We thus obtain a bootstrap sample  $\{\hat{\alpha}_{b,n}(x)'\hat{\theta}_{b}\}_{b=1}^B$. 

\subsubsection{Uniform Confidence Bands}

We can use the bootstrap sample to form uniform confidence bands. A uniform confidence
band is a collection of intervals, in our case, one for each value $x_1$, $x_2$, and $d$ in a set $\mathcal{S}$. Such a band is asymptotically valid if \textbf{every} interval contains the value of the function at the corresponding $x_1$, $x_2$, and $d$ with probability approaching the desired level. We consider $1-a$-level intervals of the form:
\[
\bigg[\hat{\alpha}_{n}(x_{1},x_{2},d)'\hat{\theta}-\frac{\hat{\sigma}(x_{1},x_{2},d)}{\sqrt{n}}\hat{c}_{1-a},\hat{\alpha}_{n}(x_{1},x_{2},d)'\hat{\theta}+\frac{\hat{\sigma}(x_{1},x_{2},d)}{\sqrt{n}}\hat{c}_{1-a}\bigg]
\]
$\hat{\sigma}(x_{1},x_{2},d)/\sqrt{n}$ is an estimate of the standard deviation of $\hat{\alpha}_{n}(x_{1},x_{2},d)'\hat{\theta}$. In practice we use the pointwise standard deviation of the bootstrap sample as our estimate of $\hat{\sigma}(x_{1},x_{2},d)/\sqrt{n}$. $\hat{c}_{1-a}$ is a uniform critical value that is equal to the smallest
scalar $c>0$ that satisfies the inequality below:
\[
\frac{1}{B}\sum_{b=1}^{B}1\big\{\sup_{(x_{1},x_{2},d)\in\mathcal{S}}\big|\frac{\hat{\alpha}_{b,n}(x_{1},x_{2},d)'\hat{\theta}_{b}-\hat{\alpha}_{n}(x_{1},x_{2},d)'\hat{\theta}}{\hat{\sigma}(x_{1},x_{2},d)/\sqrt{n}}\big|\leq c\big\}\geq1-a
\]

Again, we can adapt the procedure above to perform inference on say, $E[Y(x)]$ or $E[Y(x_1)|X=x_2]$,  simply by replacing $\hat{\alpha}_{b,n}(x_{1},x_{2},d)$ with a bootstrap estimate of $E[\phi_n(V,x,D)]$ or  $E[\phi_n(V,x_1,D)|X=x_2]$.

\subsubsection{Specification Testing}

The application of our approach to panel models, as detailed in Section 2, requires a researcher to take a stance on the degree of Markovian dependence. Generally speaking, stronger restrictions on the degree of Markovian dependence allow for more precise estimates because the researcher has more proxies to choose from and needs to condition on a smaller set of observables.

In order to inform these modeling choices we suggest a Hausman specification test (\cite{Hausman1978}). Given two alternative sets of modeling assumptions, the researcher constructs appropriate estimates $\hat{\alpha}_{n}(x_{1},x_{2},d)'\hat{\theta}$ and $\tilde{\alpha}_{n}(x_{1},x_{2},d)'\tilde{\theta}$ as specified in the previous subsections. We suppose that the first of these estimates is consistent under weaker conditions than the second. Thus the goal is to test the null hypothesis that both sets of assumptions hold against the alternative that the stronger conditions fail. It suffices to test the null that the difference between the probability limits of the two estimators is zero. 

In some cases the two specification to be tested may require two different choices of additional controls $D$. In this case researchers can instead compare corresponding estimates $\hat{\alpha}_{n}(x_{1},x_{2})'\hat{\theta}$ and $\tilde{\alpha}_{n}(x_{1},x_{2})'\tilde{\theta}$ of $E[Y(x_1)|X=x_2]$. In this case one proceeds precisely as we describe below but using these estimates and their bootstrap analogues in place of estimates of $E[Y(x_1)|X=x_2,D=d]$.

In order to carry out the test, a researcher constructs uniform confidence bands for the difference between the two estimates. We denote this difference by $\hat{\Delta}(x_1 ,x_2 ,d)$. That is:
\[
\hat{\Delta}(x_1 ,x_2 ,d)=\hat{\alpha}_{n}(x_{1},x_{2},d)'\hat{\theta}-\tilde{\alpha}_{n}(x_{1},x_{2},d)'\tilde{\theta}
\]

In order to construct confidence bands, the researcher obtains a bootstrap sample $\{\hat{\alpha}_{b,n}(x_{1},x_{2},d)'\hat{\theta}_{b}\}_{b=1}^B$ corresponding to the estimator $\hat{\alpha}_{n}(x_{1},x_{2},d)'\hat{\theta}$, as described earlier in this section. Then, using \textbf{the same} exponential weights, the researcher constructs a bootstrap sample $\{\tilde{\alpha}_{b,n}(x_{1},x_{2},d)'\tilde{\theta}_{b}\}_{b=1}^B$ that corresponds to the estimator $\tilde{\alpha}_{n}(x_{1},x_{2},d)'\tilde{\theta}$.

For a level $1-a$ test, the researcher calculates a critical value $\hat{c}_{1-a}$ is a uniform critical value that is equal to the smallest
scalar $c>0$ that satisfies the inequality below:
\[
\frac{1}{B}\sum_{b=1}^{B}1\big\{\sup_{(x_{1},x_{2},d)\in\mathcal{S}}\big|\frac{\hat{\alpha}_{b,n}(x_{1},x_{2},d)'\hat{\theta}_{b}-\tilde{\alpha}_{b,n}(x_{1},x_{2},d)'\tilde{\theta}_{b}}{\hat{\sigma}_{\Delta}(x_{1},x_{2},d)/\sqrt{n}}\big|\leq c\big\}\geq1-a
\]
$\hat{\sigma}_{\Delta}(x_{1},x_{2},d)/\sqrt{n}$ is an estimate of the pointwise variance of $\hat{\Delta}(x_1 ,x_2 ,d)$ and can itself be obtained by the pointwise variance of the bootstrap estimates of this quantity.

Uniform confidence bands are then given below. The test rejects if zero is not contained within the bands at some point.
\[
\bigg[\hat{\Delta}(x_{1},x_{2},d)-\frac{\hat{\sigma}_{\Delta}(x_{1},x_{2},d)}{\sqrt{n}}\hat{c}_{1-a},\hat{\Delta}(x_{1},x_{2},d)+\frac{\hat{\sigma}_{\Delta}(x_{1},x_{2},d)}{\sqrt{n}}\hat{c}_{1-a}\bigg]
\]

\section{Asymptotic Analysis}

In this section we provide statistical guarantees for the empirical methods in Section 3. A notable departure from existing analyses of PSMD estimators is that we do not require that there exists a smooth solution to either of the conditional moment restrictions in Theorem 1 that motivate our estimation method. This is important in our setting because solutions $\gamma$ and $\varphi$ to the equations in Theorem 1 lack  structural interpretations and need not be unique. This contrasts with analysis in NPIV settings in which case one solution to the relevant conditional moment restriction is the object of interest and has a clear structural interpretation.

A key ingredient in our asymptotic analysis is summarized in Theorem 4.1 below which establishes `well-posedness' of the estimation problem. Recall that Theorem 1.1 provides two alternative characterizations of the conditional average potential outcome based on the solutions to conditional moment restrictions. Theorem 4.1 shows that if the moment condition in 1.1.i has a solution, then the characterization of the object of interest in 1.1.ii is well-posed, and vice versa. For simplicity we omit additional confounders $D$ but the theorem extends straight-forwardly to incorporate such covariates (we prove this more general version in the appendix).

\newtheorem*{Th12}{Theorem 4.1 (Well-Posedness)}
\begin{Th12}
	Suppose there are no additional covariates $D$ and Assumptions 1-3 hold for $x=x_1,x_2$, then:
	
	i. If (\ref{eq2}) holds with $E[\varphi(Z)^{2}|X=x_1]^{1/2}\leq c<\infty$, then for any ${\gamma}$ with $E[\gamma(V)^{2}|X=x_1]<\infty$:
\begin{align*}
	 |E[Y(x_{1})|X=x_{2}]-E[{\gamma}(V)|X=x_{2}]|
	\leq  cE\big[\big(E[Y-{\gamma}(V)|Z,X]\big)^{2}\big|X=x_{1}\big]^{1/2}
\end{align*}

ii. If (\ref{eq1}) holds with $E[\gamma(V)^{2}|X=x_1]^{1/2}\leq c<\infty$, then for any ${\varphi}$ with $E[\varphi(Z)^{2}|X=x_1]<\infty$:
\begin{align*}
	& |E[Y(x_{1})|X=x_{2}]-E[Y{\varphi}(Z)|X=x_{1}]|\nonumber \\
	\leq & cE\big[\big(E[{\varphi}(Z)|V,X]-\frac{f_{V|X}(V|x_2)}{f_{V|X}(V|X)}\big)^{2}\big|X=x_{1}\big]^{1/2}
\end{align*}
\end{Th12}

To understand the implications of this result, consider Theorem 4.1.i. Suppose we find a function $\gamma$ that solves an empirical analogue of the moment restriction (\ref{eq2}) and then estimate its conditional mean $E[\gamma(V)|X=x_2]$. $\gamma$ will not satisfy the population moment restriction exactly, rather there is some error which is measured by the expectation on the right hand side of the inequality in 4.1.i. Theorem 4.1.i states that the difference between $E[\gamma(V)|X=x_2]$ and the object of interest is no greater than a factor $c$ times this error. $c$ here is the norm of a solution $\varphi$ to the moment condition (\ref{eq1}) in Theorem 1.1.i. In sum, the error in our estimate of the object of interest cannot be much greater than the error in the population moment restriction.

In our case $\phi(v,x_1)'\hat{\theta}$ is our $\gamma(v)$ and $\hat{\alpha}_n(x_1,x_2)'\hat{\theta}$ estimates its conditional mean. Theorem 4.1 shows that for good estimation of the conditional average potential outcome, $\gamma$ need not be close to an exact solution to the moment condition $\gamma_0$ (recall there may not be a unique solution), instead we only need that $\gamma$ approximately satisfies the population moment restriction. We can achieve this under weaker conditions than would be required for consistent estimation of a solution  $\gamma_0$.

In the mathematical literature `well-posedness' is usually understood, at least in part, to mean that the solutions to an integral equation is insensitive to perturbations in the function on the right-hand side. Consider Theorem 4.1.i and define a function $g(Z)$. Suppose  $E[\gamma(V)|Z,X=x_1]=g(Z)$ almost surely. Note that this condition can be understood as an integral equation. Now consider another function $\tilde{g}$ and let $\tilde{\gamma}$ satisfy $E[\tilde{\gamma}(V)|Z,X=x_1]=\tilde{g}(Z)$. From the proof of Theorem 4.1.i we see that if $g$ and $\tilde{g}$ are close in the sense that  $E[|g(Z)-\tilde{g}(Z)|^2|X=x_1]^{1/2}$ is small, then ${\gamma}$ and $\tilde{\gamma}$ are close in terms of the semi-norm  $|E[\gamma(V)-\tilde{\gamma}(V)|X=x_1]|$. In other words, a small perturbation to the function $g$ has only a small effect on the solution $\gamma$. 

Well-posedness in the sense summarized in Theorem 4.1 allows us to establish statistical guarantees without any restrictions on the `sieve measure of ill-posedness' (\cite{Chen2015}). Estimation of the `structural function' in NPIV models is typically not well-posed. While a solution $\gamma_0$ to the restriction (\ref{eq2}) is analogous to a structural function in NPIV, our interest is not in a solution $\gamma_0$ itself (which may not be unique) but rather, a linear functional thereof. \cite{Severini2012} note that estimation of a linear functional of the structural function in NPIV may be well-posed, and indeed existence of a solution to (\ref{eq1}) implies that the relevant condition in \cite{Severini2012} holds in our setting.

\subsection{Consistency}

We now state the assumptions under which we establish consistency of our estimator. It is helpful to introduce some additional notation. Recall that in Section 3 we use vectors of basis functions $\psi_{n,i}$, $\rho_{n,i}$, $\zeta_{n,i}$, and $\chi_{n,i}$. It is convenient to define the vector of product basis functions $\kappa_{n,i}=\zeta_{n,i}\otimes\chi_{n,i}$. We let $k_{\psi,n}$ be the length of $\psi_{n,i}$, $k_{\rho,n}$ the length of $\rho_{n,i}$, and similarly for $k_{\zeta,n}$, $k_{\chi,n}$, and $k_{\kappa,n}$.  We define  $\Sigma_{\psi,n}=E[\psi_{n,i}\psi_{n,i}']$, $\Sigma_{\rho,n}=E[\rho_{n,i}\rho_{n,i}']$, and similarly for the other basis functions. We define reduced-form residuals as follows:
\begin{align*}
	\epsilon_{i}&=Y_i-g_i\\
	\upsilon_{n,i}&=\rho_{n,i}-E[\rho_{n,i}|Z_{i},X_{i},D_{i}]\\
	u_{n,i}&=\rho_{n,i}-E[\rho_{n,i}|X_{i},D_{i}]
\end{align*}
  
Throughout we let `$\|\cdot\|$' denote the Euclidean norm of a vector and the Euclidean operator norm of a matrix.  In addition, we define the semi-norms $\|\cdot\|_{L_2}$ and $\|\cdot\|_{n}$ as follows. If $\delta$ is a vector-valued function of $Z$, $X$, and $D$ then we let
$\|\delta\|_{L_2}^2=E[\|\delta(Z_i,X_i,D_i)\|^2]$ and $\|\delta\|_{n}^2=\frac{1}{n}\sum_{i=1}^n\|\delta(Z_i,X_i,D_i)\|^2$. For a vector-valued function $\delta$ and vector $\beta$ of the same length, we let $\|\delta'\beta\|_{L_2}^2=E[\|\delta(Z_i,X_i,D_i)'\beta\|^2]$, and similarly for the norm $\|\cdot\|_n$.

In order to specify the approximation properties of our sieve basis functions we must introduce spaces of smooth functions. In particular, we let $\Lambda^k_s(c)$ denote the set of H{\"o}lder smoothness class $s$ functions on $\mathbb{R}^k$ with semi-norm at most $c$. For example,  $\Lambda^1_1(c)$ is the set of Lipschitz continuous functions on the real line with Lipschitz constant at most $c$. A formal definition can be found in section B.3 of the appendix. We let $\dim(X,D)$ denote the  sum of the lengths of vectors $X$ and $D$, and similarly for other collections of observables.

Finally, we let $\mu_n$ denote the smallest eigenvalue of $\Sigma_{\pi}=E[\pi_{n,i}\pi_{n,i}']$. In our setting the reciprocal of $\mu_n$ is analogous to the sieve measure of ill-posedness. As suggested above, we are able to guarantee consistency without placing any restrictions on this quantity, however it shows up in our inference results.

\theoremstyle{definition}
\newtheorem*{A51}{Assumption 4.1 (First Stage)}
\begin{A51}
For sequences $r_{g,n}$, $r_{\pi,n}$, $\bar{r}_{\pi,n}$, and $\bar{r}_{\alpha,n}$ all $o(1)$, i. $\|\hat{g}-g\|_{n}=O_p(r_{g,n})$,  ii. for any sequence nonrandom sequence $\beta_n$ with $\|\beta_n\|\leq 1$,  $\|(\hat{\pi }_n-\pi_n)'\beta_n\|_{n}=O_p(r_{\pi,n})$ iii. $\|\hat{\pi}_n-\pi_n\|_{L_2}=O_p(\bar{r}_{\pi,n})$, and iv. $\sup_{(x_1,x_2, d)\in \mathcal{S}}\|\hat{\alpha}_n(x_1,x_2,d)-\alpha_n(x_1,x_2,d)\|=O_p(\bar{r}_{\alpha,n})$.
\end{A51}

Assumption 4.1 assumes certain convergence rates for the first stage nonparametric regression estimates. Rates of convergence for nonparametric regression estimators can be found in the literature, for example in (\cite{Belloni2015}). In Lemmas C.3, C.4, and C.5 in the appendix we derive convergence rates for the first-stage estimators detailed in Section 3 under primitive conditions.

\theoremstyle{definition}
\newtheorem*{A52}{Assumption 4.2 (Bases)}
\begin{A52}
	i. The eigenvalues of $\Sigma_{\psi,n}$, $\Sigma_{\zeta,n}$, $\Sigma_{\chi,n}$, $\Sigma_{\rho,n}$, $\Sigma_{\phi,n}$, and $\Sigma_{\kappa,n}$, are bounded above and below away from zero uniformly over $n$.
	 ii. $\|\psi_{n,i}\|\leq \xi_{\psi,n}$,  $\|\rho_{n,i}\|\leq \xi_{\rho,n}$, $\|\kappa_{n,i}\|\leq \xi_{\kappa,n}$, and  $\|\chi_{n,i}\|\leq \xi_{\chi,n}$. 
	 
	 iii. For each $s>0$ there is a sequence $\ell_{\rho,n}(s)\to 0$ so that for any  $\delta\in\Lambda_{s}^{\dim(V)}(c)$:
	 \[
	 \inf_{\beta\in\mathbb{R}^{k(n)}}E\big[(\delta(V)-\rho_{n}(V)'\beta\big)^{2}\big]^{1/2}\leq c\ell_{\rho,n}(s)
	 \]
	 
	 iv.
	For each $s>0$ there is a sequence $\ell_{\psi,n}(s)\to 0$ so that for any   $\delta\in\Lambda_{s}^{\dim(X,Z,D)}(c)$,
	 letting $\beta$ minimize $E\big[(\delta(X,Z,D)-\psi_{n}(X,Z,D)'\beta\big)^{2}\big]$ we have for all $x$, $z$, and $d$:
	\[
	\big|\delta(x,z,d)-\psi_{n}(x,z,d)'\beta\big|\leq c\ell_{\psi,n}(s)
	\]
	The same holds with $\zeta$ in place of $\psi$.
	
	v. \textbf{Either} $X$ and $D$ have a finite discrete support and $\chi_{n}(X,D)$ is a vector of binary indicators (one for each possible value of X and D), \textbf{or} For each $s>0$ there is a sequence $\ell_{\chi,n}(s)\to 0$ so that for any   $\delta\in\Lambda_{s}^{\dim(X,D)}(c)$,
	letting $\beta$ minimize $E\big[(\delta(X,D)-\chi_{n}(X,D)'\beta\big)^{2}\big]$ we have:
	\[
	\big|\delta(x,d)-\chi_{n}(x,d)'\beta\big|\leq c\ell_{\chi,n}(s)
	\]

\end{A52}
\theoremstyle{definition}
\newtheorem*{A53}{Assumption 4.3 (Densities)}
\begin{A53}
	i. $V$ admits a probability density conditional on each value of $Z$, $X$, and $D$, that is bounded above and below away from zero on the the support of $V$. ii. Either $X$ and $D$ have a finite discrete support or they have a joint density that is bounded above and below away from zero on their rectangular joint support.
\end{A53}
\theoremstyle{definition}
\newtheorem*{A54}{Assumption 4.4 (Smoothness)}
\begin{A54}
	With probability $1$, i. the function that maps $v$ to $ \frac{f_{V|XDZ}(v|X_i,D_i,Z_i)}{f_{V}(v)}$ is an element of $\Lambda_{s}^{\dim(V)}(c)$,  ii.$\frac{f_{V|XDZ}(V_i|\cdot,\cdot,\cdot )}{f_{V}(V_i)}\in\Lambda_{s}^{\dim(Z,X,D)}(c)$, and 
	iii. $g\in\Lambda_{s}^{\dim(Z,X,D)}(c)$.

\end{A54}
\theoremstyle{definition}
\newtheorem*{A55}{Assumption 4.5 (Sieve Growth and Penalty)}
\begin{A55}
		i. $\frac{\xi_{\psi,n}^{2}ln(k_{\psi,n})}{n}\prec1$,
	ii. $\frac{\xi_{\chi,n}^{2}ln(k_{\chi,n})}{n}\prec1$,
 iii. $\frac{\xi_{\zeta,n}^{2}ln(k_{\zeta,n})}{n}\prec1$, iv. $\frac{\xi_{{\kappa},n}^{2}ln(k_{\kappa,n})}{n}\prec1$

\end{A55}

\newtheorem*{A46}{Assumption 4.6 (Existence)}
\begin{A46}
	i. There is a finite constant $c$ so that for each $x$ and $d$ in the joint support of $X$ and $D$, (\ref{eq2}) has a solution $\gamma(\cdot,x,d)$ with $E[\gamma(Z,x,d)^{2}|X=x,D=d]\leq c$.
	ii. There is a finite constant $c$ so that for each $(x_1,x_2,d)\in\mathcal{S}$, (\ref{eq1}) has a solution $\varphi(\cdot,x_1,x_2,d)$ so that  $E[\varphi(Z,x_1,x_2,d)^{2}|X=x_1,D=d]\leq c$.
	
\end{A46}
Assumption 4.2.i and 4.2.ii are standard. 4.2.i can be established under primitive conditions (see e.g., \cite{Belloni2015}). The rates in 4.2.ii are readily available in the literature for most basis functions used in practice. For many popular bases the supremum of the norm of the vector of basis functions grows at the same rate as the square-root of the number of functions, so for example $\xi_{\psi,n}\precsim\sqrt{k_{\psi,n}}$ (again, see \cite{Belloni2015}). Assumptions 4.2.iii-4.2.v specify the rate at which the basis functions can
approximate smooth functions. Precise bounds for particular
basis functions can be found in the approximation literature (see, for example, 
\citet{DeVore1993}). For many commonly used bases we would have  $\ell_{\psi,n}(s)\precsim k_{\psi,n}^{-s/\dim(Z,X,D)}$ and similarly for  $\ell_{\rho,n}$ and $\ell_{\chi,n}$ with the number of basis functions and dimensions adjusted accordingly.

Assumption 4.3 is a standard regularity condition on joint probability densities. Assumption 4.4 imposes that some reduced-form objects be smooth so that we can approximate them using sieve basis functions. Smoothness of $g$ in its $Z$ argument follows from 4.5.i and the existence of a $\gamma$ that satisfies the conditions of Theorem 1.1.ii. Assumption 4.5 restricts the rate at which the numbers of basis functions
can grow. This assumption allows us to apply Rudelson's matrix law
of large numbers (\citet{Rudelson1999}).

Assumption 4.6 states that the conditional moment restrictions in Theorem 1.1 admit solutions for various choices of $x_1$, $x_2$, and $d$, and moreover, that the norms of these solutions are uniformly bounded. We provide primitive conditions and further discussion in Appendix A.

The assumptions do not impose smoothness on any solutions $\gamma$ or $\varphi$ to the moment conditions in Theorem 1.1.  This is important because these functions are not unique and lack a clear structural interpretation. This presents a challenge because, in effect, we find an approximate solution $\gamma$ that is a linear combination of the basis functions $\phi_n$. However, for consistency we only need that $g_i$ is well approximated by a linear combination of the components of $\pi_{n,i}$. Lemma 4.1 below shows that such an approximation result is attainable without imposing smoothness on $\gamma$ or $\varphi$.

\theoremstyle{plain}
\newtheorem*{L39}{Lemma 4.1}
\begin{L39} 
	Suppose that for each $x$ and $d$ in the joint support of $X$ and $D$, Assumptions 1-3 hold and Assumptions 4.2, 4.3, 4.4, and 4.6.i hold. Then there is a sequence $\{\theta_{n}\}_{n=1}^{\infty}$
	with $\|\theta_{n}\|\precsim1$ and a finite constant $c$ so that:
	\[
	|g(z,x,d)-\pi_{n}(z,x,d)'\theta_{n}|\leq cr_{\theta}
	\]
	Where $r_{\theta,n}=\ell_{\rho,n}(s)$ if $X$ and $D$ have finite discrete 
	support and otherwise $r_{\theta,n}=\xi_{\chi,n}\ell_{\rho,n}(s)+\ell_{\chi,n}(1)^{-\bar{s}}$
	with $\bar{s}=\frac{\min\{s,1\}}{\min\{s,1\}+1}$.
\end{L39}

Theorem 4.2 establishes a rate of convergence for the
estimator in terms of the first stage convergence rates. This rate does not depend on any sieve measure of ill-posedness and suggests the estimator is consistent under fairly weak restrictions on the penalty $\lambda_0$.

\theoremstyle{plain}
\newtheorem*{Th31}{Theorem 4.2 (Convergence)}
\begin{Th31}
	Suppose the conditions of Lemma 4.1 hold along with Assumptions 4.5, and 4.6.ii. Assume $k_{\chi,n}\ell_{\zeta}(s)^{2}\precsim\bar{r}_{\pi,n}^{2}$ and  $\bar{r}_{\pi,n}^{2}/\lambda_{0}\to0$.  Then for
	\textbf{any} first-stage estimators $\hat{g}_{i}$, and $\hat{\alpha}_{n,i}$, 
	and any estimator $\hat{\pi}_{n,i}$ of the form $\hat{\pi}_{n,i}=\hat{\beta}_{\pi}'\kappa_{n,i}$ for some $\hat{\beta}_{\pi}$, we have:
	\begin{align*}
		&\sup_{(x_1,x_2, d)\in \mathcal{S}}\big|E[Y(x_{1})|X=x_{2},D=d]-\hat{\alpha}_{n}(x_{1},x_{2},d)'\hat{\theta}\big|\\
	\precsim_p&	(1+\frac{\bar{r}_{\alpha,n}+\bar{r}_{\pi,n}\bar{\xi}_{\kappa,n}^{1/2}}{\lambda_{0}^{1/2}})(\lambda_{0}^{1/2}+r_{\pi,n}+r_{g,n}+r_{\theta,n})
	\end{align*}
Where $r_{\theta,n}$ is as defined in Lemma 4.1.
\end{Th31}

The rate in Theorem 4.2 suggests consistency is achieved so long as $\lambda_0$ converges to zero but not too quickly compared to the convergence rates of the first-stage nonparametric regression estimates. The condition that $\bar{r}_{\pi,n}^{2}/\lambda_{0}\to0$ suggests that $\lambda_0$ must shrink strictly more slowly than $n^{-1}$. We can choose $\lambda_0$ to optimize the rate in Theorem 4.2 and this yields the rate below.
\[
	\bar{r}_{\pi,n}\bar{\xi}_{\kappa,n}^{1/2}+r_{g,n}+r_{\theta,n}+\bar{r}_{\alpha,n}+r_{\pi,n}
\]

Under standard conditions, nonparametric regression estimates converge more slowly than $n^{-1/2}$. Thus the rate in Theorem 4.2 is strictly slower than the parametric rate, even in the case of discrete $X$ and $D$. Theorem 4.3 in the next subsection shows that for certain choices of first-stage estimators it is possible to refine the rate in Theorem 4.3.  This requires that the first-stage estimates are `under-smoothed', that is, the first-stage sieve dimensions are chosen so that the bias decreases strictly more quickly than the variance. In addition, to refine the rates we utilize restrictions on $\mu_n$ which corresponds to a sieve measure of ill-posedness.

\subsection{Asymptotic Normality}

We now establish asymptotic normality of our estimator and validity
of the bootstrap inference procedure in Section 3.2. The results in
this subsection pertain to the version of our estimator that uses
the first-stage series ridge regressions as specified in Section 3.1.
In the case of continuous treatments, the relevant notion of asymptotic
normality is approximation by a sequence of Gaussian processes.

In addition, we provide potentially tighter convergence rates than
those in Theorem 4.2 when the first-stage is carried out using series
ridge regression. By restricting our attention to this particular
choice of first-stage regression method, we can carefully disentangle
the first-stage estimation errors into zero-mean components and bias.
With sufficient under-smoothing, the first-stage bias becomes second-order, 
which allows for faster convergence of the second-stage estimates.

We use Assumption 4.7 below to establish asymptotic linearity of our estimator. To be more precise, after appropriate re-scaling, the condition
helps to ensure that the estimation error is  approximately equal to a re-scaled sample
average of zero-mean iid random variables. This requires stronger restrictions on the rates at which
the penalty parameter $\lambda_{0}$ goes to zero and the rate at
which the dimension of the sieve spaces grow. 

Assumption 4.7 refers to quantities $\bar{l}_{\alpha,n}$, $\bar{l}_{g,n}$,
and $\bar{l}_{\pi,n}$, which are the rates of the linearization errors
for the first-stage estimates $\hat{\alpha}_{n}$, $\hat{g}_{i}$,
and $\hat{\pi}_{n,i}$ respectively. For more details on what these
quantities represent see Lemmas C.3, C.4, and C.5 in the appendix.
In these lemmas we establish the rates below under Assumptions 4.2,
4.4, and 4.5:
\begin{align*}
	\bar{l}_{\alpha,n} & =\sqrt{\frac{\xi_{\chi,n}^{2}ln(k_{\chi,n})k_{\chi,n}}{n}}+\sqrt{n}(\ell_{\mathcal{\chi},n}(s)+\lambda_{\alpha})\\
	\bar{l}_{g,n} & =\sqrt{\frac{\xi_{\psi,n}^{2}ln(k_{\psi,n})k_{\psi,n}}{n}}+\sqrt{n}(\ell_{\psi}(s)+\lambda_{g})\\
	\bar{l}_{\pi,n} & =\sqrt{\frac{\xi_{\zeta,n}^{2}ln(k_{\zeta,n})k_{\zeta,n}}{n}}+\sqrt{n}(\ell_{\zeta}(s)+\lambda_{\pi})
\end{align*}

Assumption 4.7 also depends on rates of convergence $\bar{r}_{\alpha,n}$
and $\bar{r}_{\pi,n}$ which satisfy Assumptions 4.1.iii and 4.1.iv.
In Lemmas C.4 and C.5 in the appendix we provide formulas for $\bar{r}_{\alpha,n}$
and $\bar{r}_{\pi,n}$  in terms
of model primitives under Assumptions 4.2, 4.4, and 4.5. These rates could perhaps be tightened under
additional smoothness restrictions on the basis functions using similar
analytical techniques to those in \cite{Belloni2015}.

\newtheorem*{A56}{Assumption 4.7 (Linearization)} \begin{A56}
	
	i. $E[\epsilon_{i}^{2}|X_{i},D_{i},Z_{i}]$ is almost surely bounded
	above and below away from zero and likewise for $\|E[\upsilon_{n,i}\upsilon_{n,i}'|Z_{i},X_{i},D_{i}]\|$
	and $\|E[u_{n,i}u_{n,i}'|X_{i},D_{i}]\|$ uniformly over $n$. ii.
	$\sqrt{\frac{\xi_{\kappa,n}^{2}(k_{\zeta,n}+k_{\psi,n})ln(k_{\kappa,n})}{n}}\leq r_{s,n}$
	and $\sqrt{\frac{k_{\rho,n}k_{\chi,n}(\xi_{\kappa,n}^{2}k_{\zeta,n}+\xi_{\psi,n}^{2}k_{\psi,n})}{n}}\leq r_{s,n}$
	where $r_{s,n}\prec0$, iii. $r_{l,n}=\sqrt{ln(n)}(\bar{l}_{\alpha,n}+\bar{l}_{g,n}+\bar{l}_{\pi,n})\prec1$,
	iv. $\sqrt{ln(n)n}r_{\theta,n}\prec1$ v. $r_{\lambda,n}=\bar{\xi}_{\kappa,n}\sqrt{\frac{\lambda_{0}n}{1+\mu_{n}/\lambda_{0}^{2}}}\prec1$,
	and vi. $r_{\mu,n}=\sqrt{\frac{(\bar{r}_{\alpha,n}^{2}+\bar{\xi}_{\kappa,n}^{2}\bar{r}_{\pi,n}^{2})(k_{\psi,n}+k_{\zeta,n})/\lambda_{0}}{1+\mu_{n}/\lambda_{0}^{2}}}\prec1$.
\end{A56}
	
	Assumption 4.7.i is standard. Assumption 4.6.ii limits the rate at
	which the number of basis functions may grow with the sample size.
	This condition ensures that some sample second moment matrices converge
	sufficiently quickly to their population counterparts. 4.7.iii requires
	that the linearization errors in the first-stage estimates shrink
	sufficiently quickly. This condition necessarily requires that the
	bias in the first stage estimates converges faster than $n^{-1/2}$
	which means the first-stage estimates are under-smoothed. The $\sqrt{ln(n)}$
	can be dropped if we only require a Gaussian approximation for our
	estimator and not for its bootstrap analogue. Assumption 4.7.iv ensures
	the second-stage sieve approximation error is negligible. Again, the
	$\sqrt{ln(n)}$ is only required for Gaussian approximation of the
	bootstrap estimator. Assumption 4.7.v requires that $\lambda_{0}$
	goes to zero sufficiently quickly with the sample size so that the
	asymptotic bias due to regularization in the second stage is negligible.
	
	Assumption 4.7.vi is more complex. It involves $\lambda_{0}$ as well
	as rates $\bar{r}_{\alpha,n}$ and $\bar{r}_{\pi,n}$. To see why
	we require Assumption 4.7.vi, recall that $\hat{\theta}$ in the second
	stage of our procedure is defined by $\hat{\theta}=(\hat{\Sigma}_{\hat{\pi}}+\lambda_0 I)^{-1}\frac{1}{n}\sum_{i=1}^{n}\hat{\pi}_{n,i}\hat{g}_{i}$.
	In order to linearize our estimator we must approximate $(\hat{\Sigma}_{\hat{\pi}}+\lambda_0 I)^{-1}$
	with its population counterpart $(\Sigma_{\pi,n}+\lambda_{0}I)^{-1}$.
	The resulting approximation error is small when $\bar{r}_{\pi,n}$
	is small, i.e., when $\hat{\pi}_{n,i}$ is close to $\pi_{n,i}$ in
	an appropriate sense. This approximation error also depends on both
	$\mu_{n}$ and $\lambda_{0}$, because for small $\mu_{n}$ and $\lambda_{0}$
	the matrix $\Sigma_{\pi,n}+\lambda_{0}I$ is close to singular.
	
	For both 4.7.v and 4.7.vi to hold, $\mu_{n}$ cannot go to zero too
	rapidly. If $\mu_{n}$ goes sufficiently quickly to zero then there
	is no sequence of penalty parameters $\lambda_{0}$ under which both
	conditions hold. The following condition on $\mu_{n}$ is sufficient
	to ensure a sequence of penalty parameters exists so that 4.7.v and
	4.7.vi are both satisfied:
	\[
	\frac{\mu_{n}^{2}}{(\bar{r}_{\alpha,n}^{2}+\bar{r}_{\pi,n}^{2})^{3}(k_{\psi,n}+k_{\zeta,n})^{3}/n}\to\infty
	\]
	
	The rate at which $\mu_{n}$ goes to zero depends on the number of
	basis functions in $\rho_{n}$. To ensure $\mu_{n}$ does not shrink
	too quickly to zero, the dimension of $\rho_{n}$ must not grow too
	rapidly. The rate at which $\mu_{n}$ decreases with $\rho_{n}$ is
	akin to a restriction on the sieve measure of ill-posedness. Thus
	there is a marked distinction between our asymptotic inference results
	and the consistency results in Theorem 4.2. For consistency we do
	not require any restriction on the rate at which $\mu_{n}$ goes to
	zero and thus we need not place any restrictions on a sieve measure
	of ill-posedness. The reason for this difference is that consistency
	only requires $n^{-1/2}r_{\lambda,n}\prec0$ and $n^{-1/2}r_{\mu,n}\prec0$,
	which can hold even if we set $\mu_{n}=0$ in the formulas for $r_{\lambda,n}$
	and $r_{\mu,n}$. We also require 4.7.v and 4.7.vi to achieve an $n^{-1/2}$-rate
	of convergence for our estimator in the case of discrete $X$ and
	$D$.
	
	Finally, the following assumptions allow us to apply Yuriskii's coupling
	(see e.g., \cite{Pollard2001} or \cite{Belloni2015}) to achieve uniform approximations of both our estimator and its
	bootstrap analogue by Gaussian processes.
	
	\newtheorem*{A57}{Assumption 4.8 (Normality)} \begin{A57}
		
		i. $E[\epsilon_{i}^{3}|x_{i},d_{i},z_{i}]$ is bounded above, ii.
		\[\frac{(k_{\psi,n}+k_{\zeta,n}+k_{\chi,n})(k_{\psi,n}\xi_{\psi}+k_{\zeta,n}\xi_{\zeta,n}\xi_{\rho,n}+k_{\chi,n}\xi_{\chi,n}\xi_{\rho,n})}{n^{1/2}/ln(n)}\prec1\]
		iii. $\sqrt{ln(n)}\|\psi_{n,i}\|\leq\xi_{\psi,n}$, $\sqrt{ln(n)}\|\rho_{n,i}\|\leq\xi_{\rho,n}$,
		$\sqrt{ln(n)}\|\kappa_{n,i}\|\leq\xi_{\kappa,n}$, and  $\sqrt{ln(n)}\|\chi_{n,i}\|\leq\xi_{\chi,n}$.
		
	\end{A57} 
	
	Assumption 4.8.i imposes a bound on conditional third moments. This
	is a standard condition that allows us to apply an appropriate central
	limit theorem under growing dimensions. Assumption 4.8.ii restricts
	the rate at which the sieve spaces grow with the sample size. This
	condition ensures that the third moment of a vector of zero-mean random
	variables, multiplied by the length of the vector, grows slowly enough
	so that we may apply Yuriskii's coupling. 4.8.iii strengthens 4.2.ii
	so that our earlier linearization arguments extend to the bootstrap
	estimator. It ensures that, not only does $\xi_{\psi,n}$ upper-bound
	the essential supremum of $\|\psi_{n,i}\|$, but that $\ensuremath{\max}_{1\leq i\leq n}\sqrt{Q_{b,i}}\psi_{n,i}\precsim_{p}\xi_{\psi,n}$
	which allows us to linearize the bootstrap analogue of the estimator.
	This follows similar arguments to those in \cite{Belloni2015}.
	
	To state the theorem below we define a vector-valued function $s_{n}$.
	First define a zero-mean random vector $\eta_{n,i}$ as follows:
	
	\[
	\eta_{n,i}=\begin{pmatrix}\Sigma_{\pi,\lambda_{0}}^{-1/2}E[\pi_{n,i}\psi_{n,i}']\Sigma_{\psi,n}^{-1}\psi_{n,i}\epsilon_{i}\\
		-\Sigma_{\pi,\lambda_{0}}^{-1/2}E[\pi_{n,i}\kappa_{n,i}'](\Sigma_{\zeta,n}^{-1}\zeta_{n,i})\otimes(\tilde{\theta}_{n}'\upsilon_{n,i})\\
		\Sigma_{\chi,n}^{-1}\chi_{n,i}\otimes(\tilde{\theta}_{n}'u_{n,i})
	\end{pmatrix}
	\]
	
	In the above, $\Sigma_{\pi,\lambda_{0}}=\Sigma_{\pi,n}+\lambda_0 I$.  $\tilde{\theta}_{n}$ is a  $k_{\rho,n}$-by-$k_{\chi,n}$ matrix obtained by partitioning the vector $\theta_{n}$ that satisfies Lemma 4.1 into $k_{\rho,n}$ contiguous length-$k_{\chi,n}$ subvectors and stacking the transposes into a matrix.  Let $\Omega_{n}=E[\eta_{n,i}\eta_{n,i}']$
	and define a vector-valued function $s_{n}$ by:
	\[
	s_{n}(x_{1},x_{2},d)=\Omega_{n}^{1/2}\begin{pmatrix}\Sigma_{\pi,\lambda_{0}}^{-1/2}\alpha_{n}(x_{1},x_{2},d)\\
		\Sigma_{\pi,\lambda_{0}}^{-1/2}\alpha_{n}(x_{1},x_{2},d)\\
		\chi_{n}(x_{2},d)\otimes\chi_{n}(x_{1},d)
	\end{pmatrix}
	\]
	
	Theorem 4.3 below shows that the distribution of the estimation error
	is approximately equal to that of a Gaussian process whose pointwise
	variance is $\|s_{n}(x_{1},x_{2},d)\|^{2}/n$. Moreover, the difference
	between the bootstrap estimator and the original estimator can be
	approximated by an identical Gaussian process that is independent
	of the data (but not the bootstrap weights).
	
	\theoremstyle{plain}
	\newtheorem*{Th33}{Theorem 4.3 (Asymptotic Normality)}
	\begin{Th33}
	Suppose Assumptions 1-3 and 4.2-4.8.ii all hold and the eigenvalues
	of $\Omega_{n}$ are bounded below away from zero uniformly over $n$.
	
	Then  $\|s_{n}(x_{1},x_{2},d)\|\precsim1+\bar{\xi}_{\kappa,n}+\xi_{\chi,n}^{2}$ and there is a sequence of mean-zero Gaussian random vectors $\text{\ensuremath{\mathcal{N}_{n}}}$
	with identity variance covariance matrices so that:
	
	\begin{align*}
		\sup_{(x_1,x_2,d)\in\mathcal{S}}\big|\frac{\hat{\alpha}_{n}(x_{1},x_{2},d)'\hat{\theta}-E[Y(x_{1})|X=x_{2},D=d]}{\|s_{n}(x_{1},x_{2},d)\|/\sqrt{n}}-\frac{s_{n}(x_{1},x_{2},d)'}{\|s_{n}(x_{1},x_{2},d)\|}\mathcal{N}_{n}\big|\prec_{p}1
	\end{align*}

	Moreover, if 4.8.iii also holds, there exists a sequence of zero-mean Gaussian
	random vectors $\mathcal{N}_{n}$ with identity variance-covariance
	matrix which are independent of the data and:
	
	\textbf{
		\begin{align*}
			\sup_{(x_1,x_2,d)\in\mathcal{S}}\big|\frac{\hat{\alpha}_{n}(x_{1},x_{2},d)'\hat{\theta}_{b}-\alpha_{n}(x_{1},x_{2},d)'\hat{\theta}}{\|s_{n}(x_{1},x_{2},d)\|/\sqrt{n}}-\frac{s_{n}(x_{1},x_{2},d)'}{\|s_{n}(x_{1},x_{2},d)\|}\mathcal{N}_{n}\big|\prec_{p}1
		\end{align*}
	}

	\end{Th33}

	Theorem 4.3 shows that the distribution of the estimation error can be well-approximated by a sequence of Gaussian processes. The theorem also provides a convergence rate for  $\|s_{n}(x_{1},x_{2},d)\|$ and thus a rate for the pointwise standard error of the approximating Gaussian process. In fact, this rate, combined with the Gaussian approximation, implies that the estimator converges to the truth at rate $(1+\bar{\xi}_{\kappa,n}+\xi_{\chi,n}^{2})/\sqrt{n}$. If $X$ and $D$ have finite support then $\chi_n$ can be chosen so that both $\bar{\xi}_{\kappa,n}$ and $\xi_{\chi,n}$ are bounded and thus the estimator converges at the parametric rate. As discussed above, this relies crucially on under-smoothing in the first-stage regressions.
	
	In addition, the theorem shows that the distribution of the estimation error can be approximated
	by its bootstrap analogue, that is, by the distribution of  $\hat{\alpha}_{n}(x_{1},x_{2},d)'\hat{\theta}_{b}-\alpha_{n}(x_{1},x_{2},d)'\hat{\theta}$
	where the data (but not the bootstrap weights) are treated as fixed.
	Validity of the bootstrap confidence bands in Section 3.2 follows
	immediately if the set $\mathcal{S}$ is finite. If $\mathcal{S}$ is infinite then asymptotically
	correct uniform coverage requires additional anti-concentration results on the suprema of Gaussian processes. Such results can be found in \citet{Chernozhukov2014}.
	
	Theorem 4.3 extends straight-forwardly
	to linear combinations of different estimators that all satisfy the
	relevant conditions. The only condition that must be strengthened
	is 4.8.ii in order to reflect the fact that the relevant normal random
	vector has greater dimension. Thus the theorem also justifies the
	specification test described in Section 3.2. 

\section{Empirical Applications}

We apply our methodology to real data. In order to emphasize the applicability
of our approach to both cross-sectional and panel models we present
two separate empirical settings. In the first application
we exploit the panel structure of the data to form proxies as suggested in Section 2. In our second application we use cross-sectional
variation to estimate causal effects

\subsection{Structural Engel Curve for Food}

A household's Engel curve for a particular class of good captures
the relationship between the share of the household's budget spent
on that class and the total expenditure of the household. An Engel
curve is `structural' if it captures the effect of an exogenous change
in total expenditure. Imagine an ideal experiment in which the household's
total expenditure is chosen by a researcher using a random number
generator and the household then chooses how to allocate that total
expenditure between different classes of goods. Then the resulting
relationship between the total expenditure and budget share is a structural
Engel curve.

Nonparametric regression of the budget share spent on food and the
total expenditure on certain classes of goods is unlikely to
represent the average structural Engel curve. This is because 
total expenditure is chosen by the household and thus depends upon
the household's underlying financial assets and consumption preferences. Household finances and 
preferences partially determine the share of expenditure that the household allocates to food.

We estimate average structural Engel curves for food eaten at home using data from
the Panel Study of Income Dynamics (PSID). The PSID study follows
US households over a number years and record expenditure on various
classes of goods. We use ten periods of data from the surveys carried
out every two years between 1999 and 2017. We drop all households
whose household heads are not married or cohabiting and drop all households
for which we lack the full ten periods of data, leaving us with $840$ households.
We take as the total expenditure the sums of expenditures on food
(both at home and away from home), housing, utilities, transportation,
education, childcare and health-care.

We apply the approach to identification with fixed-$T$ panels described in Section 2.2, allowing for time-varying preferences and unobserved assets. In this setting the condition that outcomes do not feed back to future confounders nor treatments seems reasonable. While total expenditure on non-durables directly impacts household assets, the proportion of this that is allocated to food does not.

Let $X_{t}$ denote the total expenditure on non-durables in
period $t$ and $Y_t$ the share of this expenditure allocated to food. We have $10$ periods of data and aim to estimate causal effects at time $t=4$. In line with Section 2.2, we set $V=Y_{[1:3]}$ and $Z=X_{[5:10]}$. We take $D=X_{[1:3]}$ which allows for the possibility that total expenditure in some period can directly impact expenditure at most three periods ahead and preferences/financial assets up to four periods ahead.

We apply our method using the first-stage series ridge regression procedures specified in Section 3.3. The vectors of basis functions include all squares and interactions of the variables. First-stage penalty parameters are set to zero and for the second stage penalty $\lambda_0$ we simply use the square root of the sample size.

\begin{figure}[h]
	\caption{Demand for Food}
	\subfloat[Average Engel Curve for Food]{
		
		\includegraphics[scale=0.23]{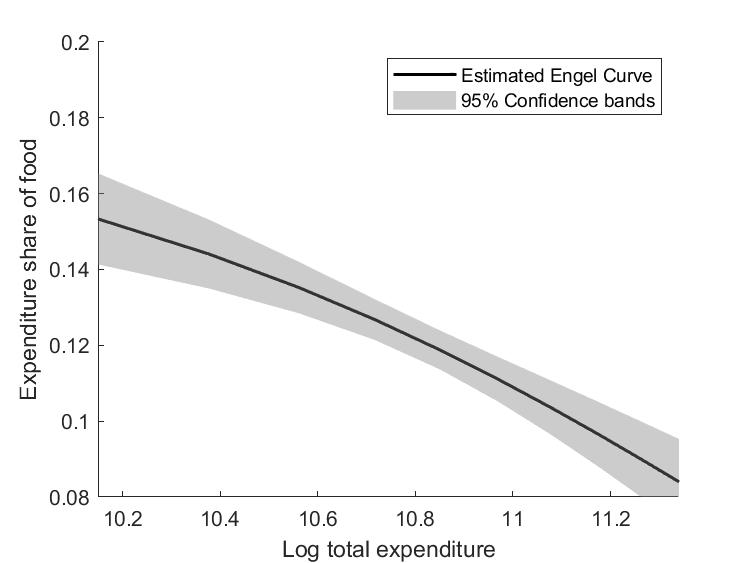}}\subfloat[Specification Test]{
		
		\includegraphics[scale=0.23]{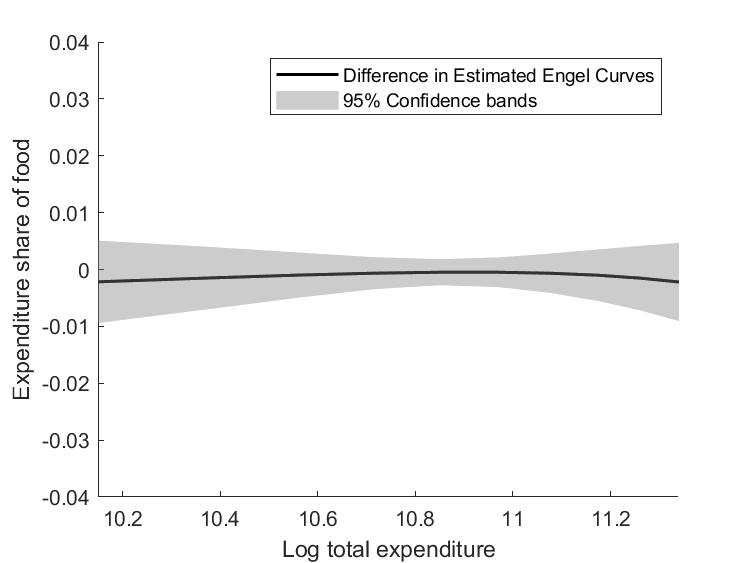}}
	
	{\scriptsize{}}%
	\noindent\begin{minipage}[t]{1\columnwidth}%
		{\scriptsize{}Estimates
			are plotted at 100 points evenly spaced (in levels not logs) between the $10\%$ and $90\%$ quantiles
			of total expenditure. The uniform confidence bands are evaluated using $10,000$ replications of the multiplier bootstrap as detailed in Section 3. The pointwise standard errors were set equal to pointwise standard deviations over the bootstrap replications.}%
	\end{minipage}{\scriptsize\par}
\end{figure}
Figure 4.a plots our nonparametric estimate of the average
 structural Engel curve for food. The figure shows a downward-sloping Engel curve that (with
a log scale for total expenditure) is subtly concave. The downward
slope of the curve suggests that food is a normal good, at least in
aggregate.

The choice of $D$ used for Figure 4.a may be overly conservative. The results in Section 2.3 suggest that under stronger restrictions on the dynamics of the data-generating process (namely $\min\{\kappa_XX,\kappa_{WX}-1\}\leq 1$), we may set $D=X_3$. To test these stronger restrictions we apply the Hausman specification method proposed in Section 3.2.3. The results are shown in Figure 4.b. The uniform confidence bands for the difference between the two alternative estimates contains zero at all values of total expenditure and so we fail to reject these restrictions at the $95\%$ level.

\subsection{Causal Impact of Grade Retention}

\cite{Fruehwirth2016} examine the causal effect of being made to repeat a particular grade level on the cognitive
development of US students. They use data from the ECLS-K panel study
which contains panel data on the early cognitive development of US
children. We use our methods to examine the effect of grade retention
on the cognitive outcomes of children in the 1998-1999 kindergarten
school year using cleaned data available with their paper. Following
\cite{Fruehwirth2016}, we take our outcome variables to be the tests
scores in reading and math when aged approximately eleven.
Also in line with \cite{Fruehwirth2016} our treatments are indicators
for retention in kindergarten, `early' (in first or second grade)
and `late' (in third or fourth grade). The cleaned data from \cite{Fruehwirth2016}
contains only students who are retained at most once in the sample
period and no students who skip a grade.

The ECLS-K dataset contains scores that measure a student's behavioral and social
skills and their scores on a range
of cognitive tests at different ages. To account for the confounding effect of unmeasured ability, 
\cite{Fruehwirth2016} estimate a latent factor model with a particular
structure. They assume that all confounding between grade retention
and potential future cognitive test scores is due entirely to the presence
of three latent factors representing different dimensions of ability. \cite{Fruehwirth2016}
then use test scores to recover the distribution of the latent factors
and their loadings. They assume a particular multiplicative
structure between the factors (which are time-invariant) and time-specific
factor loadings in both their outcome and selection equations.

Like \cite{Fruehwirth2016} we wish to use test-scores to adjust for some latent ability to perform well academically $W$. Our methods allow us to avoid any strong assumptions on the factor structure. We take the treatment to be a three-dimensional vector of binary indicators that signify whether student is held back in kindergarten, early elementary, or late elementary. Note that by defining $X$ in this way we estimate the effects of all three treatments simultaneously.

We let the set of proxies $V$ contain the student's scores tests in kindergarten and $Z$ contains
the test scores from early in elementary school (first or second grade).  Some children are held back a grade in kindergarten and therefore treatment (being held back a grade) may have a causal effect on the proxies in $Z$. This is compatible with Assumption 1 (see Figure 1.b).

In the ECLS-K dataset that we use for our empirical analysis the cognitive and behavioral scores are not shared with the students, parents nor teachers, and thus they should not determine the decision to retain a child.\footnote{This was confirmed by email with the ECLS study director.}, hence there is no causal effect of $V$ on treatments  $X$ nor any causal effect of $V$ on $Y$.

Our sample contains of $1,951$ individuals. $V$ contain the student's scores on three  cognitive and three behavioral tests in kindergarten and $Z$ contain
the scores on these tests from early elementary. Thus we have six proxies in each group. We understand to $W$ capture both underlying cognitive and behavioral characteristics. Given the discussion of Assumption 2, for identification to be plausible we need to believe that there are at most six such characteristics in all.

In this setting the full support condition (Assumption 3) requires the children of all underlying latent cognitive skills and behavioral characteristics have some probability of being held back. To support this assumption consider that children may be held back a year due to some disruption in their family life, because they are among the youngest of their peers, due to a period of illness, and any other of many reasons that are not likely to be very strongly correlated with underlying cognitive and behavioral skills. Note that no student is held back twice we understand the support of $X$ to be equal to the set of three binary indicators that sum up to weakly less than one.

The ECLS-K data also contain a number of additional observables which we include as controls $D$. These include vectors that capture family characteristics, features of each student's school, student charateristics, and the student's age.

\begin{table}[h]
	\caption{Average Effects of Treatment on the Treated}
	\small
	\subfloat[Reading]{
		\begin{tabular}{c|c|c|c|c}
			\multicolumn{1}{c}{} & \multicolumn{4}{c}{\textbf{\uline{Method:}}}\tabularnewline
			& OLS, & OLS, & Proxy & Proxy \tabularnewline
			& no proxy & kindergarten & controls,  & controls,\tabularnewline
			\textbf{\uline{Treatments:}} & scores & scores & linear & PSMD \tabularnewline
			\cline{2-5} \cline{3-5} \cline{4-5} \cline{5-5} 
			Retained kindergarten & -0.17 & -0.09 & -0.05 & -0.01\tabularnewline
			& (0.02) & (0.02) & (0.04) & (0.05)\tabularnewline
			Retained early & -0.23 & -0.14 & -0.13 & -0.08\tabularnewline
			& (0.02) & (0.02) & (0.05) & (0.06)\tabularnewline
			Retained late & -0.13 & -0.04 & -0.03 & -0.02\tabularnewline
			& (0.03) & (0.03) & (0.07) & (0.08)\tabularnewline
	\end{tabular}}
	
	\subfloat[Math]{
		\begin{tabular}{c|c|c|c|c}
			\multicolumn{1}{c}{} & \multicolumn{4}{c}{\textbf{\uline{Method:}}}\tabularnewline
			& OLS, & OLS, & Proxy & Proxy \tabularnewline
			& no proxy & kindergarten & controls,  & controls,\tabularnewline
			\textbf{\uline{Treatments:}} & scores & scores & linear & PSMD \tabularnewline
			\cline{2-5} \cline{3-5} \cline{4-5} \cline{5-5} 
			Retained kindergarten & -0.18 & -0.07 & -0.02 & -0.09\tabularnewline
			& (0.03) & (0.02) & (0.05) & (0.07)\tabularnewline
			Retained early & -0.24 & -0.14 & -0.06 & 0.09\tabularnewline
			& (0.03) & (0.02) & (0.06) & (0.07)\tabularnewline
			Retained late & -0.17 & -0.08 & -0.03 & 0.01\tabularnewline
			& (0.04) & (0.03) & (0.06) & (0.09)\tabularnewline
	\end{tabular}}
	
	\begin{spacing}{0.7}
		{\scriptsize{}}%
		\noindent\begin{minipage}[t]{1\columnwidth}%
			{\scriptsize{}Estimates of the
				average effect of treatment on the treated using various estimation methods. Numbers in parentheses
				are standard errors, calculated for the OLS estimates using the standard heteroskedasticity robust formula, and for the latter two columns using $10,000$ replications of the multiplier bootstrap method detailed in Section 3.}%
		\end{minipage}{\scriptsize\par}
	\end{spacing}
	
\end{table}

Table 2 compares estimates of the average effect of treatment on the treated under different approaches to estimation. In all cases we control for the additional observables $D$ described above. In the first column are linear least-squares estimates of the average treatment effects when we include $D$ and treatments as regressors but no other variables. The estimated effects are all strongly negative and very statistically significant. In the second column, the kindergarten cognitive and behavioral scores are included along with $D$ as regressors in an additive linear specification, note that in every case the estimated negative effects are at least halved in magnitude compared to the case in which the scores are not included.

In the third column we apply our method with linear specifications (i.e., $\rho_n$, $\psi_n$, $\zeta_n$, and $\chi_n$ return their arguments) and with the first-stage penalty parameters set to zero and the second stage penalty set to $\sqrt{n}$. In every case the the estimated ATTs less strongly negative than they are when we simply control for kindergarten scores. Finally, the last column contains the ATTs from the series ridge method specified in Section 3. The vectors of basis functions include squares and interactions of variables. Penalty parameters are set to zero for the first-stage estimates and the second stage penalty $\lambda_0$ is set to $\sqrt{n}$. Compared to the linear case the ATTs for reading are less strongly negative, the only exception being for math scores of children retained in kindergarten.

The results in Table 2 are consistent with the notion that unmeasured ability biases the estimated ATTs downwards. Including kindergarten test scores as controls mitigates some of this bias, and the proxy controls method mitigates the bias further still, resulting in mostly less negative estimated ATTs.

\bibliographystyle{authordate1}
\bibliography{proxybib}

\appendix
%dummy comment inserted by tex2lyx to ensure that this paragraph is not empty

\section{Regularity Conditions}

Below we provide sufficient conditions for existence of solutions
to the conditional moment restrictions in Theorems 1.1.i and 1.1.ii.
The conditions simply adapt Picard's criterion to this setting. Existence
then follows from Theorem 15.18 in \citet{Kress2014}.

\theoremstyle{definition} \newtheorem*{AA1}{Assumption A.1
	(Compact Operator)} \begin{AA1} $V$ and $Z$ admit a joint density
	conditional on $X$ and $D$, denoted by $f_{VZ|XD}$ that is dominated
	by the product of its marginals, and: 
	\[
	\int\int\frac{f_{VZ|XD}(v,z|x_{1},d)^{2}}{f_{V|XD}(v|x_{1},d)f_{Z|XD}(z|x_{1},d)}dvdz<\infty
	\]
\end{AA1} 
Let $L_{2}\big(f_{V|XD}(\cdot|x_{1},d)\big)$ denote the space
of functions $\delta$ that have finite square integral with respect
to the density $f_{V|XD}$, and $L_{2}\big(f_{Z|XD}(\cdot|x_{1},d)\big)$
the space of functions that have finite square integral with respect
to the density $f_{Z|XD}$. Define a linear operator $A_{x_{1},d}:L_{2}\big(f_{V|XD}(\cdot|x_{1},d)\big)\to L_{2}\big(f_{Z|XD}(\cdot|x_{1},d)\big)$
by: 
\[
A_{x_{1},d}[\delta](z)=E[\delta(V)|Z=z,X=x_{1},D=d]
\]

Under  A.1 there exists a unique singular system
$\{(u_{k},v_{k},\mu_{k})\}_{k=1}^{\infty}$ for $A_{x_{1},d}$. $\mu_{k}$
is the $k^{th}$ singular value of $A_{x_{1},d}$. $\{u_{k}\}_{k=1}^{\infty}$
and $\{v_{k}\}_{k=1}^{\infty}$ are sequences of orthonomal basis
functions for $L_{2}\big(f_{V|XD}(\cdot|x_{1},d)\big)$ and $L_{2}\big(f_{Z|XD}(\cdot|x_{1},d)\big)$
respectively and known as the singular functions of the operator $A_{x_{1},d}$.\footnote{See, e.g., \citet{Kress2014} Theorem 15.16 and associated discussion.}
Note the singular system is specific to the given value of $x_{1}$
and $d$.

\theoremstyle{definition} \newtheorem*{A4}{Assumption A.2 (Picard's
	Criterion)} \begin{A4} Assumptions A.1 holds so that $A_{x_{1},d}$
	has a singular system $\{(u_{k},v_{k},\mu_{k})\}_{k=1}^{\infty}$.
	
	i. $v\mapsto\frac{f_{V|XD}(v|x_{2},d)}{f_{V|XD}(v|x_{1},d)}\in L_{2}\big(f_{V|XD}(\cdot|x_{1},d)\big)$
	and: 
	\[
	\sum_{k=1}^{\infty}\frac{1}{\mu_{k}^{2}}E[\frac{f_{V|XD}(V|x_{2},d)}{f_{V|XD}(V|x_{1},d)}u_{k}(V)|X=x_{1},D=d]^{2}\leq c
	\]
	
	ii. Let $g(Z)=E[Y|Z,X=x_{1},D=d]$, then $g\in L_{2}\big(f_{Z|XD}(\cdot|x_{1},d)\big)$
	and: 
	\[
	\sum_{k=1}^{\infty}\frac{1}{\mu_{k}^{2}}E\big[g(Z) v_{k}(Z)\big|X=x_{1},D=d\big]^{2}\leq c
	\]
\end{A4}

Under Assumptions A.1.i and A.1.ii the equations (1.3) and (1.5) respectively
have solutions $\varphi$ and $\gamma$ so that $E[\varphi(Z)^{2}|X=x_{1},D=d]\leq c$
and $E[\gamma(V)^{2}|X=x_{1},D=d]\leq c$. This follows straight-forwardly
from an application of Picard's Theorem to this setting (see Theorem
15.18 in \citet{Kress2014}).

Conditions
of the same form are used elsewhere in the literature, for example
in \citet{Darolles}, and \citet{Miao2018}. Restrictions of this form
are sometimes described as smoothness conditions, for example by \citet{Hall2005}.

A.2.i and A.2.ii each require that generalized the Fourier coefficients go to zero sufficiently quickly. To understand why these conditions may be equated with smoothness,
note that generalized Fourier coefficients are coefficients in the
expansion of a function in a certain orthonormal basis. For example A.ii restricts decay of the coefficients in the orthonormal series expansion below: 
\[g(z)=\sum_{k=1}^{\infty}E[g(Z) v_{k}(Z)\big|X=x_{1},D=d]v_{k}(z)\]

Thus the conditions state
that the coefficients in this expansion decay sufficiently quickly and thus the function $g$ can be well-approximated by linear combinations
of a finite number of basis functions. This might reasonably be described
as a `smoothness' of $g$.

However, it is important to note that a) the basis functions $v_{k}$ are not known a priori, rather they depend upon the unknown linear operator $A_{x_1,d}$, and b) not only do the coefficients have to decay quickly, but they have to decay strictly more rapidly than the singular values $\mu_k$, which again depend on the linear operator.

Generally speaking, smoothness of the kernel of a linear integral operator is associated with rapid decay of the singular values. In our setting the kernel is a joint probability density function. For example, the singular values for a compact linear integral operator with an analytic kernel decay at least exponentially quickly (see 15.20 in \cite{Kress2014} for a more precise statement). We are not aware of a similar result for integral kernels that instead satisfy the H{\"o}lder smoothness restrictions of the kind we impose on joint densities in Section 4.

We now formally state and prove the result described in Semiparametric
Example 1.

\theoremstyle{plain} \newtheorem*{A5}{Proposition A.1 (Semiparametric
	Example 1)} \begin{A5} Suppose that, conditional on $x$, $W$ and
	$Z$ are multivariate Gaussian with a strictly positive definite variance-covariance
	matrix: 
	\[
	\begin{pmatrix}W\\
		Z
	\end{pmatrix}|X=x\sim\mathcal{N}\bigg(\begin{pmatrix}\mu_{W|x}\\
		\mu_{Z|x}
	\end{pmatrix},\begin{pmatrix}\Sigma_{WW|x} & \Sigma_{WZ|x}\\
		\Sigma_{WZ|x}' & \Sigma_{ZZ|x}
	\end{pmatrix}\bigg)
	\]
	
	Suppose that Assumptions 1-3 hold. If $\Sigma_{WW|x_{1}}=\Sigma_{WW|x_{2}}$
	then there is a scalar $\alpha$ and matrix $M$, that do not depend
	on $z$ so that (1.3) is satisfied by: 
	\begin{align*}
		\varphi(z) & =\alpha exp\big(Mz\big)
	\end{align*}
	
	In the more general case where possibly $\Sigma_{WW|x_{1}}\neq\Sigma_{WW|x_{2}}$,
	define the matrices $\Omega=\Sigma_{ZZ|x_{1}}-\Sigma_{WZ|x_{1}}'\Sigma_{WW|x_{1}}^{-1}\Sigma_{WZ|x_{1}}$
	and: 
	\[
	Q=\Omega(\Sigma_{WZ|x_{1}}'\Sigma_{WW|x_{1}}^{-2}\Sigma_{WZ|x_{1}})^{-1}\Sigma_{WZ|x_{1}}'\Sigma_{WW|x_{1}}^{-1}
	\]
	
	Suppose that the matrix $\Omega+Q(\Sigma_{WW|x_{1}}^{-1}-\Sigma_{WW|x_{2}}^{-1})Q'$
	is non-singular. Then there is a scalar $\alpha$ and matrices $M$
	and $P$ that do not depend on $z$ so that: 
	\begin{align*}
		\varphi(z) & =\alpha exp(z'Pz+Mz)
	\end{align*}
\end{A5}

\begin{proof}
	
	We will construct a function $\varphi$ with $E[\varphi(Z)^{2}|X=x_{1}]<\infty$,
	we have: 
	\[
	E[\varphi(Z)|W=w,X=x_{1}]=\frac{f_{W|X}(w|x_{2})}{f_{W|X}(w|x_{1})}
	\]
	
	Then by the reasoning in the proof of Theorem 1.1, under Assumptions
	1-3 it follows that: 
	\[
	E[\varphi(Z)|V=v,X=x_{1}]=\frac{f_{V|X}(v|x_{2})}{f_{V|X}(v|x_{1})}
	\]
	
	Under the assumption of multivariate normality $\frac{f_{W|X}(w|x_{2})}{f_{W|X}(w|x_{1})}$
	is equal to the following:
	
	\begin{align*}
		& \frac{\sqrt{det(\Sigma_{WW|x_{1}})}exp\big(-(w-\mu_{W|x_{2}})'\Sigma_{WW|x_{2}}^{-1}(w-\mu_{W|x_{2}})\big)}{\sqrt{det(\Sigma_{WW|x_{2}})}exp\big(-(w-\mu_{W|x_{1}})'\Sigma_{WW|x_{1}}^{-1}(w-\mu_{W|x_{1}})\big)}\\
		= & cexp\big(w'(\Sigma_{WW|x_{1}}^{-1}-\Sigma_{WW|x_{2}}^{-1})w+2w'(\Sigma_{WW|x_{2}}^{-1}\mu_{W|x_{2}}-\Sigma_{WW|x_{1}}^{-1}\mu_{W|x_{1}})\big)
	\end{align*}
	Where $c$ does not depend on $w$ and has formula given below: 
	\[
	\sqrt{\frac{det(\Sigma_{WW|x_{1}})}{det(\Sigma_{WW|x_{2}})}}exp(\mu_{W|x_{1}}'\Sigma_{WW|x_{1}}^{-1}\mu_{W|x_{1}}-\mu_{W|x_{2}}'\Sigma_{WW|x_{2}}^{-1}\mu_{W|x_{2}})
	\]
	
	We see then that $\int\bigg(\frac{f_{W|X}(w|x_{2})}{f_{W|X}(w|x_{1})}\bigg)^{2}f_{W|X}(w|x_{1})dw$
	is finite if and only if $\Sigma_{WW|x_{1}}^{-1}-2\Sigma_{WW|x_{2}}^{-1}$
	is strictly negative definite.
	
	Thus, under the Gaussianity assumption, the condition (1.3) is equivalent
	to the following integral equation in which $\|\cdot\|$ is the Euclidean
	norm: 
	\begin{align*}
		& \int\varphi(z)\bar{c}exp\big(-\|\Omega^{-1/2}(z-\mu_{Z|x_{1}}-\Sigma_{WZ|x_{1}}'\Sigma_{WW|x_{1}}^{-1}w)\|^{2}\big)dz\\
		= & cexp\big(w'(\Sigma_{WW|x_{1}}^{-1}-\Sigma_{WW|x_{2}}^{-1})w+2w'(\Sigma_{WW|x_{2}}^{-1}\mu_{W|x_{2}}-\Sigma_{WW|x_{1}}^{-1}\mu_{W|x_{1}})\big)
	\end{align*}
	
	Where $\Omega=\Sigma_{ZZ|x_{1}}-\Sigma_{WZ|x_{1}}'\Sigma_{WW|x_{1}}^{-1}\Sigma_{WZ|x_{1}}$
	is strictly positive definite and $\bar{c}=(2\pi)^{-d_{Z}/2}det(\Omega)^{-1/2}$.
	We can write the above more succinctly as: 
	\begin{equation}
		\int\varphi(z)\frac{\bar{c}}{c}exp\big(-l(w,z)\big)dz=1\label{eq:finreff}
	\end{equation}
	Where the function $l(w,z)$ is given by: 
	\begin{align*}
		l(w,z) & =w'Aw+2Bw-2z'\Omega^{-1}\Sigma_{WZ|x_{1}}'\Sigma_{WW|x_{1}}^{-1}w\\
		& +(z-\mu_{Z|x_{1}})'\Omega^{-1}(z-\mu_{Z|x_{1}})
	\end{align*}
	For $A$ and $B$ below: 
	\[
	A=\Sigma_{WW|x_{1}}^{-1}\Sigma_{WZ|x_{1}}\Omega^{-1}\Sigma_{WZ|x_{1}}'\Sigma_{WW|x_{1}}^{-1}+\Sigma_{WW|x_{1}}^{-1}-\Sigma_{WW|x_{2}}^{-1}
	\]
	\[
	B=\mu_{W,x_{2}}'\Sigma_{WW,x_{2}}^{-1}-\mu_{W,x_{1}}'\Sigma_{WW,x_{1}}^{-1}+\mu_{Z,x_{1}}'\Omega_{x_{1}}^{-1}\Sigma_{WZ,x_{1}}'\Sigma_{WW,x_{1}}^{-1}
	\]
	
	With some tedious algebra one can verify that: 
	\begin{align*}
		l(w,z) & =(z-Lw-R)'\Sigma^{-1}(z-Lw-R)\\
		& -z'\Sigma^{-1}z+2R\Sigma^{-1}z+(z-\mu_{Z|x_{1}})'\Omega^{-1}(z-\mu_{Z|x_{1}})
	\end{align*}
	
	The matrices $L$, $\Sigma$, and $R$ in the above can be any matrices
	that solve the following set of linear equations: 
	\begin{align*}
		\Sigma_{WZ|x_{1}}\Omega^{-1}L & =\Sigma_{WW|x_{1}}A\\
		\Sigma_{WZ|x_{1}}\Omega^{-1}\Sigma & =\Sigma_{WW|x_{1}}L'\\
		\Sigma_{WZ|x_{1}}\Omega^{-1}R & =\Sigma_{WW|x_{1}}B'
	\end{align*}
	
	In each case there must exist a solution because $\Sigma_{WZ|x_{1}}$
	has full row rank under Assumption 2. In particular, let the matrix
	$Q$ be the pseudo-inverse of $\Sigma_{WW|x_{1}}^{-1}\Sigma_{WZ|x_{1}}\Omega^{-1}$,
	let $L=QA$, $\Sigma=QAQ'$, and $R=QB'$. It remains to demonstrate
	that $\Sigma$ is non-singular. In the case of $\Sigma_{WW|x_{1}}=\Sigma_{WW|x_{2}}$,
	the first two equations above are satisfied by $L=\Sigma_{WZ|x_{1}}'\Sigma_{WW|x_{1}}^{-1}$
	and $\Sigma=\Omega$ and the latter is non-singular. In other cases
	$\Sigma$ which equals $\Omega+Q(\Sigma_{WW|x_{1}}^{-1}-\Sigma_{WW|x_{2}}^{-1})Q'$,
	is non-singular by supposition. Set $\varphi(z)$ as follows: 
	\begin{align*}
		\varphi(z) & =cdet(\Sigma)^{1/2}det(\Omega)^{-1/2}exp(\mu_{Z|x_{1}}'\Omega^{-1}\mu_{Z|x_{1}})\\
		& \times exp\big(z'(\Omega^{-1}-\Sigma^{-1})z+2(R\Sigma^{-1}-\mu_{Z|x_{1}}'\Omega^{-1})z\big)
	\end{align*}
	This function has the form given in the theorem. Substituting this
	into (\ref{eq:finreff}) the right hand side becomes:
	
	\[
	\int\frac{1}{(2\pi)^{d_{Z}/2}det(\Sigma)^{1/2}}exp\big(-(z+Lw-R)'\Sigma^{-1}(z+Lw-R)\big)dz
	\]
	
	This is the integral of a multivariate normal density and this equals
	$1$, so (\ref{eq:finreff}) is satisfied. In the case of $\Sigma_{WW|x_{1}}=\Sigma_{WW|x_{2}}$,
	$\Sigma=\Omega$ and so $\varphi(z)$ simplifies to: 
	\begin{align*}
		\varphi(z) & =cexp(\mu_{Z|x_{1}}'\Omega^{-1}\mu_{Z|x_{1}})exp\big(2(R-\mu_{Z|x_{1}}')\Omega^{-1}z\big)
	\end{align*}
	
	Note then that $E[\varphi(Z)^{2}|X=x_{1}]<\infty$ if and only if
	the following is strictly negative definite: 
	\begin{align*}
		-\Sigma_{ZZ|x_{1}}^{-1}+2\big(\Omega^{-1}-\big(\Omega+Q(\Sigma_{WW|x_{1}}^{-1}-\Sigma_{WW|x_{2}}^{-1})Q'\big)^{-1}\big)
	\end{align*}
\end{proof}

\section{Proofs}

\subsection{Proofs for Section 1}

\begin{proof}[Proof Theorem 1.i]
	
	Suppose there exists $\varphi$ so that $E[\varphi(Z)^{2}|X=x_{1},D=d]$
	is finite and the following holds for all $v$ in $\mathcal{V}_{x_{1},d}$:
	\[
	E[\varphi(Z)|X=x_{1},D=d,V=v]=\frac{f_{V|XD}(v|x_{2},d)}{f_{V|XD}(v|x_{1},d)}
	\]
	
	By iterated expectations and Assumption 1.ii we get: 
	\begin{align*}
		E[\varphi(Z)|X,D,V]= & E\big[E[\varphi(Z)|X,D,V,W]\big|X,D,V\big]\\
		= & E\big[E[\varphi(Z)|X,D,W]\big|X,D,V\big]
	\end{align*}
	
	By standard properties of densities: 
	\begin{align*}
		\frac{f_{V|XD}(v|x_{2},d)}{f_{V|XD}(v|x_{1},d)} & =\frac{\int f_{V|WXD}(v|w,x_{2},d)f_{W|XD}(w|x_{2},d)dw}{f_{V|XD}(v|x_{1},d)}
	\end{align*}
	
	By Assumption 1.ii, $f_{V|WXD}(v|w,x_{2},d)=f_{V|WXD}(v|w,x_{1},d)$
	and so from the above we get: 
	\begin{align*}
		\frac{f_{V|XD}(v|x_{2},d)}{f_{V|XD}(v|x_{1},d)} & =\frac{\int f_{V|WXD}(v|w,x_{1},d)f_{W|XD}(w|x_{2},d)dw}{f_{V|XD}(v|x_{1},d)}\\
		& =\int f_{W|VXD}(w|v,x_{1},d)\frac{f_{W|XD}(w|x_{2},d)}{f_{W|XD}(w|x_{1},d)}dw\\
		& =E[\frac{f_{W|XD}(W|x_{2},d)}{f_{W|XD}(W|x_{1},d)}|V=v,X=x_{1},D=d]
	\end{align*}
	
	Combining, we get that for all $v\in\mathcal{V}_{x_{1},d}$: 
	\begin{align*}
		& E\big[E[\varphi(Z)|X,D,W]\big|X=x_{1},D=d,V=v\big]\\
		= & E[\frac{f_{W|XD}(W|x_{2},d)}{f_{W|XD}(W|x_{1},d)}|V=v,X=x_{1},D=d]
	\end{align*}
	
	Applying Assumptions 2.ii and 3.i we have for all $w$ in the support
	of $W$: 
	\begin{equation}
		E[\varphi(Z)|X=x_{1},D=d,W=w]=\frac{f_{W|XD}(w|x_{2},d)}{f_{W|XD}(w|x_{1},d)}\label{eq:intexistq1}
	\end{equation}
	
	Note that the step above only requires 2.ii hold for all $\delta$
	so that $\delta(w)$ is of the form $E[\omega(Z)|X=x_{1},D=d_{1},W=w]$
	and with $E[\omega(Z)^{2}|X=x_{1},D=d]$ finite.
	
	Next note that by iterated expectations and Assumption 1.i: 
	\begin{align*}
		& E[Y(x_{1})|X=x_{2},D=d]\\
		= & E\big[E[Y(x_{1})|X,D,W]\big|X=x_{2},D=d\big]\\
		= & E\big[E[Y(x_{1})|X=x_{1},D=d,W]\big|X=x_{2},D=d\big]\\
		= & E\big[E[Y(x_{1})|X,D,W]\frac{f_{W|XD}(W|x_{2},d)}{f_{W|XD}(W|x_{1},d)}\big|X=x_{1},D=d\big]
	\end{align*}
	
	Where the final line follows from properties of densities. Substituting
	(\ref{eq:intexistq1}) we get: 
	\begin{align*}
		& E[Y(x_{1})|X=x_{2},D=d]\\
		= & E\big[E[Y(x_{1})|X,D,W]E[\varphi(Z)|X,D,W]\big|X=x_{1},D=d\big]\\
		= & E\big[E[Y(x_{1})\varphi(Z)|X,D,W]\big|X=x_{1},D=d\big]\\
		= & E[Y(x_{1})\varphi(Z)|X=x_{1},D=d]\\
		= & E[Y\varphi(Z)|X=x_{1},D=d]
	\end{align*}
	
	The second equality holds by Assumption 1.i which implies $Y(x_{1})$
	and $Z$ are independent given $X$ and $D$, the penultimate line
	follows by iterated expectations, and the final line $Y(X)=Y$.
	
\end{proof}

\begin{proof}[Proof Theorem 1.1.ii] Suppose that $E[\gamma(V)^{2}|X=x,D=d]$
	is finite and for all $z$ in $\mathcal{Z}_{x_{1},d}$ we have: 
	\[
	E\big[Y-\gamma(V)\big|X=x_{1},D=d,Z=z\big]=0
	\]
	By iterated expectations and Assumption 1.ii we have: 
	\begin{align*}
		E[\gamma(V)|X,D,Z]= & E\big[E[\gamma(V)|X,D,W,Z]\big|X,D,Z\big]\\
		= & E\big[E[\gamma(V)|X,D,W]\big|X,D,Z\big]
	\end{align*}
	And by iterated expectations and Assumption 1.i we get: 
	\begin{align*}
		& E[Y|X=x_{1},D=d,Z=z]\\
		= & E[Y(X)|X=x_{1},D=d,Z=z]\\
		= & E[E[Y(x_{1})|X,D,W,Z]|X=x_{1},D=d,Z=z]\\
		= & E\big[E[Y(x_{1})|X,D,W]\big|X=x_{1},D=d,Z\big]
	\end{align*}
	Where we have used that $Y(X)=Y$. Combining we get that for all $z$
	in $\mathcal{Z}_{x_{1},d}$: 
	\[
	E\big[E[Y(x_{1})-\gamma(V)|X,D,W]\big|X=x_{1},D=d,Z=z\big]=0
	\]
	Then by Assumptions 2.i and 3.i, for all $w$ in its support: 
	\[
	E[Y(x_{1})|X=x_{1},D=d,W=w]=E[\gamma(V)|X=x_{1},D=d,W=w]
	\]
	Note that for the above we only require 2.i hold for all $\delta$
	so that $\delta(w)$ is of the form $E[\omega(V)|X=x_{1},D=d_{1},W=w]$
	where $E[\omega(V)^{2}|X=x_{1},D=d]$ is finite.
	
	By Assumptions 1.i and 1.ii we then have: 
	\[
	E[Y(x_{1})|X=x_{2},D=d,W=w]=E[\gamma(V)|X=x_{2},D=d,W=w]
	\]
	Taking conditional expectations of both sides and using iterated expectations
	we get: 
	\[
	E[Y(x_{1})|X=x_{2},D=d]=E[\gamma(V)|X=x_{2},D=d]
	\]
	
\end{proof}

\subsection{Proofs for Section 2}

\begin{proof}[Proof of Proposition 2.1]
	
	Consider the directed acyclic graph associated with model (2.3). We
	first show that if $\kappa_{XY}=0$ then the set of random variables
	$X_{[t-\kappa_{XX}:t-1]}\cup\{W\}$ $d$-separates every path between
	a variable in $X_{[t:T]}$ and one in $Y_{[1:t-1]}$. The conditional
	independence in the first part of the proposition then follows by
	Theorem 1.2.4 in \citet{Pearl2009}.
	
	Suppose that the path contains a directed edge from $S\notin X_{[t:T]}$
	towards an element of $X_{[t:T]}$. Then the path contains a fork
	or chain with $S$ in its center. Because $\kappa_{XY}=0$, the set
	we must have $S\in X_{[t-\kappa_{XX}:t-1]}\cup\{W\}$, so the path
	is $d$-separated by $X_{[t-\kappa_{XX}:t-1]}\cup\{W\}$. Any other
	path between $X_{[t:T]}$ and $Y_{[1:t-1]}$ must contain an edge
	originating from an element $X_{s}$ of $X_{[t:T]}$. Because edges
	in the graph only point to contemporaneous or future variables the
	path must contain a collider at some observable in period $s$ or
	later. But $X_{[t-\kappa_{XX}:t-1]}\cup\{W\}$ contains no such elements,
	and thus $d$-separates the path.
	
	To prove the second conditional independence result we replace the
	equation for $Y_{t}$ in model (3.2) with the following:
	
	\[
	Y_{t}(x_{t})=g_{Y,t}(Y_{[t-\kappa_{YY}:t-1]},X_{[t-\kappa_{YX}:t-1]},x_{t},W,U_{t})
	\]
	In addition, we drop each equation for $Y_{s}$ for $s>t$. Since
	$\kappa_{XY}=0$ dropping these variables does not impact the rest
	of the model. Consider the graph associated with the new model. We
	will show that any path between $Y_{t}(x_{t})$ and an element of
	$X_{[t:T]}$, is $d$-separated by $X_{[t-\kappa_{XX}:t-1]}\cup\{W\}$.
	
	Again suppose that the path contains an edge from an element $S\notin X_{[t:T]}$
	to an element of $X_{[t:T]}$. By the same reasoning as before the
	path is $d$-separated by $X_{[t-\kappa_{XX}:t-1]}\cup\{W\}$. Any
	other path must contain an edge that points from some $X_{s}\in X_{[t:T]}$
	to $X_{r}$ or $Y_{r}$ for some $r>s$. Because no edge is directed
	backwards in time, we again conclude that the path contains a collider
	at an observable in a period strictly greater than $t$. and no such
	variable is in $X_{[t-\kappa_{XX}:t-1]}\cup\{W\}$. Thus the path
	is $d$-separated.
	
	Now we consider the case of $\kappa_{XY}=\kappa_{YY}=0$. Let us show
	that $X_{[t-k_{X}:t-1]}\cup\{W\}$ $d$-separates every path between
	a variable in $X_{[t:T]}\cup Y_{[t+1:T]}$ and one in $Y_{[1:t-1]}$.
	First suppose the path contains an edge from an element $S\notin X_{[t:T]}\cup Y_{[t+1:T]}$
	to an element of $X_{[t:T]}\cup Y_{[t+1:T]}$. Then $S\in X_{[t-k_{X}:t-1]}\cup\{W\}$
	and so $X_{[t-k_{X}:t-1]}\cup\{W\}$ $d$-separates the path. By the
	reasoning earlier in the proof, any other path must contain a collider
	that is not in $X_{[t-k_{X}:t-1]}\cup\{W\}$ and so the path is $d$-separated.
	
	For the final statement in the theorem, we again consider the augmented
	model with the equation for $Y_{t}$ replaced by that of the potential
	outcome. However, because $\kappa_{XY}=\kappa_{YY}=0$ we no longer
	need to drop the equations for $Y_{s}$ for $s>t$ for the model to
	remain valid. The relevant $d$-separation then follows by the same
	reasoning as in the previous paragraph. \end{proof}

\begin{proof}[Proof of Proposition 2.2]
	
	Consider the directed acyclic graph associated with model (2.3). We
	first show that the set of random variables $X_{[t-k_{X}:t-1]}\cup Y_{[t-k_{Y}:t-1]}\cup\{W\}$
	$d$-separates every path between a variable in $X_{[1:t-k_{X}-1]}\cup Y_{[1:t-k_{Y}-1]}$
	and one in $X_{[t:T]}\cup Y_{[t:T]}$. The first result in the proposition
	then follows by Theorem 1.2.4 in \citet{Pearl2009}.
	
	Consider a path between a variable in $X_{[1:t-k_{X}-1]}\cup Y_{[1:t-k_{Y}-1]}$
	and one in $X_{[t:T]}\cup Y_{[t:T]}$. Because no edge points to variables
	from prior periods, the path contains an edge from some $S\notin X_{[t:T]}\cup Y_{[t:T]}$
	towards a variable in $X_{[t:T]}\cup Y_{[t:T]}$. Then $S\in X_{[t-k_{X}:t-1]}\cup Y_{[t-k_{Y}:t-1]}\cup\{W\}$
	and so the path is blocked by $X_{[t-k_{X}:t-1]}\cup Y_{[t-k_{Y}:t-1]}\cup\{W\}$.
	
	Now we prove the second result in the proposition. As in the proof
	of Proposition 2.1 we replace the equation for $Y_{t}$ with the equation
	for $Y_{t}(x_{t})$. In addition, we drop the equations for all observables
	from periods strictly greater than $t$ to obtain a valid NSEM. Consider
	a path in the graph for the new model between $Y_{t}(x_{t})$ and
	a an element of $\{X_{t}\}\cup X_{[1:t-\kappa_{YX}-1]}\cup Y_{[1:t-\kappa_{YY}-1]}$.
	Given the absence of variables caused by $Y_{t}(x_{t})$, the path
	must contain an edge pointing from some $S$ towards $Y_{t}(x_{t})$.
	Then $S\in Y_{[t-\kappa_{YY}:t-1]}\cup X_{[t-\kappa_{YX}:t-1]}\cup\{W\}$
	and so the path is $d$-separated by $Y_{[t-\kappa_{YY}:t-1]}\cup X_{[t-\kappa_{YX}:t-1]}\cup\{W\}$.
	The result then follows by Theorem 1.2.4 in \citet{Pearl2009} \end{proof}

\begin{proof}[Proof of Proposition 2.3]
	
	Consider the directed acyclic graph associated with model (2.10).
	To prove the first independence result we show that if $\kappa_{XY}=0$
	then the set of random variables $X_{[t-k_{X}:t-1]}\cup W_{[t-k_{W}:t]}$
	$d$-separates every path between a variable in $X_{[t:T]}$ and one
	in $Y_{[1:t-1]}$. Suppose the path contains an edge from $S\notin X_{[t:T]}\cup W_{[t+1:T]}$
	into an element of $X_{[t:T]}\cup W_{[t+1:T]}$. Because $\kappa_{XY}=\kappa_{WY}=0$,
	we must have $S\in X_{[t-k_{X}:t-1]}\cup W_{[t-k_{W}:t]}$. Thus $X_{[t-k_{X}:t-1]}\cup W_{[t-k_{W}:t]}$
	$d$-separates the path. Now suppose the converse holds and the path
	does not contain an edge from an element outside of $X_{[t:T]}\cup W_{[t+1:T]}$
	into $X_{[t:T]}\cup W_{[t+1:T]}$. Then the path must contain an edge
	from an element of $X_{[t:T]}\cup W_{[t+1:T]}$ towards some $S\notin X_{[t:T]}\cup W_{[t+1:T]}$.
	Then we must have $S\in Y_{[t:T]}$. But since elements of $Y_{[1:t-1]}$
	can only be causes, and not consequences of those in $Y_{[t:T]}$,
	the path must contain a collider at some element of $Y_{[t:T]}$,
	and since this element cannot be in $X_{[t-k_{X}:t-1]}\cup W_{[t-k_{W}:t]}$,
	the path is $d$-separated.
	
	For the next result in the proposition we consider the counterfactual
	model defined in the proof of Proposition 2.1. Having dropped $Y_{[t:T]}$
	from the model the only possibility is that a path between $Y_{t}(x_{t})$
	and $X_{[t:T]}\cup W_{[t+1:T]}$ contains an edge from some $S\notin X_{[t:T]}\cup W_{[t+1:T]}$
	into an element of $X_{[t:T]}\cup W_{[t+1:T]}$. But then $S\in X_{[t-k_{X}:t-1]}\cup W_{[t-k_{W}:t]}$
	and so the path is $d$-separated.
	
	For the final two results in the proposition, we employ essentially
	the same the arguments above, replacing $X_{[t:T]}\cup W_{[t+1:T]}$
	with $X_{[t:T]}\cup Y_{[t+1:T]}\cup W_{[t+1:T]}$ and replacing $X_{[t-k_{X}:t-1]}\cup W_{[t-k_{W}:t]}$
	with $X_{[t-\bar{k}_{X}:t-1]}\cup W_{[t-\bar{k}_{W}:t]}$. Given $\kappa_{YY}=\kappa_{XY}=\kappa_{WY}=0$,
	any path between an element of $X_{[t:T]}\cup Y_{[t+1:T]}\cup W_{[t+1:T]}$
	and one not in $X_{[t:T]}\cup Y_{[t+1:T]}\cup W_{[t+1:T]}$ must contain
	an edge from an element $S\notin X_{[t:T]}\cup Y_{[t+1:T]}\cup W_{[t+1:T]}$
	into an element of $X_{[t:T]}\cup Y_{[t+1:T]}\cup W_{[t+1:T]}$. Then
	$S\in X_{[t-\bar{k}_{X}:t-1]}\cup W_{[t-\bar{k}_{W}:t]}$. So the
	path is $d$-separated and we are done. \end{proof}

The proofs of Theorems 2.1, 2.2, and 2.3 use the following three facts
about conditional independence from \citet{Dawid1979}. For any random
variables $W_{1}$, $W_{2}$, $W_{3}$, $W_{4}$, and $W_{5}$: 
\begin{align}
	W_{1}\perp\!\!\!\perp(W_{2},W_{3})|W_{4} & \implies W_{1}\perp\!\!\!\perp W_{2}|(W_{3},W_{4})\label{eq:indc2}\\
	W_{1}\perp\!\!\!\perp(W_{2},W_{3})|W_{4} & \implies W_{1}\perp\!\!\!\perp W_{2}|W_{4}\label{eq:indc3}
\end{align}
and in addition: 
\begin{align}
	& W_{1}\perp\!\!\!\perp W_{2}|(W_{3},W_{5})\,\,\land\,\,W_{1}\perp\!\!\!\perp W_{4}|(W_{2},W_{3},W_{5})\nonumber \\
	\implies & W_{1}\perp\!\!\!\perp(W_{2},W_{4})|(W_{3},W_{5})\label{eq:indc4}
\end{align}

\begin{proof}[Proof of Theorem 2.1] By supposition: 
	\begin{align*}
		Y_{[1:t-1]} & \perp\!\!\!\perp X_{[t:T]}|(X_{[t-\kappa_{XX}-1:t-1]},W)\\
		Y_{t}(x_{t}) & \indep X_{[t:T]}|(X_{[t-\kappa_{XX}-1:t-1]},W)
	\end{align*}
	
	Simply substituting $Z=X_{[t+1:T]}$, $V=Y_{[1:t-1]}$, and $D=X_{[t-\kappa_{XX}-1:t-1]}$
	yields the first result of the theorem. For the second part of the
	theorem we assume: 
	\begin{align*}
		Y_{[1:t-1]} & \perp\!\!\!\perp(X_{[t:T]},Y_{[t:T]})|(X_{[t-k_{Y}:t-1]},W)\\
		Y_{t}(x_{t}) & \indep(X_{[t:T]},Y_{[t:T]})|(X_{[t-k_{Y}:t-1]},W)
	\end{align*}
	
	Again, substituting the given choices of $Z$, $V$, and $D$ gives
	the result.\end{proof}

\begin{proof}[Proof of Theorem 2.2] By supposition:
	
	Using the definitions in the theorem, by supposition we have: 
	\begin{align}
		(Z,D_{2},X_{T},Y_{T}) & \perp\!\!\!\perp V|(D_{1},W)\label{eq:condit1}\\
		Y_{T}(x_{T}) & \perp\!\!\!\perp(X_{T},Z,D_{1},V)|(D_{2},W)\label{eq:condit2}
	\end{align}
	
	From (\ref{eq:condit1}) and (\ref{eq:indc2}) we get $(Z,X_{T},Y_{T})\perp\!\!\!\perp V|(D_{1},D_{2},W)$,
	then applying (\ref{eq:indc3}) we get $(Z,X_{T})\perp\!\!\!\perp V|(D_{1},D_{2},W)$.
	From (\ref{eq:condit2}) and (\ref{eq:indc2}) we get $Y_{T}(x_{T})\perp\!\!\!\perp(X_{T},Z,V)|(D_{1},D_{2},W)$,
	then applying (\ref{eq:indc3}) we get $Y_{T}(x_{T})\perp\!\!\!\perp(X_{T},Z)|(D_{1},D_{2},W)$.
	So $(Z,X_{T})\perp\!\!\!\perp V|(D,W)$ and $Y_{T}(x_{T})\perp\!\!\!\perp(X_{T},Z)|(D,W)$
	as required. \end{proof}

\begin{proof}[Proof of Theorem 2.3] As in the proof of Theorem
	2.1, simply substituting the $Z$, $V$, and $D$ suggested by the
	theorem into the relevant conditional independence restrictions gives
	the result. \end{proof}

\subsection{Proofs for Section 4}

Throughout the proofs in this subsection we use the notation $E_{Z}[\delta(Z,X)]$
to mean the expectation of $\delta(Z,X)$ treating $X$ as fixed and
similarly for other random variables and functions. So for example,
in the expectation below, the value of the function $\hat{\pi}_{n}(\cdot,\cdot,\cdot)$
is treated as non-random: 
\[
E_{Z,X,D}[\hat{\pi}_{n}(Z,X,D)\hat{\pi}_{n}(Z,X,D)']
\]

We also let $\Sigma_{\hat{\pi}}=E_{Z,X,D}[\hat{\pi}_{n}(Z,X,D)\hat{\pi}_{n}(Z,X,D)']$,
similarly we define $\Sigma_{\pi}=E[\pi_{n}(Z,X,D)\pi_{n}(Z,X,D)']$.
We let $\hat{\Sigma}_{\hat{\pi}}=\frac{1}{n_{g}}\sum_{i\in n_{g}}\hat{\pi}_{n,i}\hat{\pi}_{n,i}'$,
$\hat{\Sigma}_{\hat{\pi},\lambda_{0}}=\hat{\Sigma}_{\hat{\pi}}+\lambda_{0}I$,
$\Sigma_{\hat{\pi},\lambda_{0}}=\Sigma_{\hat{\pi}}+\lambda_{0}I$,
and $\Sigma_{\pi,\lambda_{0}}=\Sigma_{\pi}+\lambda_{0}I$. We also
let $\Sigma_{\psi,\lambda_{g}}=\Sigma_{\psi,n}+\lambda_{g}I$, and
similarly for the other basis functions and penalty parameters.

The space $\Lambda_{s}^{k}(c)$ is defined as follows. For each $q=(q_{1}q_{2},...,q_{k})\in\mathbb{N}_{0}^{k}$
, and $x\in\mathbb{R}^{k}$, let $\Delta_{q}[\delta](x)$ return the
partial derivatives as follows:
\[
\Delta_{q}[\delta](x)=\frac{\partial^{\|q\|_{1}}}{(\partial x_{1})^{q_{1}}(\partial x_{2})^{q_{2}}...(\partial x_{k})^{q_{k}}}\delta(x)
\]

We say $\delta\in\Lambda_{s}^{k}(c)$ if for each $q$ with $\|q\|_{1}<\lfloor s\rfloor$,
the partial derivative $\Delta_{q}[\delta]$ exists and is uniformly
bounded in magnitude by $c$. Moreover, for any $q\in\mathbb{N}_{0}^{k}$
whose entries sum to $\lfloor s\rfloor$, we must have that for any
$x,x'\in\mathbb{R}^{k}$:
\[
|\Delta_{q}[\delta](x)-\Delta_{q}[\delta](x')|\leq c\|x-x'\|^{s-\lfloor s\rfloor}
\]

\begin{proof}[Proof Theorem 4.1.i] By the properties of probability
	densities, for any $\gamma$ with $E[\gamma(V)^{2}|X=x_{1},D=d]$
	finite we have: 
	\begin{align*}
		E[\gamma(V)|D=d,X=x_{2}]= & E[\gamma(V)\frac{f_{V|XD}(V|x_{2},d)}{f_{V|XD}(V|x_{1},d)}|D=d,X=x_{1}]\\
		= & E\big[\gamma(V)E[\varphi(Z)|X,D,V]\big|D=d,X=x_{1}\big]
	\end{align*}
	Where $\varphi$ satisfies the conditions of Theorem 1.1.i. By iterated
	expectations: 
	\begin{align*}
		& E\big[\gamma(V)E[\varphi(Z)|X,D,V]\big|D=d,X=x_{1}\big]\\
		= & E\big[E[\gamma(V)|X,D,Z]\varphi(Z)\big|D=d,X=x_{1}\big]
	\end{align*}
	From the conclusion of Theorem 1.1.i and the equation above, we get:
	\begin{align*}
		& E[Y(x_{1})|X=x_{2},D=d]-E[\gamma(V)|X=x_{2},D=d]\\
		= & E\big[\big(Y-E[\gamma(V)|X,D,Z]\big)\varphi(Z)\big|X=x_{1},D=d\big]
	\end{align*}
	Applying Cauchy-Schwartz: 
	\begin{align*}
		& (E[Y(x_{1})|X=x_{2},D=d]-E[\gamma(V)|X=x_{2},D=d])^{2}\\
		\leq & E[\varphi(Z)^{2}|X=x_{1},D=d]E\big[\big(Y-E[\gamma(V)|X,D,Z]\big)^{2}\big|X=x_{1},D=d\big]
	\end{align*}
	And by supposition $E[\varphi(Z)^{2}|X=x_{1},D=d]^{1/2}\leq C$.
	
\end{proof}

\begin{proof}[Proof Theorem 4.1.ii] By iterated expectations for
	any function $\varphi$ that satisfies $E[\varphi(Z)^{2}|D=d,X=x_{1}]\leq\infty$
	we have: 
	\[
	E[\varphi(Z)Y|X=x_{1},D=d]=E\big[\varphi(Z)E[Y|X,D,Z]\big|X=x_{1},D=d\big]
	\]
	Letting $\gamma$ satisfy the conditions in Theorem 1.1.ii we can
	substitute for $E[Y|X,D,Z]$ in the above to get: 
	\begin{align}
		E[\varphi(Z)Y|X=x_{1},D=d]= & E\big[\varphi(Z)E[\gamma(V)|X,D,Z]\big|X=x_{1},D=d\big]\nonumber \\
		= & E\big[E[\varphi(Z)|X,D,V]\gamma(V)\big|X=x_{1},D=d\big]\label{eq:crr}
	\end{align}
	Where the second equality follows by iterated expectations. Using
	Theorem 1.1.ii and the properties of densities: 
	\begin{align*}
		E[Y(x_{1})|X=x_{2},D=d] & =E[\gamma(V)|X=x_{2},D=d]\\
		& =E\big[\gamma(V)\frac{f_{V|XD}(V|x_{2},d)}{f_{V|XD}(V|x_{1},d)}\big|X=x_{1},D=d\big]
	\end{align*}
	Combining with (\ref{eq:crr}) we get: the following: 
	\begin{align*}
		& E[Y(x_{1})|X=x_{2},D=d]-E[\varphi(Z)Y|X=x_{1},D=d]\\
		= & E\big[\gamma(V)\big(\frac{f_{V|XD}(V|x_{2},d)}{f_{V|XD}(V|x_{1},d)}-E[\varphi(Z)|X=x_{1},D=d,V]\big)\big|X=x_{1},D=d\big]
	\end{align*}
	By Cauchy-Schwartz: 
	\begin{align*}
		& \big(E[Y(x_{1})|X=x_{2},D=d]-E[\varphi(Z)Y|X=x_{1},D=d]\big)^{2}\\
		\leq & E[\gamma(V)^{2}|X=x_{1},D=d]\\
		\times & E\big[\big(\frac{f_{V|XD}(V|x_{2},d)}{f_{V|XD}(V|x_{1},d)}-E[\varphi(Z)|X=x_{1},D=d,V]\big)^{2}\big|X=x_{1},D=d\big]
	\end{align*}
	And by supposition $E\big[\gamma(V)^{2}|X=x_{1},D=d\big]^{1/2}\leq C$
	which yields the result. \end{proof}

\begin{proof}[Proof of Lemma 4.1]
	
	By Assumption 4.6.i, for each $x$ and $d$ there is a function $\gamma(\cdot,x,d)$
	with $E[\gamma(V,x,d)^{2}|X=x,D=d]^{1/2}\leq C$ so that: 
	\[
	g(z,x,d)=E[\gamma(V,x,d)|Z=z,X=x,D=d]
	\]
	
	In the case of $X$ and $D$ with finite discrete support, by Assumption
	4.2.v, $\chi_{n}(X,D)$ is a vector of binary indicators (one for
	each possible value of $X$ and $D$). Thus we can set $\gamma^{*}(v)$
	to be the vector whose $k^{th}$ component equals $\gamma(v,x,d)$
	for the $(x,d)$ pair that corresponds to the indicator in the $k^{th}$
	component of $\chi_{n}(X,D)$. Then we have $\chi_{n}(x,d)'\gamma^{*}(v)=\gamma(v,x,d)$
	and so: 
	\[
	g(z,x,d)-\chi_{n}(x,d)'E[\gamma^{*}(V)|Z=z,X=x,D=d]=0
	\]
	
	For each $x$ and $d$ in the support of $X$ and $D$, let $k(x,d)$
	be the index of the component of $\chi_{n}(x,d)$ that equals $1$
	and let $\gamma_{k}^{*}$ be the $k_{th}$ component of $\gamma^{*}$.
	By Assumption 4.3.i there is a constant $c_{f}$ so that $E[\gamma_{k(x,d)}^{*}(V)^{2}]\leq c_{f}E[\gamma_{k(x,d)}^{*}(V)^{2}|X=x,D=d]$
	and note that 
	\[
	E[\gamma_{k(x,d)}^{*}(V)^{2}|X=x,D=d]=E[\gamma(V,x,d)^{2}|X=x,D=d]
	\]
	Thus from the definition of $\gamma^{*}$ and the bound on $E[\gamma(V,x,d)^{2}|X=x,D=d]^{1/2}$,
	we have that $E[\|\gamma^{*}(V)\|^{2}]$ is bounded above.
	
	In the non-discrete case we can apply Lemma C.1 to get: 
	\[
	|g(z,x,d)-\chi_{n}(x,d)'E[\gamma^{*}(V)|Z=z,X=x,D=d]|\leq c_{1}\ell_{\chi,n}(1)^{\bar{s}}
	\]
	
	Where $E[\|\gamma^{*}(V)\|^{2}]$ is bounded above, and $c_{1}$ is
	a finite constant.
	
	Now we show that there exists a matrix $\beta_{n}$ whose Frobenius
	norm (the square root of the sum of squared entries of $\beta_{n}$)
	is bounded above, so that for some finite constant $c_{2}$: 
	\[
	|\chi_{n}(x,d)'E[\gamma^{*}(V)-\beta_{n}'\rho_{n}(V)|X=x,D=d,Z=z]|\leq c_{2}\xi_{\chi,n}\ell_{\rho,n}(s)
	\]
	
	We can then conclude that in the non-discrete case: 
	\begin{align*}
		& |g(x,d,z)-\chi_{n}(x,d)'\beta_{n}E[\rho_{n}(V)|X=x,D=d,Z=z]|\\
		\leq & c_{2}\xi_{\chi,n}\ell_{\rho,n}(s)+c_{1}\ell_{\chi,n}(1)^{\bar{s}}
	\end{align*}
	In the discrete case we can drop the second term and note that in
	this case $\xi_{\chi,n}$ is bounded above and so we can ignore this
	factor. The Euclidean norm of the vectorization of $\beta_{n}$ is
	equal to the Frobenius norm of $\beta_{n}$, and so, taking $\theta_{n}$
	to be the vectorization of $\beta_{n}$ gives the result.
	
	By properties of Kronecker products: 
	\begin{align*}
		& \chi_{n}(x,d)'\beta_{n}E[\rho_{n}(V)|X=x,D=d,Z=z]\\
		= & E[\rho_{n}(V)\otimes\chi_{n}(x,d)|X=x,D=d,Z=z]'\theta_{n}\\
		= & \pi_{n}(z,x,d)'\theta_{n}
	\end{align*}
	
	Where $\theta_{n}=vec(\beta_{n})$ and thus has norm $\|\theta_{n}\|$
	equal to the Frobenius norm of $\beta_{n}$. Combining Assumptions
	4.2.ii and 4.4.i, for each $z$, $x$, and $d$ in the joint support
	of $Z$, $X$, and $D$: 
	\[
	\inf_{B}E\big[\big(\frac{f_{V|ZXD}(V|z,x,d)}{f_{V}(V)}-\rho_{n}(V)'B\big)^{2}\big]^{1/2}\precsim\ell_{\rho,n}(s)
	\]
	This implies that there exist sequences of functions $B_{n}$ and
	$r_{n}$ so that for all $x$, $d$, $z$, and $v$ in the joint support
	of $X$, $D$, $Z$, and $V$: 
	\begin{equation}
		\frac{f_{V|ZXD}(v|z,x,d)}{f_{V}(v)}=\rho_{n}(v)'B_{n}(z,x,d)+r_{n}(x,d,z,v)\label{eq:densbound-1-1}
	\end{equation}
	Where $E\big[r_{n}(x,z,d,V)^{2}\big]^{1/2}\leq c_{2}\ell_{\rho,n}(s)$
	for some finite constant $c_{2}$. For any matrix of conformable dimensions
	$\beta_{n}$, by elementary properties of probability densities we
	have: 
	\begin{align*}
		& E[\gamma^{*}(V)-\beta_{n}'\rho_{n}(V)|X=x,D=d,Z=z]\\
		= & E\big[\big(\gamma^{*}(V)-\beta_{n}'\rho_{n}(V)\big)\frac{f_{V|XDZ}(V|x,d,z)}{f_{V}(V)}\big]
	\end{align*}
	
	Substituting (\ref{eq:densbound-1-1}), the RHS above becomes: 
	\begin{align}
		& \big(E[\gamma^{*}(V)\rho_{n}(V)']B_{n}(z,x,d)-\beta_{n}'E[\rho_{n}(V)\rho_{n}(V)']B_{n}(z,x,d)\label{eq:repe-1-1}\\
		+ & E\big[\big(\gamma^{*}(V)-\beta_{n}'\rho_{n}(V)\big)r_{n}(x,d,z,V)\big]\nonumber 
	\end{align}
	
	By Assumption 4.2.i $E[\rho_{n}(V)\rho_{n}(V)']$ is non-singular,
	and so we can set $\beta_{n}$ so that: 
	\[
	\beta_{n}=E[\rho_{n}(V)\rho_{n}(V)']^{-1}E[\rho_{n}(V)\gamma^{*}(V)']
	\]
	Substituting into (\ref{eq:repe-1-1}) the first term disappears and
	we get: 
	\begin{align*}
		& E[\gamma^{*}(V)-\beta_{n}'\rho_{n}(V)|X=x,D=d,Z=z]\\
		= & E\big[\big(\gamma^{*}(V)-\beta_{n}'\rho_{n}(V)\big)r_{n}(x,d,z,V)\big]
	\end{align*}
	
	By Cauchy-Schwartz and Assumption 4.2.ii: 
	\begin{align*}
		& \big|\chi_{n}(x,d)'E\big[\big(\gamma^{*}(V)-\beta_{n}'\rho_{n}(V)\big)r_{n}(x,d,z,V)\big]\big|\\
		\leq & \|\chi_{n}(x,d)\|E\big[\|\gamma^{*}(V)-\beta_{n}'\rho_{n}(V)\|^{2}\big]^{1/2}E[r_{n}(x,z,V)^{2}]^{1/2}\\
		\leq & c_{2}\xi_{\chi,n}\ell_{\rho,n}(s)E\big[\|\gamma^{*}(V)-\beta_{n}'\rho_{n}(V)\|^{2}\big]^{1/2}
	\end{align*}
	
	By properties of least squares projections:
	
	\[
	E\big[\|\gamma^{*}(V)-\beta_{n}'\rho_{n}(V)\|^{2}\big]^{1/2}\leq E[\|\gamma^{*}(V)\|^{2}]
	\]
	
	So in all we get that for some finite constant $c$: 
	\[
	|\chi_{n}(x,d)'E[\gamma^{*}(V)-\beta_{n}'\rho_{n}(V)|X=x,D=d,Z=z]|\leq c\xi_{\chi,n}\ell_{\rho,n}(s)
	\]
	
	Finally, note that by properties of least-squares projections: 
	\[
	\|\beta_{n}\|_{F}^{2}\leq\|\Sigma_{\rho,n}^{-1/2}\|E[\|\gamma^{*}(V)\|^{2}]
	\]
	
	Where $\|\cdot\|_{F}$ is the Frobenius norm. By Assumption 4.2.i
	$\|\Sigma_{\rho,n}^{-1/2}\|$ is bounded above, and we already established
	that $E[\|\gamma^{*}(V)\|^{2}]$ is bounded above, and so we are done.
\end{proof}

\begin{proof}[Proof Theorem 4.2]
	
	First note that by Lemma 4.1 there exists a sequence $\{\theta_{n}\}_{n=1}^{\infty}$
	so that $\|\theta_{n}\|$ is bounded above (by a constant that does
	not vary with $n$) and for some finite constant $c$ and sequence
	$r_{\theta,n}\to0$, for all $z$, $x$, and $d$ in the joint support
	of $Z$, $X$, and $D$: 
	\[
	|g(z,x,d)-\pi_{n}(z,x,d)'\theta_{n}|\leq cr_{\theta,n}
	\]
	Throughout $\theta_{n}$ denotes this particular sequence of parameter
	vectors.
	
	By Assumption 4.6.ii, for any $(x_{1},x_{2},d)\in\mathcal{S}$, there
	is a solution $\varphi(Z)$ to the conditional moment restriction
	in Theorem 1.1.i so that $E[\varphi(Z)^{2}|X=x_{1},D=d]^{1/2}$ is
	less than a finite constant $C$. We make the dependence on $x_{1}$,
	$x_{2}$, and $d$ explicit so that the solution that corresponds
	to a particular value of these variables if denoted by $\varphi(\cdot,x_{1},x_{2},d)$.
	
	From the steps in the proof of Theorem 4.1.i, for any function $\delta$
	that satisfies the condition $E[\delta(V)^{2}|X=x,D=d]<\infty$ we
	have: 
	\begin{align*}
		& E[\delta(V)|X=x_{2},D=d]\\
		= & E\big[E[\delta(V)|X=x_{1},D=d,Z]\varphi(Z,x_{1},x_{2},d)\big|X=x_{1},D=d\big]
	\end{align*}
	
	Using the definitions of $\alpha_{n}$ and $\pi_{n}$, we have: 
	\begin{equation}
		\alpha_{n}(x_{1},x_{2},d)=E[\varphi(Z,x_{1},x_{2},d)\pi_{n}(Z,x_{1},d)|X=x_{1},D=d]\label{eq:alphdec}
	\end{equation}
	By Theorem 1.1.i and iterated expectations: 
	\begin{align}
		& E[Y(x_{1})|X=x_{2},D=d]\nonumber \\
		= & E\big[\varphi(Z,x_{1},x_{2},d)Y\big|X=x_{1},D=d\big]\nonumber \\
		= & E\big[\varphi(Z,x_{1},x_{2},d)g(Z,x_{1},d)\big|X=x_{1},D=d\big]\label{eq:Ydec}
	\end{align}
	Using (\ref{eq:alphdec}) and \eqref{eq:Ydec} and adding and subtracting
	terms we get: 
	\begin{align*}
		& E[Y(x_{1})|X=x_{2},D=d]-\hat{\alpha}_{n}(x_{1},x_{2},d)'\hat{\theta}\\
		= & E\big[\varphi(Z,x_{1},x_{2},d)\big(g(Z,x_{1},d)-\pi_{n}(Z,x_{1},d)'\theta_{n}\big)\big|X=x_{1},D=d\big]\\
		+ & \big(\alpha_{n}(x_{1},x_{2},d)-\hat{\alpha}_{n}(x_{1},x_{2},d)\big)'\hat{\theta}\\
		+ & \alpha_{n}(x_{1},x_{2},d)'(\theta_{n}-\hat{\theta})
	\end{align*}
	
	By supposition $E\big[\varphi(Z,x_{1},x_{2},d)^{2}|X=x_{1},D=d\big]^{1/2}$
	is less than $C$, so applying Cauchy-Schwartz and the triangle inequality:
	\begin{align}
		& \big|E[Y(x_{1})|X=x_{2},D=d]-\hat{\alpha}_{n}(x_{1},x_{2},d)'\hat{\theta}\big|\nonumber \\
		\leq & CE\big[|g_{n}(Z,x_{1},d)-\pi_{n}(Z,x_{1},d)'\theta_{n}|^{2}\big|X=x_{1},D=d\big]^{1/2}\nonumber \\
		+ & \|\hat{\alpha}_{n}(x_{1},x_{2},d)-\alpha_{n}(x_{1},x_{2},d)\|\|\hat{\theta}\|\nonumber \\
		+ & |\alpha_{n}(x_{1},x_{2},d)'(\theta_{n}-\hat{\theta})|\nonumber \\
		\precsim_{p} & r_{\theta,n}+\bar{r}_{\alpha,n}\|\hat{\theta}\|\label{eq:full bound}
	\end{align}
	
	Where the last line uses Assumption 4.1.iv. So it remains to derive
	rates for $\|\hat{\theta}\|$ and $|\alpha_{n}(x_{1},x_{2},d)'(\theta_{n}-\hat{\theta})|$.
	
	\subsubsection*{Deriving a rate for $\|\hat{\theta}\|$}
	
	Recall that $\hat{\theta}=\hat{\Sigma}_{\hat{\pi},\lambda_{0}}^{-1}\frac{1}{n}\sum_{i=1}^{n}\hat{\pi}_{n,i}\hat{g}_{i}$.
	Adding and subtracting terms we get that $\hat{\theta}-\theta_{n}$
	is equal to the following: 
	\[
	\hat{\Sigma}_{\hat{\pi},\lambda_{0}}^{-1}\frac{1}{n}\sum_{i=1}^{n}\hat{\pi}_{n,i}\big((\pi_{n,i}-\hat{\pi}_{n,i})'\theta_{n}+(\hat{g}_{i}-g_{i})+g_{i}-\pi_{n,i}'\theta_{n}\big)-\lambda_{0}\hat{\Sigma}_{\hat{\pi},\lambda_{0}}^{-1}\theta_{n}
	\]
	
	By the triangle inequality and properties of the matrix norm, we get:
	\begin{align*}
		& \|\hat{\Sigma}_{\hat{\pi},\lambda_{0}}^{1/2}(\hat{\theta}-\theta_{n})\|\\
		\leq & \|\hat{\Sigma}_{\hat{\pi},\lambda_{0}}^{-1/2}\hat{\Sigma}_{\hat{\pi}}^{1/2}\|\|\hat{\Sigma}_{\hat{\pi}}^{-1/2}\frac{1}{n}\sum_{i=1}^{n}\hat{\pi}_{n,i}\big((\pi_{n,i}-\hat{\pi}_{n,i})'\theta_{n}+(\hat{g}_{i}-g_{i})+g_{i}-\pi_{n,i}'\theta_{n}\big)\|\\
		+ & \lambda_{0}\|\hat{\Sigma}_{\hat{\pi},\lambda_{0}}^{-1/2}\|\|\theta_{n}\|
	\end{align*}
	
	By standard results for ridge regularized matrices we must have that
	$\|\hat{\Sigma}_{\hat{\pi},\lambda_{0}}^{-1/2}\|\leq\lambda_{0}^{-1/2}$,
	and $\|\hat{\Sigma}_{\hat{\pi},\lambda_{0}}^{-1/2}\hat{\Sigma}_{\hat{\pi}}^{1/2}\|\leq1$.
	Using the triangle inequality and the properties of least squares
	projections we get that:
	
	\begin{align}
		\|\hat{\Sigma}_{\hat{\pi},\lambda_{0}}^{1/2}(\hat{\theta}-\theta_{n})\|\leq & \lambda_{0}^{1/2}\|\theta_{n}\|+\big(\frac{1}{n}\sum_{i=1}^{n}|(\pi_{n,i}-\hat{\pi}_{n,i})'\theta_{n}|^{2}\big)^{1/2}\nonumber \\
		+ & \big(\frac{1}{n}\sum_{i=1}^{n}(\hat{g}_{i}-g_{i})^{2}\big)^{1/2}+\big(\frac{1}{n}\sum_{i=1}^{n}(g_{i}-\pi_{n,i}'\theta_{n})^{2}\big)^{1/2}\nonumber \\
		\precsim_{p} & \lambda_{0}^{1/2}+(r_{\pi,n}+r_{g,n}+r_{\theta,n})\label{sigth}
	\end{align}
	
	Where the final line uses Assumption 4.1. Then noting that $\|\hat{\theta}-\theta_{n}\|\leq\lambda_{0}^{-1/2}\|\hat{\Sigma}_{\hat{\pi},\lambda_{0}}^{1/2}(\hat{\theta}-\theta_{n})\|$
	we get: 
	\begin{align}
		\|\hat{\theta}-\theta_{n}\| & \precsim_{p}1+\lambda_{0}^{-1/2}(r_{\pi,n}+r_{g,n}+r_{\theta,n})\label{eq:thetnorm-1}
	\end{align}
	
	By the reverse triangle inequality and $\|\theta_{n}\|\precsim1$
	we get: 
	\begin{align}
		\|\hat{\theta}\| & \precsim_{p}1+\lambda_{0}^{-1/2}(r_{\pi,n}+r_{g,n}+r_{\theta,n})\label{eq:thetnorm}
	\end{align}

	\subsubsection*{Deriving a rate for $|\alpha_{n}(x_{1},x_{2},d)'(\theta_{n}-\hat{\theta})|$}
	
	By the properties of the matrix norm: 
	\[
	|\alpha_{n}(x_{1},x_{2},d)'(\hat{\theta}-\theta_{n})|\leq\|\alpha_{n}(x_{1},x_{2},d)'\hat{\Sigma}_{\hat{\pi},\lambda_{0}}^{-1/2}\|\|\hat{\Sigma}_{\hat{\pi}}^{1/2}(\hat{\theta}-\theta_{n})\|
	\]
	
	We bounded the second term in the product above in (\ref{sigth}),
	so it remains to bound $\|\alpha_{n}(x_{1},x_{2},d)'\hat{\Sigma}_{\hat{\pi},\lambda_{0}}^{-1/2}\|$.
	By Lemma C.6 we have: 
	\[
	\|\alpha_{n}(x_{1},x_{2},d)'\hat{\Sigma}_{\hat{\pi},\lambda_{0}}^{-1/2}\|=O_{p}\bigg(1+\sqrt{\frac{\lambda_{0}\bar{r}_{\pi,n}^{2}\bar{\xi}_{\kappa,n}}{\mu_{n}+\lambda_{0}^{2}}}\bigg)
	\]
	
	And thus: 
	\[
	|\alpha_{n}(x_{1},x_{2},d)'(\hat{\theta}-\theta_{n})|\precsim_{p}\bigg(1+\sqrt{\frac{\lambda_{0}\bar{r}_{\pi,n}^{2}\bar{\xi}_{\psi,n}}{\mu_{n}+\lambda_{0}^{2}}}\bigg)(r_{\pi,n}+r_{g,n}+r_{\theta,n}+\lambda_{0}^{1/2})
	\]
	
	\textbf{Combining everything}
	
	We can apply the rates on the rate of $\|\hat{\theta}\|$ and $|\alpha_{n}(x_{1},x_{2},d)'(\hat{\theta}-\theta_{n})|$
	to \eqref{eq:full bound} and we get:
	
	\begin{align*}
		& \big|E[Y(x_{1})|X=x_{2},D=d]-\hat{\alpha}_{n}(x_{1},x_{2},d)'\hat{\theta}\big|\\
		\precsim_{p} & \bigg(1+\frac{\bar{r}_{\alpha,n}}{\lambda_{0}^{1/2}}+\sqrt{\frac{\lambda_{0}\bar{r}_{\pi,n}^{2}\bar{\xi}_{\psi,n}}{\mu_{n}+\lambda_{0}^{2}}}\bigg)(\lambda_{0}^{1/2}+r_{\pi,n}+r_{g,n}+r_{\theta,n})
	\end{align*}
	
	Noting $\mu_{n}$ is weakly positive gives the result.
	
\end{proof}

\theoremstyle{plain} \newtheorem*{L42}{Lemma 4.2 (Linearization)}
\begin{L42} Suppose Assumptions 1, 2, 3, and 4.2-4.7 hold. Let $r_{\theta,n}$
	be defined as in Lemma 4.1. Suppose that the smallest eigenvalue of
	$\Omega_{n}$ is bounded below away from zero and $\|s_{n}(x_{1},x_{2},d)\|>0$.
	Then we have: 
	\begin{align*}
		\text{sup}_{(x_{1},x_{2},d)\in\mathcal{S}} & |\frac{\sqrt{n}\big(\hat{\alpha}_{n}(x_{1},x_{2},d)'\hat{\theta}-E[Y(x_{1})|X=x_{2},D=d]\big)}{\|s_{n}(x_{1},x_{2},d)\|}\\
		& -\frac{s_{n}(x_{1},x_{2},d)'\frac{1}{\sqrt{n}}\sum_{i=1}^{n}\Omega_{n}^{-1/2}\eta_{n,i}}{\|s_{n}(x_{1},x_{2},d)\|}|\\
		\precsim_{p} & r_{l,n}+\sqrt{n}r_{\theta,n}+r_{s,n}+r_{\lambda,n}+r_{\mu,n}
	\end{align*}
\end{L42} \begin{proof}
	
	Define $R_{g,i}$, $R_{\pi,n,i}$, and $R_{\alpha,n,i}$ as in the
	proofs of Lemmas C.3 and C.4. Let $\{\theta_{n}\}_{n=1}^{\infty}$
	be the sequence of vectors shown to exist in Lemma 4.1. Let $\tilde{\theta}_{n}$
	denote the $k_{\rho,n}$-by-$k_{\chi,n}$ matrix obtained by partitioning
	the vector $\theta_{n}$ into contiguous length-$k_{\chi,n}$ subvectors
	and staking the transposes into a matrix. Let $vec(\cdot)$ be the
	vectorization operator so that for a $k_{1}$-by-$k_{2}$ matrix $A$,
	$vec(A)$ is the vector obtained by transposing the rows of $A$ and
	stacking them into a vector, thus $vec(\tilde{\theta}_{n})=\theta_{n}$.
	
	\textbf{Step 1: Linearize $\alpha_{n}(x_{1},x_{2},d)'(\hat{\theta}-\theta_{n})$
		and obtain a rate for $\|\Sigma_{\pi,\lambda_{0}}^{1/2}(\hat{\theta}-\theta_{n})\|$.}
	
	Using the definitions above and mixed-product properties of the Kronecker
	product we get:
	
	\begin{align*}
		\hat{\theta}-\theta_{n}
		= & \Sigma_{\pi,\lambda_{0}}^{-1}E[\pi_{n,i}\psi_{n,i}']\Sigma_{\psi,\lambda_{g}}^{-1}\frac{1}{n}\sum_{i=1}^{n}\psi_{n,i}\epsilon_{i}\\
		- & \Sigma_{\pi,\lambda_{0}}^{-1}E[\pi_{n,i}\kappa_{n,i}']vec(\Sigma_{\zeta,\lambda_{\pi}}^{-1}\frac{1}{n}\sum_{i=1}^{n}\zeta_{n,i}\upsilon_{n,i}'\tilde{\theta}_{n})\\
		+ & R_{\theta,n}
	\end{align*}
	
	Define $l_{g,n,i}$ and $l_{\pi,n,i}$ as in the statements of Lemmas
	C.3 and C.4. The remainder term $R_{\theta,n}$ is given below: 
	\begin{align}
		 R_{\theta,n}\nonumber 
		= & \hat{\Sigma}_{\hat{\pi},\lambda_{0}}^{-1}\frac{1}{n}\sum_{i=1}^{n}\hat{\pi}_{n,i}(l_{g,n,i}-l_{\pi,n,i}'\theta_{n})\nonumber \\
		+ & \hat{\Sigma}_{\hat{\pi},\lambda_{0}}^{-1}\frac{1}{n}\sum_{i=1}^{n}\hat{\pi}_{n,i}(g_{i}-\pi_{n,i}'\theta_{n})\nonumber \\
		+ & \big(\hat{\Sigma}_{\hat{\pi},\lambda_{0}}^{-1}\frac{1}{n}\sum_{i=1}^{n}\hat{\pi}_{n,i}\psi_{n,i}'-\Sigma_{\pi,\lambda_{0}}^{-1}E[\pi_{n,i}\psi_{n,i}']\big)\Sigma_{\psi,\lambda_{g}}^{-1}\frac{1}{n}\sum_{i=1}^{n}\psi_{n,i}\epsilon_{i}\nonumber \\
		+ & \big(\Sigma_{\pi,\lambda_{0}}^{-1}E[\pi_{n,i}\kappa_{n,i}']-\hat{\Sigma}_{\hat{\pi},\lambda_{0}}^{-1}\frac{1}{n}\sum_{i=1}^{n}\hat{\pi}_{n,i}\kappa_{n,i}'\big)vec\big(\Sigma_{\zeta,\lambda_{\pi}}^{-1}\frac{1}{n}\sum_{i=1}^{n}\zeta_{n,i}\upsilon_{n,i}'\tilde{\theta}_{n}\big)\nonumber \\
		- & \lambda_{0}\hat{\Sigma}_{\hat{\pi},\lambda_{0}}^{-1}\theta_{n}\label{eq:remainder}
	\end{align}
	
	We proceed to derive a rate for $\sqrt{n}\|\Sigma_{\pi,\lambda_{0}}^{1/2}R_{\theta,n}\|$.
	By the triangle inequality we can bound this quantity as follows:
	\begin{align*}
		\sqrt{n}\|\Sigma_{\hat{\pi},\lambda_{0}}^{1/2}R_{\theta,n}\| & \leq\sqrt{n}\|\Sigma_{\pi,\lambda_{0}}^{1/2}\hat{\Sigma}_{\hat{\pi},\lambda_{0}}^{-1/2}\|(\tilde{r}_{1}+\tilde{r}_{2}+\tilde{r}_{3}+\tilde{r}_{4}+\tilde{r}_{5})
	\end{align*}
	
	$\tilde{r}_{1}$, $\tilde{r}_{2}$, $\tilde{r}_{3}$, $\tilde{r}_{4}$,
	and $\tilde{r}_{5}$ respectively refer to the norm of $\hat{\Sigma}_{\hat{\pi},\lambda_{0}}^{1/2}$
	times the corresponding term on the RHS of \eqref{eq:remainder}.
	By Lemma C.6 $\|\Sigma_{\pi,\lambda_{0}}^{1/2}\hat{\Sigma}_{\hat{\pi},\lambda_{0}}^{-1/2}\|\precsim_{p}1$.
	We derive upper bounds on each sequentially. We start with $\tilde{r}_{1}$.
	Define $\bar{l}_{g,n}$ and $\bar{l}_{\pi,n}$ as in Lemmas C.3 and
	C.4. 
	\begin{align*}
		\sqrt{n}\tilde{r}_{1} & =\sqrt{n}\|\hat{\Sigma}_{\hat{\pi},\lambda_{0}}^{-1/2}\frac{1}{n}\sum_{i=1}^{n}\hat{\pi}_{n,i}(l_{g,n,i}+l_{\pi,n,i}'\theta_{n})\|\\
		& \le\sqrt{n}\|\hat{\Sigma}_{\hat{\pi},\lambda_{0}}^{-1/2}\hat{\Sigma}_{\hat{\pi}}^{1/2}\|\|\hat{\Sigma}_{\hat{\pi}}^{-1/2}\frac{1}{n}\sum_{i=1}^{n}\hat{\pi}_{n,i}(l_{g,n,i}+l_{\pi,n,i}'\theta_{n})\|\\
		& \leq\sqrt{n}\big(\frac{1}{n}\sum_{i=1}^{n}l_{g,n,i}^{2}\big)^{1/2}+\sqrt{n}\big(\frac{1}{n}\sum_{i=1}^{n}(l_{\pi,n,i}'\theta_{n})^{2}\big)^{1/2}\\
		& \precsim_{p}\bar{l}_{g,n}+\bar{l}_{\pi,n}
	\end{align*}
	
	Where the second line follows by properties of the matrix norm, the
	third follows by $\|\hat{\Sigma}_{\hat{\pi},\lambda_{0}}^{-1/2}\hat{\Sigma}_{\hat{\pi}}^{1/2}\|\leq1$,
	the properties of least squares projections, and the triangle inequality.
	The final line follows from Lemmas C.3 and C.4. Next we bound $\tilde{r}_{2}$.
	Again using that $\|\hat{\Sigma}_{\hat{\pi},\lambda_{0}}^{-1/2}\hat{\Sigma}_{\hat{\pi}}^{1/2}\|\leq1$
	and the properties of least squares projections, we get the following:
	\begin{align*}
		\tilde{r}_{2} & =\|\hat{\Sigma}_{\hat{\pi},\lambda_{0}}^{-1/2}\frac{1}{n}\sum_{i=1}^{n}\hat{\pi}_{n,i}(g_{i}-\pi_{n,i}'\theta_{n})\|\\
		& \leq\big(\frac{1}{n}\sum_{i=1}^{n}(g_{i}-\pi_{n,i}'\theta_{n})^{2}\big)^{1/2}
	\end{align*}
	
	Then by Lemma 4.1, we have $\tilde{r}_{2}\precsim_{p}r_{\theta,n}$.
	We now derive a rate for $\tilde{r}_{3}$. Applying the properties
	of the matrix norm and the triangle inequality we get the following:
	
	\begin{align}
		\tilde{r}_{3} & =\|\hat{\Sigma}_{\hat{\pi},\lambda_{0}}^{1/2}\big(\hat{\Sigma}_{\hat{\pi},\lambda_{0}}^{-1}\frac{1}{n}\sum_{i=1}^{n}\hat{\pi}_{n,i}\psi_{n,i}'-\Sigma_{\pi,\lambda_{0}}^{-1}E[\pi_{n,i}\psi_{n,i}']\big)\Sigma_{\psi,\lambda_{g}}^{-1}\frac{1}{n}\sum_{i=1}^{n}\psi_{n,i}\epsilon_{i}\|\nonumber \\
		& \leq\|\hat{\Sigma}_{\hat{\pi},\lambda_{0}}^{-1/2}\big(\frac{1}{n}\sum_{i=1}^{n}\hat{\pi}_{n,i}\psi_{n,i}'-E[\pi_{n,i}\psi_{n,i}']\big)\|\|\Sigma_{\psi,\lambda_{g}}^{-1}\|\|\frac{1}{n}\sum_{i=1}^{n}\psi_{n,i}\epsilon_{i}\|\nonumber \\
		& +\|\hat{\Sigma}_{\hat{\pi},\lambda_{0}}^{-1/2}(\Sigma_{\pi}-\hat{\Sigma}_{\hat{\pi}})\Sigma_{\pi,\lambda_{0}}^{-1/2}\|\|\frac{1}{n}\sum_{i=1}^{n}\Sigma_{\pi,\lambda_{0}}^{-1/2}E[\pi_{n,i}\psi_{n,i}']\Sigma_{\psi,\lambda_{g}}^{-1}\psi_{n,i}\epsilon_{i}\|\label{eq:compbound}
	\end{align}
	
	By Assumption 4.2.i, $\|\Sigma_{\psi,n}^{-1}\|$ is bounded and by
	the law of large numbers $\|\frac{1}{n}\sum_{i=1}^{n}\psi_{n,i}\epsilon_{i}\|\precsim_{p}\sqrt{\frac{k_{\psi,n}}{n}}$
	(see the proof of Lemma C.3). In addition, by the triangle inequality:
	\begin{align*}
		& \|\hat{\Sigma}_{\hat{\pi},\lambda_{0}}^{-1/2}\big(\frac{1}{n}\sum_{i=1}^{n}\hat{\pi}_{n,i}\psi_{n,i}'-E[\pi_{n,i}\psi_{n,i}']\big)\|\\
		\leq & \|\hat{\Sigma}_{\hat{\pi},\lambda_{0}}^{-1/2}\|\|\frac{1}{n}\sum_{i=1}^{n}(\hat{\pi}_{n,i}-\pi_{n,i})\psi_{n,i}'\|\\
		+ & \|\hat{\Sigma}_{\hat{\pi},\lambda_{0}}^{-1/2}\Sigma_{\pi,\lambda_{0}}^{1/2}\|\|\frac{1}{n}\sum_{i=1}^{n}\Sigma_{\pi,\lambda_{0}}^{-1/2}\pi_{n,i}\psi_{n,i}'-E[\Sigma_{\pi,\lambda_{0}}^{-1/2}\pi_{n,i}\psi_{n,i}']\|
	\end{align*}
	
	By Lemma C.6 $\|\hat{\Sigma}_{\hat{\pi},\lambda_{0}}^{-1/2}\Sigma_{\pi,\lambda_{0}}^{1/2}\|\precsim_{p}1$.
	By properties of least squares we have:
	
	\begin{align*}
		\|\frac{1}{n}\sum_{i=1}^{n}(\hat{\pi}_{n,i}-\pi_{n,i})\psi_{n,i}'\| & \leq\|\hat{\Sigma}_{\psi}\|^{1/2}\big(\frac{1}{n}\sum_{i=1}^{n}\|\hat{\pi}_{n,i}-\pi_{n,i}\|^{2}\big)^{1/2}
	\end{align*}
	
	Note that $\|\hat{\Sigma}_{\psi}\|\precsim_{p}1$ (see the proof of
	Lemma C.3), and by Lemma C.4 $\frac{1}{n}\sum_{i=1}^{n}\|\hat{\pi}_{n,i}-\pi_{n,i}\|^{2}\precsim_{p}\bar{r}_{\pi,n}^{2}$,
	and so $\|\frac{1}{n}\sum_{i=1}^{n}(\hat{\pi}_{n,i}-\pi_{n,i})\psi_{n,i}'\|\precsim_{p}\bar{r}_{\pi,n}$.
	Since $\|\Sigma_{\pi,\lambda_{0}}^{1/2}\hat{\Sigma}_{\hat{\pi},\lambda_{0}}^{-1/2}\|\precsim_{p}1$
	and $\|\Sigma_{\pi,\lambda_{0}}^{-1/2}\|\leq\sqrt{\frac{\lambda_{0}}{\mu_{n}+\lambda_{0}^{2}}}$
	we have $\|\hat{\Sigma}_{\hat{\pi},\lambda_{0}}^{-1/2}\|\leq\sqrt{\frac{\lambda_{0}}{\mu_{n}+\lambda_{0}^{2}}}$,
	and so:
	
	\[
	\|\hat{\Sigma}_{\hat{\pi},\lambda_{0}}^{-1/2}\|\|\frac{1}{n}\sum_{i=1}^{n}(\hat{\pi}_{n,i}-\pi_{n,i})\psi_{n,i}'\|\precsim_{p}\sqrt{\frac{\lambda_{0}\bar{r}_{\pi,n}^{2}}{\mu_{n}+\lambda_{0}^{2}}}
	\]
	
	Using properties of the Frobenious norm (denoted by `$\|\cdot\|_{F}$')
	we have: 
	\begin{align*}
		E[\|\Sigma_{\pi,\lambda_{0}}^{-1/2}\pi_{n,i}\psi_{n,i}'\|_{F}^{2}] & \leq E[\|\Sigma_{\pi}^{-1/2}\pi_{n,i}\|^{2}]\text{ess sup\ensuremath{\|\psi_{n,i}\|^{2}}}\\
		& \precsim\xi_{\psi,n}^{2}k_{\rho,n}k_{\chi,n}
	\end{align*}
	
	Where the final line follows by Assumption 4.2.ii. Then by the law
	of large numbers: 
	\[
	\|\frac{1}{n}\sum_{i=1}^{n}\Sigma_{\pi,\lambda_{0}}^{-1/2}\pi_{n,i}\psi_{n,i}'-E[\Sigma_{\pi,\lambda_{0}}^{-1/2}\pi_{n,i}\psi_{n,i}']\|\precsim_{p}\sqrt{\frac{\xi_{\psi,n}^{2}k_{\rho,n}k_{\chi,n}}{n}}
	\]
	
	Combining we get:
	
	\[
	\|\hat{\Sigma}_{\hat{\pi},\lambda_{0}}^{-1/2}\big(\frac{1}{n}\sum_{i=1}^{n}\hat{\pi}_{n,i}\psi_{n,i}'-E[\pi_{n,i}\psi_{n,i}']\big)\|\precsim_{p}\sqrt{\frac{\lambda_{0}\bar{r}_{\pi,n}^{2}}{\mu_{n}+\lambda_{0}^{2}}}+\sqrt{\frac{\xi_{\psi,n}^{2}k_{\rho,n}k_{\chi,n}}{n}}
	\]
	
	Note that $E\big[\Sigma_{\pi,\lambda_{0}}^{-1/2}E[\pi_{n,i}\psi_{n,i}']\Sigma_{\psi,n}^{-1}\psi_{n,i}\epsilon_{i}\big]=0$
	and by properties of least squares projections:
	
	\begin{align*}
		& E[\|\Sigma_{\pi,\lambda_{0}}^{-1/2}E[\pi_{n,i}\psi_{n,i}']\Sigma_{\psi,n}^{-1}\psi_{n,i}\epsilon_{i}\|^{2}]\\
		\leq & E[\|\Sigma_{\pi,\lambda_{0}}^{-1/2}E[\pi_{n,i}\psi_{n,i}']\Sigma_{\psi,n}^{-1}\psi_{n,i}\|^{2}]\text{ess sup}E[\|\epsilon_{i}\|^{2}|z_{i},x_{i},d_{i}]\\
		= & \|\Sigma_{\pi,\lambda_{0}}^{-1/2}E[\pi_{n,i}\psi_{n,i}']\Sigma_{\psi,n}^{-1/2}\|_{F}^{2}\bar{\sigma}_{\epsilon}^{2}\\
		\leq & E[\|\Sigma_{\psi,n}^{-1/2}\psi_{n,i}\|^{2}]\bar{\sigma}_{\epsilon}^{2}\\
		\precsim & k_{\psi,n}
	\end{align*}
	
	Where in the above $\bar{\sigma}_{\epsilon}^{2}$ is the essential
	supremum of $E[\epsilon_{i}^{2}|x_{i},d_{i},z_{i}]$ which is bounded
	by Assumption 4.7.i. And so, by the law of large numbers: 
	\[
	\|\frac{1}{n}\sum_{i=1}^{n}\Sigma_{\pi,\lambda_{0}}^{-1/2}E[\pi_{n,i}\psi_{n,i}']\Sigma_{\psi,n}^{-1}\psi_{n,i}(\epsilon_{i}+R_{g,i})\|\precsim_{p}\sqrt{\frac{k_{\psi,n}}{n}}
	\]
	
	By the properties fo the matrix norm and Lemma C.6 we have:
	
	\begin{align*}
		& \|\hat{\Sigma}_{\hat{\pi},\lambda_{0}}^{-1/2}(\Sigma_{\pi}-\hat{\Sigma}_{\hat{\pi}})\Sigma_{\pi,\lambda_{0}}^{-1/2}\|\\
		\leq & \|\hat{\Sigma}_{\hat{\pi},\lambda_{0}}^{-1/2}\Sigma_{\pi,\lambda_{0}}^{1/2}\|\|\Sigma_{\pi,\lambda_{0}}^{-1/2}(\Sigma_{\pi}-\hat{\Sigma}_{\hat{\pi}})\Sigma_{\pi,\lambda_{0}}^{-1/2}\|\\
		\precsim_{p} & \sqrt{\frac{\xi_{\kappa,n}^{2}ln(k_{\kappa,n})}{n}}+\sqrt{\frac{\lambda_{0}\bar{r}_{\pi,n}^{2}}{\mu_{n}+\lambda_{0}^{2}}}
	\end{align*}
	
	Combining the various rates above with \eqref{eq:compbound} and simplifying
	we get:
	
	\begin{align*}
		& \sqrt{n}\tilde{r}_{3}\\
		\precsim_{p} & \sqrt{\frac{\lambda_{0}\bar{r}_{\pi,n}^{2}k_{\psi,n}}{\mu_{n}+\lambda_{0}^{2}}}+\sqrt{\frac{\big(\xi_{\kappa,n}^{2}ln(k_{\kappa,n})+\xi_{\psi,n}^{2}k_{\rho,n}k_{\chi,n}\big)k_{\psi,n}}{n}}
	\end{align*}
	
	Next we consider $\tilde{r}_{4}$. By the triangle inequality and
	properties of the matrix norm: 
	\begin{align*}
		\tilde{r}_{4} & \leq\|\hat{\Sigma}_{\hat{\pi},\lambda_{0}}^{1/2}\big(\Sigma_{\pi,\lambda_{0}}^{-1}E[\pi_{n,i}\kappa_{n,i}']-\hat{\Sigma}_{\hat{\pi},\lambda_{0}}^{-1}\frac{1}{n}\sum_{i=1}^{n}\hat{\pi}_{n,i}\kappa_{n,i}'\big)vec\big(\Sigma_{\zeta,n}^{-1}\frac{1}{n}\sum_{i=1}^{n}\zeta_{n,i}\upsilon_{n,i}'\tilde{\theta}_{n}\big)\|\\
		& \leq\|\hat{\Sigma}_{\hat{\pi},\lambda_{0}}^{-1/2}\big(E[\pi_{n,i}\kappa_{n,i}']-\frac{1}{n}\sum_{i=1}^{n}\hat{\pi}_{n,i}\kappa_{n,i}'\big)\|\|\Sigma_{\zeta,n}^{-1}\|\|\frac{1}{n}\sum_{i=1}^{n}\zeta_{n,i}\upsilon_{n,i}'\tilde{\theta}_{n}\|_{F}\\
		& +\|\hat{\Sigma}_{\hat{\pi},\lambda_{0}}^{-1/2}(\Sigma_{\pi}-\hat{\Sigma}_{\hat{\pi}})\Sigma_{\pi,\lambda_{0}}^{-1/2}\|\|\frac{1}{n}\sum_{i=1}^{n}\Sigma_{\pi,\lambda_{0}}^{-1/2}E[\pi_{n,i}\kappa_{n,i}']vec\big(\Sigma_{\zeta,n}^{-1}\zeta_{n,i}\upsilon_{n,i}'\tilde{\theta}_{n}\big)\|
	\end{align*}
	
	By Assumption 4.2.i $\|\Sigma_{\zeta,n}^{-1}\|$ is bounded. Using
	the same steps as in the proof of Lemma C.4 we have $\|\frac{1}{n}\sum_{i=1}^{n}\zeta_{n,i}\upsilon_{n,i}'\tilde{\theta}_{n}\|_{F}\precsim_{p}\sqrt{\frac{k_{\zeta,n}}{n}}$.
	Using similar steps to those used to obtain the rate of $\tilde{r}_{3}$
	we get:
	
	\[
	\|\hat{\Sigma}_{\hat{\pi},\lambda_{0}}^{-1/2}\big(\frac{1}{n}\sum_{i=1}^{n}\hat{\pi}_{n,i}\kappa_{n,i}'-E[\pi_{n,i}\kappa_{n,i}']\big)\|\precsim_{p}\sqrt{\frac{\lambda_{0}\bar{r}_{\pi,n}^{2}}{\mu_{n}+\lambda_{0}^{2}}}+\sqrt{\frac{\xi_{\kappa,n}^{2}k_{\rho,n}k_{\chi,n}}{n}}
	\]
	
	Next note that: 
	\[
	E\big[\Sigma_{\pi,\lambda_{0}}^{-1/2}E[\pi_{n,i}\kappa_{n,i}']vec\big(\Sigma_{\zeta,n}^{-1}\zeta_{n,i}\upsilon_{n,i}'\tilde{\theta}_{n}\big]=0
	\]
	
	Note that for a vector $\theta$, by properties of least squares projections
	we have: 
	\[
	\|\Sigma_{\pi,\lambda_{0}}^{-1/2}E[\pi_{n,i}\kappa_{n,i}']\theta\|^{2}\leq E[\|\kappa_{n,i}'\theta\|^{2}]=\|\Sigma_{\kappa,n}^{1/2}\theta\|
	\]
	
	Using the above we see that: 
	\begin{align*}
		& E\big[\|\Sigma_{\pi,\lambda_{0}}^{-1/2}E[\pi_{n,i}\kappa_{n,i}']vec\big(\Sigma_{\zeta,n}^{-1}\zeta_{n,i}\upsilon_{n,i}'\tilde{\theta}_{n}\big)\|^{2}\big]\\
		\leq & E\big[\|\Sigma_{\kappa,n}^{1/2}vec\big(\Sigma_{\zeta,n}^{-1}\zeta_{n,i}\upsilon_{n,i}'\tilde{\theta}_{n}\big)\|^{2}\big]\\
		\leq & \|\Sigma_{\kappa,n}\|\|\Sigma_{\zeta,n}^{-1}\|^{2}E[\|\zeta_{n,i}\upsilon_{n,i}'\tilde{\theta}_{n}\|_{F}^{2}]\\
		\precsim & \sqrt{k_{\zeta,n}}
	\end{align*}
	
	Where the penultimate line follows by properties of the matrix norm
	and the final line follows by Assumption 4.2.i and the reasoning in
	the proof of Lemma C.4. Applying the law of large numbers we then
	get: 
	\[
	\|\frac{1}{n}\sum_{i=1}^{n}\Sigma_{\pi,\lambda_{0}}^{-1/2}E[\pi_{n,i}\kappa_{n,i}']vec\big(\Sigma_{\zeta,n}^{-1}\zeta_{n,i}\upsilon_{n,i}'\tilde{\theta}_{n}\big)\|\precsim_{p}\sqrt{\frac{k_{\zeta,n}}{n}}
	\]
	
	Combining and using the earlier rate for $\|\hat{\Sigma}_{\hat{\pi},\lambda_{0}}^{-1/2}(\Sigma_{\pi}-\hat{\Sigma}_{\hat{\pi}})\Sigma_{\pi,\lambda_{0}}^{-1/2}\|$,
	we get: 
	\begin{align*}
		\sqrt{n}\tilde{r}_{4} & \precsim_{p}\sqrt{\frac{\lambda_{0}\bar{r}_{\pi,n}^{2}k_{\zeta,n}}{\mu_{n}+\lambda_{0}^{2}}}+\sqrt{\frac{\big(ln(k_{\kappa,n})+k_{\rho,n}k_{\chi,n}\big)k_{\zeta,n}\xi_{\kappa,n}^{2}}{n}}
	\end{align*}
	
	Finally, we consider $\tilde{r}_{5}$. Using the properties of the
	matrix norm and our earlier results: 
	\begin{align*}
		\sqrt{n}\tilde{r}_{5} & =\sqrt{n}\lambda_{0}\|\hat{\Sigma}_{\hat{\pi},\lambda_{0}}^{-1/2}\theta_{n}\|\\
		& \leq\sqrt{n}\lambda_{0}\|\Sigma_{\pi,\lambda_{0}}^{1/2}\hat{\Sigma}_{\hat{\pi},\lambda_{0}}^{-1/2}\|\|\Sigma_{\pi,\lambda_{0}}^{-1/2}\|\|\theta_{n}\|\\
		& \precsim_{p}\sqrt{\frac{\lambda_{0}^{3}n}{\mu_{n}+\lambda_{0}^{2}}}
	\end{align*}
	
	Combining, we get 
	\begin{align*}
		\sqrt{n}\|\Sigma_{\pi,\lambda_{0}}^{1/2}R_{\theta,n}\| & \precsim_{p}\bar{l}_{g,n}+\bar{l}_{\pi,n}+\sqrt{n}r_{\theta,n}+r_{s,n}\\
		& +\sqrt{\frac{\bar{r}_{\pi,n}^{2}(k_{\zeta,n}+k_{\psi,n})/\lambda_{0}}{1+\mu_{n}/\lambda_{0}^{2}}}+\sqrt{\frac{\lambda_{0}n}{1+\mu_{n}/\lambda_{0}^{2}}}
	\end{align*}
	
	Now, from our earlier decomposition:
	
	\begin{align*}
		& \alpha_{n}(x_{1},x_{2},d)'\hat{\theta}-E[Y(x_{1})|X=x_{2},D=d]\\
		= & \alpha_{n}(x_{1},x_{2},d)'\Sigma_{\pi,\lambda_{0}}^{-1}E[\pi_{n,i}\psi_{n,i}']\Sigma_{\psi,\lambda_{\psi}}^{-1}\frac{1}{n}\sum_{i=1}^{n}\psi_{n,i}\epsilon_{i}\\
		- & \alpha_{n}(x_{1},x_{2},d)'\Sigma_{\pi,\lambda_{0}}^{-1}E[\pi_{n,i}\kappa_{n,i}']vec\big(\Sigma_{\zeta,\lambda_{\pi}}^{-1}\frac{1}{n}\sum_{i=1}^{n}\zeta_{n,i}\upsilon_{n,i}'\tilde{\theta}_{n}\big)\\
		+ & \alpha_{n}(x_{1},x_{2},d)'R_{\theta,n}\\
		+ & E[Y(x_{1})|X=x_{2},D=d]-\alpha_{n}(x_{1},x_{2},d)'\theta_{n}
	\end{align*}
	
	By the properties of the matrix norm and the definition of $s_{n}(x_{1},x_{2},d)$:
	
	\begin{align*}
		\|\alpha_{n}(x_{1},x_{2},d)'R_{\theta,n}\| & \leq\|\alpha_{n}(x_{1},x_{2},d)'\Sigma_{\pi,\lambda_{0}}^{-1/2}\|\|\Sigma_{\hat{\pi},\lambda_{0}}^{1/2}R_{\theta,n}\|\\
		& \leq\|\Omega_{n}^{-1/2}\|\|s_{n}(x_{1},x_{2},d)\|\|\Sigma_{\hat{\pi},\lambda_{0}}^{1/2}R_{\theta,n}\|
	\end{align*}
	
	By supposition, the smallest eigenvalue of $\Omega_{n}$ is bounded
	away from zero and so $\|\Omega_{n}^{-1/2}\|\precsim1$. So we get
	that unfirmly over $x_{1}$, $x_{2}$, and $d$: 
	\begin{align*}
		\|\alpha_{n}(x_{1},x_{2},d)'R_{\theta,n}\| & \precsim\|s_{n}(x_{1},x_{2},d)\|\|\hat{\Sigma}_{\hat{\pi},\lambda_{0}}^{1/2}R_{\theta,n}\|
	\end{align*}
	
	By Lemma 4.1: 
	\[
	\|E[Y(x_{1})|X=x_{2},D=d]-\alpha_{n}(x_{1},x_{2},d)'\theta_{n}\|\precsim r_{\theta,n}
	\]
	
	Combining we get that uniformly over $(x_{1},x_{2},d)\in\mathcal{S}$:
	\begin{align*}
		& \frac{\sqrt{n}}{\|s_{n}(x_{1},x_{2},d)\|}\bigg(\alpha_{n}(x_{1},x_{2},d)'\hat{\theta}-E[Y(x_{1})|X=x_{2},D=d]\\
		- & \alpha_{n}(x_{1},x_{2},d)'\Sigma_{\pi,\lambda_{0}}^{-1}E[\pi_{n,i}\psi_{n,i}']\Sigma_{\psi,n}^{-1}\frac{1}{n}\sum_{i=1}^{n}\psi_{n,i}\epsilon_{i}\\
		+ & \alpha_{n}(x_{1},x_{2},d)'\Sigma_{\pi,\lambda_{0}}^{-1}E[\pi_{n,i}\kappa_{n,i}']vec\big(\Sigma_{\zeta,n}^{-1}\frac{1}{n}\sum_{i=1}^{n}\zeta_{n,i}\upsilon_{n,i}'\tilde{\theta}_{n}\big)\bigg)\\
		\precsim_{p} & \bigg(\bar{l}_{g,n}+\bar{l}_{\pi,n}+\sqrt{n}r_{\theta,n}+r_{s,n}\bigg)\\
		+ & \sqrt{\frac{\lambda_{0}n+\bar{r}_{\pi,n}^{2}(k_{\zeta,n}+k_{\psi,n})/\lambda_{0}}{1+\mu_{n}/\lambda_{0}^{2}}}
	\end{align*}
	
	Now, we obtain a rate for $\|\Sigma_{\pi,\lambda_{0}}^{1/2}(\hat{\theta}-\theta_{n})\|$.
	By the triangle inequality and properties of the matrix norm: 
	\begin{align*}
		& \|\Sigma_{\pi,\lambda_{0}}^{1/2}(\hat{\theta}-\theta_{n})\|\\
		\leq & \|\Sigma_{\pi,\lambda_{0}}^{-1/2}E[\pi_{n,i}\psi_{n,i}']\Sigma_{\psi,n}^{-1}\frac{1}{n}\sum_{i=1}^{n}\psi_{n,i}\epsilon_{i}\|\\
		+ & \|\Sigma_{\pi,\lambda_{0}}^{-1/2}E[\pi_{n,i}\kappa_{n,i}']vec\big(\Sigma_{\zeta,n}^{-1}\frac{1}{n}\sum_{i=1}^{n}\zeta_{n,i}\upsilon_{n,i}'\tilde{\theta}_{n}\big)\|\\
		+ & \|\Sigma_{\pi,\lambda_{0}}^{1/2}R_{\theta,n}\|
	\end{align*}
	
	We have already derived rates for each term on the RHS. Using our
	earlier results we get: 
	\begin{align*}
		\|\Sigma_{\pi,\lambda_{0}}^{1/2}(\hat{\theta}-\theta_{n})\| & \precsim_{p}\sqrt{\frac{k_{\psi,n}+k_{\zeta,n}}{n}}+\|\Sigma_{\pi,\lambda_{0}}^{1/2}R_{\theta,n}\|
	\end{align*}
	
	\textbf{Step 2: Derive a rate for $\|\big(\hat{\alpha}_{n}(x_{1},x_{2},d)-\alpha_{n}(x_{1},x_{2},d)\big)'(\hat{\theta}-\theta_{n})\|$.}
	
	\begin{align*}
		& \|\big(\hat{\alpha}_{n}(x_{1},x_{2},d)-\alpha_{n}(x_{1},x_{2},d)\big)'(\hat{\theta}-\theta_{n})\|\\
		\leq & \|\hat{\alpha}_{n}(x_{1},x_{2},d)-\alpha_{n}(x_{1},x_{2},d)\|\|\Sigma_{\pi,\lambda_{0}}^{-1/2}||\|\Sigma_{\pi,\lambda_{0}}^{1/2}(\hat{\theta}-\theta_{n})\|
	\end{align*}
	
	From Lemma C.5 we have:
	
	\[
	\sup_{(x_{1},x_{2},d)\in\mathcal{S}}\|\hat{\alpha}_{n}(x_{1},x_{2},d)-\alpha_{n}(x_{1},x_{2},d)\|\precsim_{p}\bar{r}_{\alpha,n}
	\]
	
	Where $\bar{r}_{\alpha,n}$ is given in the statement of that lemma.
	Step 1 provides a rate for $\|\Sigma_{\pi,\lambda_{0}}^{1/2}(\hat{\theta}-\theta_{n})\|$,
	and note that $\|\Sigma_{\pi,\lambda_{0}}^{1/2}\|\precsim_{p}\sqrt{\frac{\lambda_{0}}{\mu_{n}+\lambda_{0}^{2}}}$.
	
	In all: 
	\begin{align*}
		& \sqrt{n}\|\big(\hat{\alpha}_{n}(x_{1},x_{2},d)-\alpha_{n}(x_{1},x_{2},d)\big)'(\hat{\theta}-\theta_{n})\|\\
		\leq & \sqrt{\frac{\bar{r}_{\alpha,n}^{2}(k_{\psi,n}+k_{\zeta,n})/\lambda_{0}}{1+\mu_{n}/\lambda_{0}^{2}}}+\sqrt{\frac{\bar{r}_{\alpha,n}^{2}/\lambda_{0}}{1+\mu_{n}/\lambda_{0}^{2}}}\sqrt{n}\|\hat{\Sigma}_{\hat{\pi},\lambda_{0}}^{1/2}R_{\theta,n}\|
	\end{align*}
	
	\textbf{Step 3: Linearize $\big(\hat{\alpha}_{n}(x_{1},x_{2},d)-\alpha_{n}(x_{1},x_{2},d)\big)'\theta_{n}$.}
	
	By Lemma C.5 we have that uniformly over $(x_{1},x_{2},d)$ in $\mathcal{S}$:
	\begin{align*}
		& \frac{\sqrt{n}}{\|\chi_{n}(x_{1},d)\otimes\chi_{n}(x_{2},d)\|}\bigg(\big(\hat{\alpha}_{n}(x_{1},x_{2},d)-\alpha_{n}(x_{1},x_{2},d)\big)'\theta_{n}\\
		- & \chi_{n}(x_{2},d)'\Sigma_{\chi,\lambda_{\alpha}}^{-1}\frac{1}{\sqrt{n}}\sum_{i=1}^{n}\chi_{n,i}u_{n,i}'\tilde{\theta}_{n}\chi_{n}(x_{1},d)\bigg)\\
		\precsim_{p} & \bar{l}_{\alpha,n}
	\end{align*}
	
	By the definition of $s_{n}$, 
	\begin{align*}
		\|\chi_{n}(x_{1},d)\otimes\chi_{n}(x_{2},d)\| & \leq\|\Omega_{n}^{-1/2}s_{n}(x_{1},x_{2},d)\|\\
		& \precsim\|s_{n}(x_{1},x_{2},d)\|
	\end{align*}
	
	And so we have: 
	\begin{align*}
		& \frac{\sqrt{n}}{\|s_{n}(x_{1},x_{2},d)\|}\bigg(\big(\hat{\alpha}_{n}(x_{1},x_{2},d)-\alpha_{n}(x_{1},x_{2},d)\big)'\theta_{n}\\
		- & \chi_{n}(x_{2},d)'\Sigma_{\chi,\lambda_{\alpha}}^{-1}\frac{1}{\sqrt{n}}\sum_{i=1}^{n}\chi_{n,i}u_{n,i}'\tilde{\theta}_{n}\chi_{n}(x_{1},d)\bigg)\\
		\precsim_{p} & \bar{l}_{\alpha,n}
	\end{align*}
	
	\textbf{Step 4: Linearize $\hat{\alpha}_{n}(x_{1},x_{2},d)'\hat{\theta}-\alpha_{n}(x_{1},x_{2},d)'\theta_{n}$.}
	
	Combining the results of the previous steps, uniformly over $x_{1}$,
	$x_{2}$, and $d$:
	
	\textbf{ 
		\begin{align*}
			& \frac{\sqrt{n}}{\|s_{n}(x_{1},x_{2},d)\|}\bigg(\hat{\alpha}_{n}(x_{1},x_{2},d)'\hat{\theta}-\alpha_{n}(x_{1},x_{2},d)'\theta_{n}\\
			- & \alpha_{n}(x_{1},x_{2},d)'\Sigma_{\pi,\lambda_{0}}^{-1}E[\pi_{n,i}\psi_{n,i}']\frac{1}{n}\sum_{i=1}^{n}\Sigma_{\psi,n}^{-1}\psi_{n,i}\epsilon_{i}\\
			+ & \alpha_{n}(x_{1},x_{2},d)'\Sigma_{\pi,\lambda_{0}}^{-1}E[\pi_{n,i}\kappa_{n,i}']\frac{1}{n}\sum_{i=1}^{n}vec(\Sigma_{\zeta,n}^{-1}\zeta_{n,i}\upsilon_{n,i}'\tilde{\theta}_{n})\\
			- & \chi_{n}(x_{2},d)'\Sigma_{\chi,\lambda_{\alpha}}^{-1}\frac{1}{\sqrt{n}}\sum_{i=1}^{n}\chi_{n,i}u_{n,i}'\tilde{\theta}_{n}\chi_{n}(x_{1},d)\bigg)\\
			\precsim_{p} & \bar{l}_{\alpha,n}+\bar{l}_{g,n}+\bar{l}_{\pi,n}+\sqrt{n}r_{\theta,n}+r_{s,n}\\
			+ & \sqrt{\frac{\lambda_{0}n+(\bar{r}_{\alpha,n}^{2}+\bar{r}_{\pi,n}^{2})(k_{\psi,n}+k_{\zeta,n})/\lambda_{0}}{1+\mu_{n}/\lambda_{0}^{2}}}
		\end{align*}
	}
	
	Using the definitions of $s_{n}$ and $\eta_{n,i}$ we can write this
	more succinctly as:\textbf{ 
		\begin{align*}
			& \frac{\sqrt{n}\big(\hat{\alpha}_{n}(x_{1},x_{2},d)'\hat{\theta}-\alpha_{n}(x_{1},x_{2},d)'\theta_{n}\big)}{\|s_{n}(x_{1},x_{2},d)\|}-\frac{s_{n}(x_{1},x_{2},d)'\Omega_{n}^{-1/2}\frac{1}{\sqrt{n}}\sum_{i=1}^{n}\eta_{n,i}}{\|s_{n}(x_{1},x_{2},d)\|}\\
			\precsim_{p} & \bar{l}_{\alpha,n}+\bar{l}_{g,n}+\bar{l}_{\pi,n}+\sqrt{n}r_{\theta,n}+r_{s,n}\\
			+ & \sqrt{\frac{\lambda_{0}n+(\bar{r}_{\alpha,n}^{2}+\bar{r}_{\pi,n}^{2})(k_{\psi,n}+k_{\zeta,n})/\lambda_{0}}{1+\mu_{n}/\lambda_{0}^{2}}}
		\end{align*}
	}
	
	By Assumption 4.7, the RHS tends to zero with $n$. \end{proof}

\begin{proof}[Proof of Theorem 4.3]
	
	\textbf{Step 1: Asymptotic Normality}
	
	We apply Yurinskii's coupling (see e.g., Theorem 10 in \citet{Pollard2001}
	or \citet{Belloni2015}). This states that for any $\delta>0$ and
	sequence of independent, zero-mean length-$k_{n}$ vectors $a_{n,i}$
	with finite third moments, there is length-$k_{n}$ multivariate Guassian
	$\mathcal{N}_{n}$ with mean zero and same covariance matrix as $\frac{1}{\sqrt{n}}\sum_{i=1}^{n}a_{n,i}$
	so that: 
	\[
	P\big[\|\frac{1}{\sqrt{n}}\sum_{i=1}^{n}a_{n,i}-\mathcal{N}_{n}\|>3\delta\big]\leq C_{0}k_{n}q_{n}n^{-1/2}\delta^{-3}\big(1+\frac{|-ln(k_{n}q_{n}\delta^{-3}n^{-1/2})|}{k_{n}}\big)
	\]
	$C_{0}$ is a finite constant and $q_{n}=E[\frac{1}{n}\sum_{i=1}^{n}\|a_{n,i}\|^{3}]$.
	To apply this in our case we let $a_{n,i}=\frac{1}{\sqrt{n}}\sum_{i=1}^{n}\eta_{n,i}$
	and note that $E[a_{n,i}]=0$. Furthermore $k_{n}$ is replaced by
	$k_{\psi,n}+k_{\zeta,n}+k_{\chi,n}$ and by the triangle inequality
	and properties of the kronecker product, the average third moment
	$q_{n}$ satisfies: 
	\begin{align*}
		q_{n}^{2/3} & \leq\|\Sigma_{\psi,\lambda_{\psi}}^{-1}\|^{2}E[\|\psi_{n,i}\epsilon_{i}\|^{3}]^{2/3}+\|\Sigma_{\zeta,\lambda_{\psi}}^{-1}\|^{2}E[\|\zeta_{n,i}\upsilon_{n,i}'\tilde{\theta}_{n}\|_{F}^{3}]^{2/3}\\
		& +E[\|\chi_{n,i}u_{n,i}'\tilde{\theta}_{n}\|_{F}^{3}]^{2/3}
	\end{align*}
	
	Applying Assumption 4.2.ii: 
	\begin{align*}
		E[\|\zeta_{n,i}\upsilon_{n,i}'\tilde{\theta}_{n}\|_{F}^{3}]\leq & \text{ess sup}\|\zeta_{n,i}\upsilon_{n,i}'\tilde{\theta}_{n}\|_{F}E[\|\zeta_{n,i}\upsilon_{n,i}'\tilde{\theta}_{n}\|_{F}^{2}]\\
		\leq & \xi_{\zeta,n}\xi_{\rho,n}\|\theta_{n}\|E[\|\zeta_{n,i}\upsilon_{n,i}'\tilde{\theta}_{n}\|_{F}^{2}]\\
		\precsim & \xi_{\zeta,n}\xi_{\rho,n}k_{\zeta,n}
	\end{align*}
	
	Where the final inequality uses that $E[\|\zeta_{n,i}\upsilon_{n,i}'\tilde{\theta}_{n}\|_{F}^{2}]\precsim k_{\zeta,n}$which
	we establish in the proof of Lemma C.4. By similar reasoning we have
	$E[\|\chi_{n,i}u_{n,i}'\tilde{\theta}_{n}\|_{F}^{3}]\precsim\xi_{\chi,n}\xi_{\rho,n}k_{\chi,n}$
	and $E[\|\psi_{n,i}\epsilon_{i}\|^{3}]\precsim\xi_{\psi}k_{\psi,n}$,
	and so:
	
	\[
	q_{n}\precsim\xi_{\psi}k_{\psi,n}+\xi_{\zeta,n}\xi_{\rho,n}k_{\zeta,n}+\xi_{\chi,n}\xi_{\rho,n}k_{\chi,n}
	\]
	
	Assumption 4.8.ii implies that for some $r_{\mathcal{N},n}\to0$:
	\[
	\frac{(k_{\psi,n}+k_{\zeta,n}+k_{\chi,n})(k_{\psi,n}\xi_{\psi}+k_{\zeta,n}\xi_{\zeta,n}\xi_{\rho,n}+k_{\chi,n}\xi_{\chi,n}\xi_{\rho,n})}{ln(n)^{-1}n^{1/2}r_{\mathcal{N},n}^{3}}\to0
	\]
	And so, defining: 
	\[
	\tilde{q}_{n}=\frac{(k_{\psi,n}+k_{\zeta,n}+k_{\chi,n})(\xi_{\psi}k_{\psi,n}+\xi_{\zeta,n}\xi_{\rho,n}k_{\zeta,n}+\xi_{\chi,n}\xi_{\rho,n}k_{\chi,n})}{n^{1/2}}
	\]
	we get that for any fixed $\delta$, for a sufficiently large $n$:
	\begin{align*}
		& \tilde{q}_{n}r_{\mathcal{N},n}^{-3}\delta^{-3}\big(1+\frac{|-ln(\tilde{q}_{n}r_{\mathcal{N},n}^{-3}\delta^{-3})|}{k_{n}}\big)\\
		\leq & \tilde{q}_{n}r_{\mathcal{N},n}^{-3}\delta^{-3}\big(1+\frac{|ln(log(n)\delta^{-3})|}{k_{\phi,n}}\big)\prec1
	\end{align*}
	
	So using the Yurinskii coupling gives us that for all fixed $\delta>0$:
	\begin{align*}
		P\big[||\frac{1}{\sqrt{n}}\sum_{i=1}^{n}\eta_{n,i}-\mathcal{N}_{n}||>3r_{\mathcal{N},n}\delta\big] & \prec_{p}1
	\end{align*}
	Then using $\|\Omega_{n}^{-1/2}\|\precsim1$ and Cauchy-Schwartz,
	we have that:
	
	\begin{align*}
		& |\frac{s_{n}(x_{1},x_{2},d)}{\|s_{n}(x_{1},x_{2},d)\|}'\frac{1}{\sqrt{n}}\sum_{i=1}^{n}\Omega_{n}^{-1/2}\eta_{n,i}-\frac{s_{n}(x_{1},x_{2},d)'}{\|s_{n}(x_{1},x_{2},d)\|}\tilde{\mathcal{N}}_{n}|\\
		\precsim_{p} & r_{\mathcal{N},n}
	\end{align*}
	Where $\tilde{\mathcal{N}}_{n}=\Omega_{n}^{-1/2}\mathcal{N}_{n}$
	and thus has identity vairance covariance matrix. Combining with Lemma
	4.2 gives the first result of the Theorem.
	
	\textbf{Step 2: Bound $\|s_{n}(x_{1},x_{2},d)\|$}
	
	From the definition of $s_{n}$: 
	\begin{align*}
		& \|s_{n}(x_{1},x_{2},d)\|^{2}\\
		= & E\bigg[\|\alpha_{n}(x_{1},x_{2},d)'\Sigma_{\pi,\lambda_{0}}^{-1}E[\pi_{n,i}\psi_{n,i}']\Sigma_{\psi,\lambda_{\psi}}^{-1}\psi_{n,i}\epsilon_{i}\\
		- & \alpha_{n}(x_{1},x_{2},d)'\Sigma_{\pi,\lambda_{0}}^{-1}E[\pi_{n,i}\kappa_{n,i}']\big((\Sigma_{\zeta,n}^{-1}\zeta_{n,i})\otimes(\tilde{\theta}_{n}'\upsilon_{n,i})\big)\\
		+ & \big(\chi_{n}(x_{2},d)\otimes\chi_{n}(x_{1},d)\big)'\Sigma_{\chi,\lambda_{\alpha}}^{-1}\chi_{n,i}\otimes(\tilde{\theta}_{n}'u_{n,i})\|^{2}\bigg]
	\end{align*}
	
	So by the triangle inequality and properties of the matrix norm and
	Kronecker products: 
	\begin{align*}
		& \|s_{n}(x_{1},x_{2},d)\|\\
		\leq & \|\alpha_{n}(x_{1},x_{2},d)\Sigma_{\pi,\lambda_{0}}^{-1/2}\|\|\Sigma_{\pi,\lambda_{0}}^{-1/2}E[\pi_{n,i}\psi_{n,i}']\|\|\Sigma_{\psi,\lambda_{\psi}}^{-1}\|\|E[\psi_{n,i}\psi_{n,i}'\epsilon_{i}^{2}]^{1/2}\|\\
		+ & \|\alpha_{n}(x_{1},x_{2},d)\Sigma_{\pi,\lambda_{0}}^{-1/2}\|\|\Sigma_{\pi,\lambda_{0}}^{-1/2}E[\pi_{n,i}\kappa_{n,i}']\|\|E[(\Sigma_{\zeta,n}^{-1}\zeta_{n,i}\zeta_{n,i}'\Sigma_{\zeta,n}^{-1})\otimes(\tilde{\theta}_{n}'\upsilon_{n,i}\upsilon_{n,i}\tilde{\theta}_{n})]^{1/2}\|\\
		+ & \|\chi_{n}(x_{2},d)\|\|\chi_{n}(x_{1},d)\|\|\Sigma_{\chi,\lambda_{\alpha}}^{-1}\|\|E[(\chi_{n,i}\chi_{n,i}')\otimes(\tilde{\theta}_{n}'u_{n,i}u_{n,i}'\tilde{\theta}_{n})]^{1/2}\|
	\end{align*}
	
	By Lemma C.6 $\|\alpha_{n}(x_{1},x_{2},d)\Sigma_{\pi,\lambda_{0}}^{-1/2}\|\precsim1+\bar{\xi}_{\kappa,n}$,
	by Assumption 4.2.ii $\|\chi_{n}(x_{2},d)\|\|\chi_{n}(x_{1},d)\|\leq\xi_{\chi,n}^{2}$.
	By the properties of least-squares projections $\|\Sigma_{\pi,\lambda_{0}}^{-1/2}E[\pi_{n,i}\psi_{n,i}']\|\leq\|\Sigma_{\psi,n}\|^{1/2}$
	and $\|\Sigma_{\pi,\lambda_{0}}^{-1/2}E[\pi_{n,i}\kappa_{n,i}']\|\leq\|\Sigma_{\kappa,n}\|^{1/2}$.
	By Assumption 4.2.i $\|\Sigma_{\psi,n}\|\precsim1$, $\|\Sigma_{\kappa,n}\|\precsim1$,
	$\|\Sigma_{\psi,\lambda_{\psi}}^{-1}\|\precsim1$, and $\|\Sigma_{\chi,\lambda_{\alpha}}^{-1}\|\precsim1$.
	By iterated expectations, Assumption 4.2.i, and 4.7.i: 
	\begin{align*}
		\|E[\psi_{n,i}\psi_{n,i}'\epsilon_{i}^{2}]^{1/2}\| & \leq\|\Sigma_{\psi,n}\|^{1/2}\text{ess sup}E[\epsilon_{i}^{2}|Z_{i},X_{i},D_{i}]\\
		& \precsim1
	\end{align*}
	
	Next note that for any vector $b$ of appropriate length with $\|b\|\leq1$
	we have:
	
	\begin{align*}
		& \|E\big[\big((\Sigma_{\zeta,n}^{-1}\zeta_{n,i})\otimes(\tilde{\theta}_{n}'\upsilon_{n,i})\big)\big((\Sigma_{\zeta,n}^{-1}\zeta_{n,i})\otimes(\tilde{\theta}_{n}'\upsilon_{n,i})\big)'\big]^{1/2}b\|^{2}\\
		= & E[\|E[\upsilon_{n,i}\upsilon_{n,i}'|Z_{i},X_{i},D_{i}]^{1/2}\tilde{\theta}_{n}\tilde{b}\Sigma_{\zeta,n}^{-1}\zeta_{n,i}\|^{2}]\\
		\leq & \text{ess sup}\|E[\upsilon_{n,i}\upsilon_{n,i}'|Z_{i},X_{i},D_{i}]\|E[\|\tilde{\theta}_{n}\tilde{b}\Sigma_{\zeta,n}^{-1}\zeta_{n,i}\|^{2}]\\
		\leq & k_{\chi,n}\text{ess sup}\|E[\upsilon_{n,i}\upsilon_{n,i}'|Z_{i},X_{i},D_{i}]\|\|\Sigma_{\zeta,n}^{-1}\|\|\tilde{b}\|\|\tilde{\theta}_{n}\|\\
		\leq & k_{\chi,n}\text{ess sup}\|E[\upsilon_{n,i}\upsilon_{n,i}'|Z_{i},X_{i},D_{i}]\|\|\Sigma_{\zeta,n}^{-1}\|\|\tilde{\theta}_{n}\|
	\end{align*}
	
	Where $\tilde{b}$ is $b$ stacked into a matrix in the same manner
	as $\tilde{\theta}_{n}$ for $\theta_{n}$. And so weh have:
	\[
	\|E[(\Sigma_{\zeta,n}^{-1}\zeta_{n,i}\zeta_{n,i}'\Sigma_{\zeta,n}^{-1})\otimes(\tilde{\theta}_{n}'\upsilon_{n,i}\upsilon_{n,i}\tilde{\theta}_{n})]^{1/2}\|\precsim k_{\chi,n}
	\]
	
	And similarly for $\|E[(\chi_{n,i}\chi_{n,i}')\otimes(\tilde{\theta}_{n}'u_{n,i}u_{n,i}'\tilde{\theta}_{n})]^{1/2}\|$.
	So in all, using that $k_{\chi,n}\precsim\xi_{\chi,n}^{2}$ we have:
	
	\[
	\|s_{n}(x_{1},x_{2},d)\|\precsim1+\bar{\xi}_{\kappa,n}+\xi_{\chi,n}^{2}
	\]
	
	\textbf{Step 3: Bootstrap Validity}
	
	We can extend the results from the original estimator to the bootstrap
	estimator. First consider the bootstrap first-stage estimates. To
	extend Lemmas C.3, C.4, and C.5 to the case in which the bootstrap
	first-stage estimates we employ a trick from \citet{Belloni2015}.
	Note that the relevant linearization errors are not defined using
	exponentially weighted sums and bootstrap estimates. For example,
	$\sqrt{n}\big(\frac{1}{n}\sum_{i=1}^{n}l_{b,g,n,i}^{2}\big)^{1/2}$
	in place of $\sqrt{n}\big(\frac{1}{n}\sum_{i=1}^{n}l_{g,n,i}^{2}\big)^{1/2}$
	in Lemma C.3, where $l_{b,g,n,i}$ is given by: 
	\[
	l_{b,g,n,i}=\sqrt{Q_{b,i}}\hat{g}_{b,i}-\sqrt{Q_{b,i}}g_{i}-\sqrt{Q_{b,i}}\psi_{n,i}'\Sigma_{\psi,\lambda_{g}}^{-1}\frac{1}{n}\sum_{i=1}^{n}Q_{b,i}\psi_{n,i}\epsilon_{i}
	\]
	
	The bootstrap coefficient estimates are identical to the original
	estimates except that $\psi_{n,i}$, $\chi_{n,i}$, $\zeta_{n,i}$,
	$\rho_{n,i}$, and the outcome $Y_{i}$, are pre-multiplied by the
	square root of the exponential weight $Q_{b,i}$. Because $E[Q_{b,i}]=1$
	and $Q_{b,i}$ is independent of the sample, the second moments, and
	cross moments of the basis functions and outcomes are unaffected by
	this reweighting. $\max_{1\leq i\leq n}\sqrt{Q_{b,i}}\precsim_{p}\sqrt{ln(n)}$,
	and so, given condition 4.7.iii, Rudelson's LLN also applies when
	the basis functions are reweighted by $\sqrt{Q_{b,i}}$. Exponential
	weighting may reduce the speed at which the approximation error converges
	to zero, which we account for in the $\sqrt{ln(n)}$ term in Assumption
	4.7.iii. Consider the expansion in Lemma C.3. Let $\beta_{g,n}=\Sigma_{\psi,n}^{-1}E[\psi_{n,i}Y_{i}]$
	and $R_{g,n,i}=g_{i}-\psi_{n,i}'\beta_{g,n}$. In Lemma C.3 we use
	the following decomposition of the outcome:
	
	\[
	Y_{i}=\psi_{n,i}'\beta_{g,n}+R_{g,n,i}+\epsilon_{i}
	\]
	
	In the bootstrap case, we simply replace this with an exponentially
	weighted version:
	
	\[
	\sqrt{Q_{b,i}}Y_{i}=\sqrt{Q_{b,i}}\psi_{n,i}'\beta_{g,n}+\sqrt{Q_{b,i}}R_{g,n,i}+\sqrt{Q_{b,i}}\epsilon_{i}
	\]
	
	The properties random exponentials ensures that $\max_{1\leq i\leq n}\sqrt{Q_{b,i}}|R_{g,n,i}|\precsim\sqrt{ln(n)}\max_{1\leq i\leq n}|R_{g,n,i}|$.
	Thus the linearization error for the bootstrap estimator has a rate
	at most $\sqrt{ln(n)}$ times that of the original estimator, but
	the multiplication by $\sqrt{ln(n)}$ in Assumption 4.7.iii ensures
	that the linearization error is still asymptotically negligible. Similar
	reasoning applies for Lemma C.4 and C.5.
	
	For the second-stage estimates one can expand the difference between
	the boostrap estimate $\hat{\theta}_{b}$ and the original estimate
	$\theta_{n}$ analogously to Lemma 4.2: 
	\begin{align*}
		& \hat{\theta}_{b}-\theta_{n}\\
		= & \Sigma_{\pi,\lambda_{0}}^{-1}E[\pi_{n,i}\psi_{n,i}']\Sigma_{\psi,\lambda_{g}}^{-1}\frac{1}{n}\sum_{i=1}^{n}Q_{b,i}\psi_{n,i}\epsilon_{i}\\
		- & \Sigma_{\pi,\lambda_{0}}^{-1}E[\pi_{n,i}\kappa_{n,i}']vec(\Sigma_{\zeta,\lambda_{\pi}}^{-1}\frac{1}{n}\sum_{i=1}^{n}Q_{b,i}\zeta_{n,i}\upsilon_{n,i}'\tilde{\theta}_{n})\\
		+ & R_{b,\theta,n}
	\end{align*}
	
	The remainder term $R_{b,\theta,n}$ has a form that is similar to
	$R_{\theta,n}$ in the original estimator. The difference is that
	sample averages in the original formula are replaced with exponentially
	weighted sample averages. The approximation error $r_{\theta,n}$
	is again blown up by a factor with rate $\sqrt{ln(n)}$ which we account
	for in Assumption 4.7.iv. Again, the fact that $E[Q_{b,i}]=1$ and
	$Q_{b,i}$ is independent of the sample, and the strengthening of
	Assumption 4.2.ii to 4.7.i, ensure our original steps go through.
	In all, we get that:\textbf{ 
		\begin{align*}
			& \frac{\sqrt{n}\big(\hat{\alpha}_{n}(x_{1},x_{2},d)'\hat{\theta}_{b}-\alpha_{n}(x_{1},x_{2},d)'\theta_{n}\big)}{\|s_{n}(x_{1},x_{2},d)\|}-\frac{s_{n}(x_{1},x_{2},d)'\Omega_{n}^{-1/2}\frac{1}{\sqrt{n}}\sum_{i=1}^{n}Q_{b,i}\eta_{n,i}}{\|s_{n}(x_{1},x_{2},d)\|}\\
			\precsim_{p} & ln(n)(\bar{l}_{\alpha,n}+\bar{l}_{g,n}+\bar{l}_{\pi,n}+\sqrt{n}r_{\theta,n})+r_{s,n}\\
			+ & \sqrt{\frac{\lambda_{0}n+(\bar{r}_{\alpha,n}^{2}+\bar{r}_{\pi,n}^{2})(k_{\psi,n}+k_{\zeta,n})/\lambda_{0}}{1+\mu_{n}/\lambda_{0}^{2}}}
		\end{align*}
	}
	
	Differencing from the result in Lemma 4.2 we get by the triangle inequality:
	\begin{align*}
		& \frac{\sqrt{n}\big(\hat{\alpha}_{n}(x_{1},x_{2},d)'\hat{\theta}_{b}-\alpha_{n}(x_{1},x_{2},d)'\hat{\theta}\big)}{\|s_{n}(x_{1},x_{2},d)\|}\\
		- & \frac{s_{n}(x_{1},x_{2},d)'\frac{1}{\sqrt{n}}\sum_{i=1}^{n}(Q_{b,i}-1)\Omega_{n}^{-1/2}\eta_{n,i}}{\|s_{n}(x_{1},x_{2},d)\|}\\
		\precsim_{p} & ln(n)(\bar{l}_{\alpha,n}+\bar{l}_{g,n}+\bar{l}_{\pi,n}+\sqrt{n}r_{\theta,n})+r_{s,n}\\
		+ & \sqrt{\frac{\lambda_{0}n+(\bar{r}_{\alpha,n}^{2}+\bar{r}_{\pi,n}^{2})(k_{\psi,n}+k_{\zeta,n})/\lambda_{0}}{1+\mu_{n}/\lambda_{0}^{2}}}
	\end{align*}
	
	Finally, we apply the Yuriskii coupling argument for $\frac{1}{\sqrt{n}}\sum_{i=1}^{n}(Q_{b,i}-1)\eta_{n,i}$
	conditional on the original data. Note that $E[Q_{b,i}-1]=0$, and
	so $(Q_{b,i}-1)\eta_{n,i}$ is zero mean, even if we condition on
	the original data and treat only $Q_{b,i}$ as random. $E[(Q_{b,i}-1)^{2}]=1$
	and $\{Q_{b,i}\}_{i=1}^{n}$ are iid and thus the second moment matrix
	of $\frac{1}{\sqrt{n}}\sum_{i=1}^{n}(Q_{b,i}-1)\Omega_{n}^{-1/2}\eta_{n,i}$
	conditional on the data is the sample second moment matrix of $\eta_{n,i}$.
	Moreover, $E[(Q_{b,i}-1)^{3}]\precsim1$ and so the third moments
	of $\frac{1}{\sqrt{n}}\sum_{i=1}^{n}(Q_{b,i}-1)\eta_{n,i}$ grow no
	more quickly than the third moments of $\eta_{n,i}$. Thus we can
	apply the same Yuriskii coupling argument above to the weighted average
	$\frac{1}{\sqrt{n}}\sum_{i=1}^{n}(Q_{b,i}-1)\eta_{n,i}$ conditional
	on the original data. Denoting the original data by $data_{n}$ we
	get that for some sequence of zero-mean normals $\mathcal{N}_{n}$
	with identity variance-covariance matrix: 
	\begin{align*}
		P\big[||\frac{1}{\sqrt{n}}\sum_{i=1}^{n}(Q_{b,i}-1)\Omega_{n}^{-1/2}\eta_{n,i}-\mathcal{N}_{n}||>3r_{\mathcal{N},n}\delta|data_{n}\big] & \prec_{p}1
	\end{align*}
	
	Since we can define such a $\mathcal{N}_{n}$ for each realization
	of the data, the above also holds unconditionally on the data for
	a $\mathcal{N}_{n}$ that is independent of the data. Combining yields
	the result. \end{proof}

\section{Additional Supporting Lemmas}

\theoremstyle{plain} \newtheorem*{LAP}{Lemma C.1} \begin{LAP}
	Suppose that for each $x$ and $d$ in the joint support of $X$ and
	$D$ Assumptions 1, 2, 3, hold and equation (1.5) is satisfied by
	some $\gamma(\cdot,x,d)$ with $E[\gamma(V,x,d)^{2}|X=x,D=d]^{1/2}\leq C$.
	In addition support Assumptions 4.2.i, 4.2.v, 4.3.i, 4.3.ii, 4.4.ii,
	and 4.4.iii hold. Then there is some $c<\infty$ so that for each
	$n$ there is a vector-valued function $\gamma_{n}^{*}$ with $E[\|\gamma_{n}^{*}(V)\|^{2}]\leq c$
	and with probabiltiy $1$:
	
	\[
	|g(X,D,Z)-\chi_{n}(X,D)'E[\gamma_{n}^{*}(V)|X,D,Z]|\leq c\ell_{\chi,n}(1)^{\bar{s}}
	\]
	
	Where $\bar{s}=\frac{\min\{1,s\}}{1+\min\{1,s\}}$. \end{LAP}

\begin{proof}
	
	By supposition, for each $x$ and $d$ there is a function $\gamma(\cdot,x,d)$
	with $E[\gamma(V,x,d)^{2}|X=x,D=d]^{1/2}\leq C$ so that: 
	\[
	g(z,x,d)=E[\gamma(V,x,d)|Z=z,X=x,D=d]
	\]
	
	Define the function $g^{*}$ as follows:
	
	\begin{align*}
		g^{*}(x,d,z,x',d') & =E[\gamma(x,d,V)|Z=z,X=x',D=d']
	\end{align*}
	
	Note that $g^{*}(x,d,z,x,d)=g(x,d,z)$. By the triangle inequality:
	
	\begin{align*}
		|g^{*}(x,d,z,x,d)-g^{*}(x',d',z,x,d)| & \leq|g(x,d,z)-g(x',d',z)|\\
		& +|g^{*}(x',d',z,x,d)-g^{*}(x',d',z,x',d')|
	\end{align*}
	
	By Assumption 4.4.iii there is a constant finite $c_{g}$ so that:
	\[
	|g(x,d,z)-g(x',d',z)|\leq c_{g}(\|x-x\|^{2}+\|d-d'\|^{2})^{\min\{1,s\}/2}
	\]
	By Cauchy-Schwartz for some finite constant $c_{f}$: 
	\begin{align*}
		& |g^{*}(x',d',z,x,d)-g^{*}(x',d',z,x',d')|\\
		= & |E[\gamma(x',d',V)\bigg(\frac{f_{V|XDZ}(V|x,d,z)}{f_{V}(V)}-\frac{f_{V|XDZ}(V|x',d',z)}{f_{V}(V)}\bigg)]|\\
		\leq & \text{ess sup}\frac{f_{V}(V)}{f_{V|XD}(V|x',d')}E[\gamma(x',d',V)^{2}|X=x',D=d']^{1/2}\\
		& \times E\bigg[\bigg(\frac{f_{V|XDZ}(V|x,d,z)}{f_{V}(V)}-\frac{f_{V|XDZ}(V|x',d',z)}{f_{V}(V)}\bigg)^{2}\bigg]^{1/2}\\
		\leq & c_{f}C(\|x-x\|^{2}+\|d-d'\|^{2})^{\min\{1,s\}/2}
	\end{align*}
	
	Where the final inequality holds for some $c_{f}$ by Assumptions
	4.3.i and 4.4.ii. And so in all there is a constant $c$ so that:
	
	\begin{align}
		|g^{*}(x,d,z,x,d)-g^{*}(x',d',z,x,d)| & \leq c(\|x-x\|^{2}+\|d-d'\|^{2})^{\min\{1,s\}/2}\label{eq:distb}
	\end{align}
	
	Now, we will use the above to show that we can approximate $\gamma(x,d,v)$
	with a function that is smooth in $x$ and $d$. Let $b_{n}\to0$
	with $b_{n}>0$. Let $B_{(x,d),b_{n}}$ be the Euclidean ball in $\mathbb{R}^{\dim(X)+\dim(D)}$
	of radius $b_{n}$ centered at $(x,d)$, and define a kernel smoothing
	operator $M_{n}$ as follows:
	
	\[
	M_{n}[\delta](x,d)=\frac{\int_{\mathcal{XD}\cap B_{(x,d),b_{n}}}\delta(x',d')d(x',d')}{\int_{\mathcal{XD}\cap B_{(x,d),b_{n}}}d(x',d')}
	\]
	
	Where $\mathcal{X\mathcal{D}}$ denotes the joint support of $X$
	and $D$. We define $\tilde{\gamma}(x,d,v)=M_{n}[\gamma(\cdot,\cdot,v)](x,d)$
	and a function $\tilde{g}^{*}$ by: 
	\begin{align*}
		\tilde{g}^{*}(x,d,z,x',d') & =E[\tilde{\gamma}(x,d,V)|Z=z,X=x',D=d']\\
		& =M_{n}[g^{*}(\cdot,\cdot,z,x',d')](x,d)
	\end{align*}
	
	It is not too hard to see that from \eqref{eq:distb} we get: 
	\[
	|g^{*}(x,d,z,x,d)-\tilde{g}^{*}(x,d,z,x,d)|\leq cb_{n}^{\min\{1,s\}}
	\]
	And so: 
	\[
	|g(x,d,z)-\tilde{g}^{*}(x,d,z,x,d)|\leq cb_{n}^{\min\{1,s\}}
	\]
	
	Moreover, note that there exists a $\bar{c}$ with $\frac{f_{V|XD}(v|x,d)}{f_{V|XD}(v|x',d')}\leq\bar{c}$
	from Assumption 4.3.i, using this and Jensen's inequality we get:
	\begin{align*}
		& E[\tilde{\gamma}(x,d,V)^{2}|X=x,D=d]\\
		\leq & E\big[M[\gamma(\cdot,\cdot,V)^{2}](x,d)\big|X=x,D=d\big]\\
		= & \int_{\mathcal{V}}\frac{\int_{\mathcal{XD}\cap B_{(x,d),b_{n}}}\gamma(x',d',v)^{2}d(x',d')}{\int_{\mathcal{XD}\cap B_{(x,d),b_{n}}}d(x',d')}f_{V|XD}(v|x,d)dv\\
		\leq & \bar{c}\int_{\mathcal{V}}\frac{\int_{\mathcal{XD}\cap B_{(x,d),b_{n}}}\gamma(x',d',v)^{2}d(x',d')}{\int_{\mathcal{XD}\cap B_{(x,d),b_{n}}}d(x',d')}f_{V|XD}(v|x',d')dv\\
		= & \bar{c}\frac{\int_{\mathcal{XD}\cap B_{(x,d),b_{n}}}E[\gamma(x',d',V)^{2}|X=x',D=d']d(x',d')}{\int_{\mathcal{XD}\cap B_{(x,d),b_{n}}}d(x',d')}\\
		\leq & \bar{c}C
	\end{align*}
	
	Next we show that $\tilde{g}^{*}(x',d',z,x,d)$ is smooth in $d'$
	and $x'$. Note that: 
	\begin{align*}
		& \tilde{g}^{*}(x'',d'',z,x,d)-\tilde{g}^{*}(x',d',z,x,d)\\
		= & M[g^{*}(\cdot,\cdot,z,x,d)](x'',d'')-M[g^{*}(\cdot,\cdot,z,x,d)](x',d')
	\end{align*}
	Using Assumption 4.3.ii there is a constant $\underline{r}>0$ so
	that $\frac{\int_{\mathcal{X}\cap B_{x,b_{n}}}dx'}{\int_{B_{0,b_{n}}}dx'}\geq\underline{r}$.
	Using this, and some geometry to lower bound the volume of two Euclidean
	balls and a rectangle, one can show that for any function $\delta$
	with $|\delta(X,D)|\leq c$ almost surely: 
	\begin{equation}
		|M_{n}[\delta](x_{1},d_{1})-M_{n}[\delta](x_{2},d_{2})|\leq\frac{2}{\underline{r}}c\frac{\dim(X)+\dim(D)}{b_{n}}(\|x_{1}-x_{2}\|^{2}+\|d_{1}-d_{2}\|^{2})^{1/2}\label{eq:lipzres-1}
	\end{equation}
	
	Now, in our setting the constant $c$ in the above is equal to the
	essential supremum of $|g^{*}(x',d',z,x,d)|$ which is given below:
	\begin{align*}
		& |g^{*}(x',d',z,x,d)|\\
		= & E[\gamma(x',d',V)|Z=z,X=x,D=d]|\\
		\leq & \frac{f_{V|ZXD}(v|z,x,d)}{f_{V|XD}(v|x'',d'')}|E[\gamma(x',d',V)|X=x',D=d']|\\
		\leq & \bar{c}C
	\end{align*}
	
	So we conclude that $\tilde{g}^{*}(x',d',z,x,d)$ is Lipschitz continuous
	in $x'$ and $d'$ with constant at most $b_{n}^{-1}c^{*}$ where
	$c^{*}=\frac{2}{\underline{r}}\bar{c}C\big(\dim(X)+\dim(D)\big)$.
	So by Assumption 4.2.v: 
	\[
	|\tilde{g}^{*}(x',d',z,x,d)-\chi_{n}(x',d')\Sigma_{\chi,n}^{-1}E[\chi(X,D)\tilde{g}^{*}(X,D,z,x,d)]|\leq\frac{c^{*}}{b_{n}}\ell_{\chi,n}(1)
	\]
	
	Define $\gamma_{n}^{*}(v)$ by: 
	\[
	\gamma^{*}(v)=\Sigma_{\chi,n}^{-1}E[\chi_{n}(X,D)\tilde{\gamma}(v,X,D)]
	\]
	
	Then we have:
	
	\begin{align*}
		& \chi_{n}(x',d')'\Sigma_{\chi,n}^{-1}E[\chi_{n}(X,D)\tilde{g}^{*}(X,D,z,x,d)]\\
		= & \chi_{n}(x',d')'E[\gamma_{n}^{*}(V)|Z=z,X=x,D=d]
	\end{align*}
	
	By properties of least squares we have:
	
	\begin{align*}
		\|\gamma^{*}(v)\| & \leq\|\Sigma_{\chi,n}^{-1/2}\|\|\Sigma_{\chi,n}^{-1/2}E[\chi_{n}(X,D)\tilde{\gamma}(v,X,D)]\|\\
		& \leq\|\Sigma_{\chi,n}^{-1/2}\|E[\tilde{\gamma}(v,X,D)^{2}]^{1/2}
	\end{align*}
	
	And so, using Assumptions 4.3.i and 4.2.i we have for some finite
	constants $c_{f}$ and $c_{\chi}$:
	
	\begin{align*}
		& E[\|\gamma^{*}(V)\|^{2}]\\
		\leq & \|\Sigma_{\chi,n}^{-1}\|\int_{\mathcal{V}}\int_{\mathcal{X}\mathcal{D}}\tilde{\gamma}(v,x,d)^{2}f_{XD}(x,d)d(x,d)f_{V}(v)dv\\
		= & \|\Sigma_{\chi,n}^{-1}\|\int_{\mathcal{V}}\int_{\mathcal{X}\mathcal{D}}\tilde{\gamma}(v,x,d)^{2}\frac{f_{V}(v)}{f_{V|XD}(v|x,d)}f_{V|XD}(v|x,d)dvf_{XD}(x,d)d(x,d)\\
		\leq & c_{f}\|\Sigma_{\chi,n}^{-1}\|\int_{\mathcal{V}}\int_{\mathcal{X}\mathcal{D}}\tilde{\gamma}(v,x,d)^{2}f_{V|XD}(v|x,d)dvf_{XD}(x,d)d(x,d)\\
		= & c_{f}\|\Sigma_{\chi,n}^{-1}\|E[\tilde{\gamma}(V,X,D)^{2}]\\
		\leq & c_{f}c_{\chi}C
	\end{align*}
	
	Combining, and using the triangle inequality we get: 
	\begin{align*}
		& |g(x,d,z)-\chi_{n}(x,d)'E[\gamma^{*}(V)|Z=z,X=x,D=d]|\\
		\leq & cb_{n}^{\min\{1,s\}}+\frac{c^{*}\tilde{c}}{b_{n}}\ell_{\chi,n}(1)
	\end{align*}
	
	Optimizing over $b_{n}$ we get the that for some constant $c^{\circ}$:
	\begin{align*}
		& |g(x,d,z)-\chi_{n}(x,d)'E[\gamma^{*}(V)|Z=z,X=x,D=d]|\\
		\leq & c^{\circ}\ell_{\chi,n}(1)^{\min\{1,s\}/(1+\min\{1,s\})}
	\end{align*}
	
\end{proof}

\theoremstyle{plain} \newtheorem*{LA2}{Lemma C.2} \begin{LA2}
	
	Suppose Assumptions 4.2.i, 4.2.iv, 4.2.v, and 4.4.ii hold. Then for
	some constant $c<\infty$: 
	\begin{align*}
		\|E[\rho_{n,i}'|X_{i},D_{i},Z_{i}]-\zeta_{n,i}'\Sigma_{\zeta,n}^{-1}E[\zeta_{n,i}\rho_{n,i}']\| & \leq c\ell_{\mathcal{\zeta},n}(s)
	\end{align*}
	
	And: 
	\[
	\|E[\rho_{n,i}'|X_{i},D_{i}]-\chi_{n,i}'\Sigma_{\chi,n}^{-1}E[\chi_{n,i}\rho_{n,i}']\|\leq c\ell_{\mathcal{\chi},n}(s)
	\]
\end{LA2} \begin{proof}
	
	By Assumption 4.4.ii, with probability $1$, $\frac{f_{V|XDZ}(V_{i}|\cdot,\cdot,\cdot)}{f_{V}(V_{i})}\in\Lambda_{s}^{\dim(X,D,Z)}(c)$.
	This means that for any vector $q\in\mathbb{N}_{0}^{\dim(X,D,Z)}$
	whose entries sum to less than $s$, the partial derivative $\Delta_{q}[\frac{f_{V|X}(V_{i}|\cdot,\cdot,\cdot)}{f_{V}(V_{i})}](x,d,z)$
	exists and has magnitude less than some constant $c$ (see the definition
	at the beginning of B.3). Let $\theta\in\mathbb{R}^{k_{\rho,n}}$.
	By the dominated convergence theorem we can differentiate under the
	integral to get: 
	\begin{align}
		& \big|\Delta_{q}\big[E[\rho_{n,i}'\theta|X=\cdot,D=\cdot,Z=\cdot]\big](x,d,z)\big|\nonumber \\
		= & \big|E\big[\rho_{n,i}'\theta\Delta_{q}[\frac{f_{V|XDZ}(V_{i}|\cdot,\cdot,\cdot)}{f_{V}(V_{i})}](x,d,z)\big]\big|\nonumber \\
		\leq & cE[|\rho_{n,i}'\theta|^{2}]^{1/2}\nonumber \\
		= & c\|\Sigma_{\rho,n}^{1/2}\theta\|\label{eq:smooth1-1}
	\end{align}
	Moreover, for any $q\in\mathbb{N}_{0}^{\dim(X,D,Z)}$ whose entries
	sum to $\lfloor s\rfloor$, we have that with probability $1$: 
	\begin{align*}
		& \big|\Delta_{q}[\frac{f_{V|XDZ}(V_{i}|\cdot,\cdot,\cdot)}{f_{V}(V_{i})}](x_{1},d_{1},z_{1})-\Delta_{q}[\frac{f_{V|XDZ}(V_{i}|\cdot,\cdot,\cdot)}{f_{V}(V_{i})}](x_{2},d_{2},z_{2})\big|\\
		\leq & c(\|x_{1}-x_{2}\|^{2}+\|d_{1}-d_{2}\|^{2}+\|z_{1}-z_{2}\|^{2})^{\frac{s-\lfloor s\rfloor}{2}}
	\end{align*}
	Again, differentiating under the integral and using Cauchy-Schwartz
	we get: 
	\begin{align}
		& \big|\Delta_{q}\big[E[\rho_{n,i}'\theta|X=\cdot,D=\cdot,Z=\cdot]\big](x_{1},d_{1},z_{1})\nonumber \\
		& -\Delta_{q}\big[E[\rho_{n,i}'\theta|X=\cdot,D=\cdot,Z=\cdot]\big](x_{2},d_{2},z_{2})\big|\nonumber \\
		\leq & c\|\Sigma_{\rho,n}^{1/2}\theta\|(\|x_{1}-x_{2}\|^{2}+\|d_{1}-d_{2}\|^{2}+\|z_{1}-z_{2}\|^{2})^{\frac{s-\lfloor s\rfloor}{2}}\label{eq:smooth2-1}
	\end{align}
	(\ref{eq:smooth1-1}) and (\ref{eq:smooth2-1}) together imply: 
	\[
	E[\rho_{n,i}'\theta|X=\cdot,D=\cdot,Z=\cdot]\in\Lambda_{s}^{\dim(X,D,Z)}(c\|\Sigma_{\rho,n}^{1/2}\theta\|)
	\]
	
	And so, re-scaling: 
	\[
	\frac{E[\rho_{n}(V)'\theta|X=\cdot,D=\cdot,Z=\cdot]}{\|\Sigma_{\rho,n}^{1/2}\theta\|}\in\Lambda_{s}^{\dim(X,D,Z)}(c)
	\]
	
	By Assumption 4.2.iv we then have:
	
	\begin{align*}
		|\big(E[\rho_{n,i}'|X_{i},D_{i},Z_{i}]-\zeta_{n,i}'\Sigma_{\zeta,n}^{-1}E[\zeta_{n,i}\rho_{n,i}']\big)'\theta| & \leq c\|\Sigma_{\rho,n}^{1/2}\theta\|\ell_{\mathcal{\zeta},n}(s)\\
		& \leq c\|\theta\|\|\Sigma_{\rho,n}\|^{1/2}\ell_{\mathcal{\zeta},n}(s)
	\end{align*}
	
	Since this holds for any $\theta$, we can optimize over $\theta$
	with $\|\theta\|=1$ to get: 
	\begin{align*}
		\|E[\rho_{n,i}'|X_{i},D_{i},Z_{i}]-\zeta_{n,i}'\Sigma_{\zeta,n}^{-1}E[\zeta_{n,i}\rho_{n,i}']\| & \leq c\|\Sigma_{\rho,n}\|^{1/2}\ell_{\mathcal{\zeta},n}(s)
	\end{align*}
	
	Assumption 4.4.ii also implies that $\frac{f_{V|XD}(v|\cdot,\cdot)}{f_{V}(v)}\in\Lambda_{s}^{\dim(X,D)}(c)$
	and so we can use nearly identical reasoning (with Assumption 4.2.v)
	to get: 
	\[
	\|E[\rho_{n,i}'|X_{i},D_{i}]-\chi_{n,i}'\Sigma_{\chi,n}^{-1}E[\chi_{n,i}\rho_{n,i}']\|\leq c\|\Sigma_{\rho,n}\|^{1/2}\ell_{\mathcal{\chi},n}(s)
	\]
	
	4.2.i implies that $\|\Sigma_{\rho,n}\|$ is bounded above for all
	$n$ which yeilds the result. \end{proof}

\theoremstyle{plain} \newtheorem*{LA3}{Lemma C.3} \begin{LA3}
	
	Suppose Assumptions 4.2.i, 4.2.ii, 4.2.iv, 4.4.iii, 4.5.i, and 4.7.i
	hold. Let $g_{i}$ and $\hat{g}_{i}$ be defined as in Section 3.1.
	Define $l_{g,n,i}$ as follows: 
	\[
	l_{g,n,i}=\hat{g}_{i}-g_{i}-\psi_{n,i}'\Sigma_{\psi,\lambda_{g}}^{-1}\frac{1}{n}\sum_{i=1}^{n}\psi_{n,i}\epsilon_{i}
	\]
	
	We have: 
	\[
	\sqrt{n}\big(\frac{1}{n}\sum_{i=1}^{n}l_{g,n,i}^{2}\big)^{1/2}\precsim_{p}\bar{l}_{g,n}=\sqrt{\frac{\xi_{\psi,n}^{2}k_{\psi,n}ln(k_{\psi,n})}{n}}+\sqrt{n}(\ell_{\psi}(s)+\lambda_{g})
	\]
	
	Moreover, $\|\hat{g}_{n}-g_{n}\|_{n}\precsim_{p}\sqrt{\frac{k_{\psi,n}}{n}}+\frac{\bar{l}_{g,n}}{\sqrt{n}}$.
\end{LA3} \begin{proof}
	
	Recall that $\hat{g}_{i}=\psi_{n,i}'\hat{\beta}_{g}$ where $\hat{\beta}_{g}=\hat{\Sigma}_{\psi,\lambda_{g}}^{-1}\frac{1}{n}\sum_{i=1}^{n}\psi_{n,i}Y_{i}$.
	Let $\beta_{g,n}=\Sigma_{\psi,n}^{-1}E[\psi_{n,i}Y_{i}]$ and $R_{g,n,i}=g_{i}-\psi_{n,i}'\beta_{g,n}$.
	Using this definition we can decompose: 
	\begin{align*}
		\hat{g}_{i}-g_{i} & =\psi_{n,i}'\Sigma_{\psi,\lambda_{g}}^{-1}\frac{1}{n}\sum_{i=1}^{n}\psi_{n,i}\epsilon_{i}+l_{g,n,i}
	\end{align*}
	
	Where the linearization error $l_{g,n,i}$ is given by:
	
	\begin{align*}
		l_{g,n,i} & =\psi_{n,i}'(\hat{\Sigma}_{\psi,\lambda_{g}}^{-1}-\Sigma_{\psi,\lambda_{g}}^{-1})\frac{1}{n}\sum_{i=1}^{n}\psi_{n,i}(\epsilon_{i}+R_{g,n,i})\\
		& +\psi_{n,i}'\Sigma_{\psi,\lambda_{g}}^{-1}\frac{1}{n}\sum_{i=1}^{n}\psi_{n,i}R_{g,n,i}\\
		& -R_{g,i}-\lambda_{g}\psi_{n,i}'\hat{\Sigma}_{\psi,\lambda_{g}}^{-1}\beta_{g,n}
	\end{align*}
	
	Using the triangle inequality, the properties of the matrix norm we
	get: 
	\begin{align*}
		& \big(\frac{1}{n}\sum_{i=1}^{n}l_{g,n,i}^{2}\big)^{1/2}\\
		\leq & \|\hat{\Sigma}_{\psi}\|^{1/2}\|\hat{\Sigma}_{\psi,\lambda_{g}}^{-1}-\Sigma_{\psi,\lambda_{g}}^{-1}\|\big(\|\frac{1}{n}\sum_{i=1}^{n}\psi_{n,i}\epsilon_{i}\|+\|\frac{1}{n}\sum_{i=1}^{n}\psi_{n,i}R_{g,n,i}\|\big)\\
		+ & \|\hat{\Sigma}_{\psi}\|^{1/2}\|\Sigma_{\psi,\lambda_{g}}^{-1}\|\|\frac{1}{n}\sum_{i=1}^{n}\psi_{n,i}R_{g,n,i}\|\\
		+ & \big(\frac{1}{n}\sum_{i=1}^{n}R_{g,n,i}^{2}\big)^{1/2}+\lambda_{g}\|\hat{\Sigma}_{\psi}\|^{1/2}\|\Sigma_{\psi,n}^{-1}\|^{1/2}\|\hat{\Sigma}_{\psi,\lambda_{g}}^{-1}\|\|\Sigma_{\psi,n}^{1/2}\beta_{g,n}\|
	\end{align*}
	
	By Assumptions 4.2.i, 4.2.ii, and 4.5.i we can apply Rudelson's matrix
	law of large numbers so that $\|\hat{\Sigma}_{\psi}-\Sigma_{\psi,n}\|\precsim_{p}\sqrt{\frac{\xi_{\psi,n}^{2}ln(k_{\psi,n})}{n}}$
	which goes to zero. By Assumption 4.2.i $\|\Sigma_{\psi,n}\|$ and
	$\|\Sigma_{\psi,n}^{-1}\|$ are both bounded, and so $\|\hat{\Sigma}_{\psi}^{-1}-\Sigma_{\psi,n}^{-1}\|\precsim_{p}\sqrt{\frac{\xi_{\psi,n}^{2}ln(k_{\psi,n})}{n}}$
	and $\|\hat{\Sigma}_{\psi,\lambda_{g}}^{-1}-\Sigma_{\psi,\lambda_{g}}^{-1}\|\precsim_{p}\sqrt{\frac{\xi_{\psi,n}^{2}ln(k_{\psi,n})}{n}}$.
	Thus $\|\hat{\Sigma}_{\psi}\|\precsim_{p}1$, and $\|\hat{\Sigma}_{\psi}^{-1}\|\precsim_{p}1$.
	Note that $\|\hat{\Sigma}_{\psi,\lambda_{g}}^{-1}\|\leq\|\hat{\Sigma}_{\psi}^{-1}\|$
	and the latter is asymptotically bounded. Moreover, by the properties
	of least squares $\|\Sigma_{\psi,n}^{1/2}\beta_{g,n}\|\leq E[g_{i}^{2}]^{1/2}\precsim1$
	and by Assumptions 4.4.iii and 4.2.iv, $\big(\frac{1}{n}\sum_{i=1}^{n}R_{g,n,i}^{2}\big)^{1/2}\precsim_{p}\ell_{\psi}(s)$.
	By construction $E[\psi_{n,i}\epsilon_{i}]=0$ and $E[\psi_{n,i}R_{g,n,i}]=0$,
	then by Assumption 4.7.i: 
	\begin{align*}
		E[\|\psi_{n,i}\epsilon_{i}\|^{2}] & \leq E\big[\|\psi_{n,i}\|^{2}E[\epsilon_{i}^{2}|X_{i},D_{i},Z_{i}]\big]\\
		& \precsim E[\|\psi_{n,i}\|^{2}]\bar{\sigma}_{\epsilon}^{2}\\
		& \precsim k_{\psi,n}
	\end{align*}
	
	Where in the above $\bar{\sigma}_{\epsilon}^{2}$ is a bound on the
	essential supremum of $E[\epsilon_{i}^{2}|X_{i},D_{i},Z_{i}]$, such
	a finite bound exists by Assumption 4.7.i. The rate of $E[\|\psi_{n,i}\|^{2}]$
	follows from Assumption 4.2.i. And similarly:
	
	\begin{align*}
		E[\|\psi_{n,i}R_{g,n,i}\|^{2}] & \precsim E[\|\psi_{n,i}\|^{2}]\ell_{\psi}(s)^{2}\\
		& \precsim k_{\psi,n}\ell_{\psi}(s)^{2}
	\end{align*}
	
	Applying the law of large numbers we get $\|\frac{1}{n}\sum_{i=1}^{n}\psi_{n,i}\epsilon_{i}\|\precsim_{p}\sqrt{\frac{k_{\psi,n}}{n}}$
	and $\|\frac{1}{n}\sum_{i=1}^{n}\psi_{n,i}R_{g,n,i}\|\precsim_{p}\sqrt{\frac{k_{\psi,n}\ell_{\psi}(s)^{2}}{n}}$
	which is dominated by $\sqrt{\frac{k_{\psi,n}}{n}}$.
	
	So in all (and using $\sqrt{\frac{k_{\psi,n}}{n}}\to0$ which follows
	from Assumptions 4.2.i , 4.2.ii, and 4.5.i) we get: 
	\begin{align*}
		& \big(\frac{1}{n}\sum_{i=1}^{n}l_{g,n,i}^{2}\big)^{1/2}\precsim_{p}\frac{\sqrt{\xi_{\psi,n}^{2}k_{\psi,n}ln(k_{\psi,n})}}{n}+\ell_{\psi}(s)+\lambda_{g}
	\end{align*}
	
	Now for the final statement in the lemma. By the triangle inequality:
	\[
	\big(\frac{1}{n}\sum_{i=1}^{n}(\hat{g}_{n.i}-g_{n,i})^{2}\big)^{1/2}\leq\|\hat{\Sigma}_{\psi,\lambda_{g}}^{-1/2}\frac{1}{n}\sum_{i=1}^{n}\psi_{n,i}\epsilon_{i}\|+\big(\frac{1}{n}\sum_{i=1}^{n}l_{g,n,i}^{2}\big)^{1/2}
	\]
	
	Using the rates derived above we can apply the properties of the matrix
	norm to get: 
	\begin{align*}
		\|\hat{\Sigma}_{\psi,\lambda_{g}}^{-1/2}\frac{1}{n}\sum_{i=1}^{n}\psi_{n,i}\epsilon_{i}\| & \leq\|\hat{\Sigma}_{\psi}^{-1}\|^{1/2}\|\frac{1}{n}\sum_{i=1}^{n}\psi_{n,i}\epsilon_{i}\|\\
		& \precsim_{p}\sqrt{\frac{k_{\psi,n}}{n}}
	\end{align*}
	
	Combining gives the result. \end{proof} \theoremstyle{plain} \newtheorem*{LA4}{Lemma
	C.4} \begin{LA4}
	
	Suppose Assumptions 4.2.i, 4.2.ii, 4.2.iv, 4.2.v, 4.4.ii, 4.5.iii,
	4.5.iv, hold. Let $\hat{\pi}_{n,i}$ and $\pi_{n,i}$ be defined as
	in Section 3.1. Define $l_{\pi,n,i}$ by: 
	\[
	l_{\pi,n,i}=\hat{\pi}_{n,i}-\pi_{n,i}-\bigg(\big(\frac{1}{n}\sum_{i=1}^{n}\upsilon_{n,i}\zeta_{n,i}'\Sigma_{\zeta,\lambda_{\pi}}^{-1}\big)\zeta_{n,i}'\bigg)\otimes\chi_{n,i}
	\]
	
	Then for any non-random sequence $\{\theta_{n}\}_{n=1}^{\infty}$
	where $\theta_{n}$ is a length-$k_{\rho,n}k_{\chi,n}$ column vector
	and $\|\theta_{n}\|\precsim1$ we have: 
	\[
	\sqrt{n}\big(\frac{1}{n}\sum_{i=1}^{n}(l_{\pi,n,i}'\theta_{n})^{2}\big)^{1/2}\precsim_{p}\bar{l}_{\pi,n}
	\]
	
	Where $\bar{l}_{\pi,n}=\sqrt{\frac{\xi_{\zeta,n}^{2}ln(k_{\zeta,n})k_{\zeta,n}}{n}}+\sqrt{n}(\ell_{\zeta}(s)+\lambda_{\pi})$.
	Moreover: 
	\[
	\big(\frac{1}{n}\sum_{i=1}^{n}|(\hat{\pi}_{n,i}-\pi_{n,i})'\theta_{n}|^{2}\big)^{1/2}\precsim_{p}\sqrt{\frac{k_{\zeta,n}}{n}}+\frac{\bar{l}_{\pi,n}}{\sqrt{n}}
	\]
	
	In addition $\|\hat{\pi}_{n}-\pi_{n}\|_{L_{2}}\precsim_{p}\bar{r}_{\pi,n}$
	and $\frac{1}{n}\sum_{i=1}^{n}\|\hat{\pi}_{n,i}-\pi_{n,i}\|\precsim_{p}\bar{r}_{\pi,n}$
	where $\bar{r}_{\pi,n}$ is given by:
	
	\[
	\bar{r}_{\pi,n}=\sqrt{k_{\rho,n}k_{\chi,n}}\bigg(\sqrt{\frac{k_{\zeta,n}}{n}}+\frac{\ell_{\zeta}(s)}{\sqrt{k_{\rho,n}}}+\lambda_{\pi}\bigg)
	\]
\end{LA4} \begin{proof} Recall that $\hat{\pi}_{n,i}=(\hat{\beta}_{\pi}'\zeta_{n,i})\otimes\chi_{n,i}$
	where $\hat{\beta}_{\pi}=\hat{\Sigma}_{\zeta,\lambda_{\pi}}^{-1}\frac{1}{n}\sum_{i=1}^{n}\zeta_{n,i}\rho_{n,i}'$.
	Let $\beta_{\pi,n}=\Sigma_{\zeta,n}^{-1}E[\zeta_{n,i}\rho_{n,i}']$
	and define $R_{\pi,n,i}=E[\rho_{n,i}|Z_{i},X_{i},D_{i}]-\beta_{\pi,n}'\zeta_{n,i}$.
	Let $\tilde{\theta}_{n}$ denote the $k_{\rho,n}$-by-$k_{\chi,n}$
	matrix obtained by partitioning the vector $\theta_{n}$ into $k_{\rho,n}$
	contiguous length-$k_{\chi,n}$ subvectors and stacking the transposes
	into a matrix. Let $vec(\cdot)$ be the vectorization operator so
	that for a $k_{1}$-by-$k_{2}$ matrix $A$, $vec(A)$ is the vector
	obtained by transposing the rows of $A$ and stacking them into a
	vector. Note that $vec(\tilde{\theta}_{n})=\theta_{n}$. We let $\|\cdot\|_{F}$
	denote the Frobenius norm, and note that $\|\tilde{\theta}_{n}\|_{F}=\|\theta_{n}\|$.
	
	Using these definitions we have:
	
	\begin{align*}
		\hat{\pi}_{n,i}-\pi_{n,i} & =\bigg(\big(\frac{1}{n}\sum_{i=1}^{n}(\upsilon_{n,i}+R_{\pi,n,i})\zeta_{n,i}'\Sigma_{\zeta,\lambda_{\pi}}^{-1}\big)\zeta_{n,i}'\bigg)\otimes\chi_{n,i}+l_{\pi,n,i}
	\end{align*}
	
	Where the linearization error $l_{\pi,n,i}$ is defined as follows:
	
	\begin{align*}
		l_{\pi,n,i} & =\bigg(\big((\hat{\Sigma}_{\zeta,\lambda_{\pi}}^{-1}-\Sigma_{\zeta,\lambda_{\pi}}^{-1})\frac{1}{n}\sum_{i=1}^{n}\zeta_{n,i}\upsilon_{n,i}'\big)\otimes I_{\chi}\bigg)'\kappa_{n,i}\\
		& +\big((\Sigma_{\zeta,\lambda_{\pi}}^{-1}\frac{1}{n}\sum_{i=1}^{n}\zeta_{n,i}R_{\pi,n,i}')\otimes I_{\chi}\big)'\kappa_{n,i}\\
		& -R_{\pi,n,i}\otimes\chi_{n,i}-\lambda_{\pi}\big((\hat{\Sigma}_{\zeta,\lambda_{\pi}}^{-1}\beta_{\pi,n})\otimes I_{\chi}\big)'\kappa_{n,i}
	\end{align*}
	
	And so, using properties of the Kronecker product: 
	\begin{align*}
		l_{\pi,n,i}'\theta_{n} & =vec\big((\hat{\Sigma}_{\zeta,\lambda_{\pi}}^{-1}-\Sigma_{\zeta,\lambda_{\pi}}^{-1})\frac{1}{n}\sum_{i=1}^{n}\zeta_{n,i}\upsilon_{n,i}'\tilde{\theta}_{n}\big)'\kappa_{n,i}\\
		& +vec\big(\Sigma_{\zeta,\lambda_{\pi}}^{-1}\frac{1}{n}\sum_{i=1}^{n}\zeta_{n,i}R_{\pi,n,i}'\tilde{\theta}_{n}\big)'\kappa_{n,i}\\
		& -R_{\pi,n,i}'\tilde{\theta}_{n}\chi_{n,i}-\lambda_{\pi}vec(\hat{\Sigma}_{\zeta,\lambda_{\pi}}^{-1}\beta_{\pi,n}\tilde{\theta}_{n})'\kappa_{n,i}
	\end{align*}
	
	Using the triangle inequality and the properties of the matrix norms,
	we get the following:
	
	\begin{align*}
		& \big(\frac{1}{n}\sum_{i=1}^{n}(l_{\pi,n,i}'\theta_{n})^{2}\big)^{1/2}\\
		\leq & \|\hat{\Sigma}_{\kappa}\|^{1/2}\|\hat{\Sigma}_{\zeta,\lambda_{\pi}}^{-1}-\Sigma_{\zeta,\lambda_{\pi}}^{-1}\|\|\frac{1}{n}\sum_{i=1}^{n}\zeta_{n,i}\upsilon_{n,i}'\tilde{\theta}_{n}\|_{F}\\
		+ & \|\hat{\Sigma}_{\kappa}\|^{1/2}\|\Sigma_{\zeta,\lambda_{\pi}}^{-1}\|\|\frac{1}{n}\sum_{i=1}^{n}\zeta_{n,i}R_{\pi,n,i}'\tilde{\theta}_{n}\|_{F}\\
		+ & \big(\frac{1}{n}\sum_{i=1}^{n}(R_{\pi,n,i}'\tilde{\theta}_{n}\chi_{n,i})^{2}\big)^{1/2}+\lambda_{\pi}\|\hat{\Sigma}_{\kappa}\|^{1/2}\|\hat{\Sigma}_{\zeta,\lambda_{\pi}}^{-1}\|\|\Sigma_{\zeta,n}^{-1}\|^{1/2}\|\Sigma_{\zeta,n}^{1/2}\beta_{\pi,n}\tilde{\theta}_{n}\|_{F}
	\end{align*}
	
	$\|\Sigma_{\kappa,n}\|$ is bounded by Assumption 4.2.i and by Assumptions
	4.2.ii and 4.5.iv we can apply Rudelson's law of large numbers so
	we get $\|\hat{\Sigma}_{\kappa}\|\to^{p}\|\Sigma_{\kappa,n}\|$ and
	so $\|\hat{\Sigma}_{\kappa}\|\precsim_{p}1$. By Assumption 4.2.i
	$\|\Sigma_{\zeta,n}\|$ and $\|\Sigma_{\zeta,n}^{-1}\|$ are bounded
	and then by Rudelson's matrix LLN and Assumption 4.5.iii we get $\|\hat{\Sigma}_{\zeta}^{-1}-\Sigma_{\zeta,n}^{-1}\|\precsim_{p}\sqrt{\frac{\xi_{\zeta,n}^{2}ln(k_{\zeta,n})}{n}}$
	and $\|\hat{\Sigma}_{\zeta,\lambda_{\pi}}^{-1}-\Sigma_{\zeta,\lambda_{\pi}}^{-1}\|$,
	and so $\|\hat{\Sigma}_{\zeta}^{-1}\|\precsim_{p}1$. Since $\|\hat{\Sigma}_{\zeta,\lambda_{\pi}}^{-1}\hat{\Sigma}_{\zeta}\|\leq1$
	we then get $\|\hat{\Sigma}_{\zeta,\lambda_{\pi}}^{-1}\|\precsim_{p}1$.
	By the definition of $\beta_{\pi,n}$, and the properties of least
	squares and of the Frobenius norm we have: 
	\begin{align*}
		\|\Sigma_{\zeta,n}^{1/2}\beta_{\pi,n}\tilde{\theta}_{n}\|_{F} & =\|\Sigma_{\zeta,n}^{-1/2}E[\zeta_{n,i}\rho_{n,i}'\tilde{\theta}_{n}]\|_{F}\\
		& \leq E[\|\rho_{n,i}'\tilde{\theta}_{n}\|_{F}^{2}]^{1/2}\\
		& =\|\Sigma_{\rho,n}^{1/2}\tilde{\theta}_{n}\|_{F}\\
		& \leq\|\Sigma_{\rho,n}\|^{1/2}\|\theta_{n}\|\\
		& \precsim1
	\end{align*}
	
	Where for the final line we have used Assumption 4.2.i. Moreover,
	using Lemma C.2 we get:
	
	\begin{align*}
		\big(\frac{1}{n}\sum_{i=1}^{n}(R_{\pi,n,i}'\tilde{\theta}_{n}\chi_{n,i})^{2}\big)^{1/2} & \leq\max_{1\leq i\leq n}\|R_{\pi,n,i}\|\big(\frac{1}{n}\sum_{i=1}^{n}\|\tilde{\theta}_{n}\chi_{n,i}\|^{2}\big)^{1/2}\\
		& \leq c\ell_{\zeta}(s)\|\Sigma_{\zeta,n}\|^{1/2}\|\hat{\Sigma}_{\chi}\|^{1/2}\|\tilde{\theta}_{n}\|_{F}\\
		& \precsim_{p}\ell_{\zeta}(s)
	\end{align*}
	
	Where we have used that $\|\Sigma_{\zeta,n}\|$ is bounded by Assumption
	4.2.i and by 4.2.i, 4.2.ii and 4.5.ii we can apply Rudelson's matrix
	LLN to get $\|\hat{\Sigma}_{\chi}\|\precsim_{p}1$. Next note that
	$E[\zeta_{n,i}\upsilon_{n,i}'\tilde{\theta}_{n}]=0$ and $E[\zeta_{n,i}R_{\pi,n,i}'\tilde{\theta}_{n}]=0$
	and we have:
	
	\begin{align*}
		E[\|\zeta_{n,i}\upsilon_{n,i}'\tilde{\theta}_{n}\|_{F}^{2}]^{1/2} & =E[\|\zeta_{n,i}\|^{2}\|\upsilon_{n,i}'\tilde{\theta}_{n}\|^{2}]^{1/2}\\
		& \leq E[\|\zeta_{n,i}\|^{2}]^{1/2}\text{ess sup }E[\|\upsilon_{n,i}'\tilde{\theta}_{n}\|^{2}|X_{i},D_{i},Z_{i}]^{1/2}\\
		& \leq E[\|\zeta_{n,i}\|^{2}]^{1/2}\text{ess sup }\|E[\upsilon_{n,i}\upsilon_{n,i}'|X_{i},D_{i},Z_{i}]\|^{1/2}\|\tilde{\theta}_{n}\|_{F}\\
		& \precsim_{p}\sqrt{k_{\zeta,n}}
	\end{align*}
	
	Where the final line uses Assumption 4.7.i. Similarly:
	
	\begin{align*}
		E[\|\zeta_{n,i}R_{\pi,n,i}'\tilde{\theta}_{n}\|_{F}^{2}]^{1/2} & =E[\|\zeta_{n,i}\|^{2}\|R_{\pi,n,i}'\tilde{\theta}_{n}\|^{2}]^{1/2}\\
		& +E[\|\zeta_{n,i}\|^{2}]^{1/2}\text{ess sup }\|R_{\pi,n,i}\|\|\tilde{\theta}_{n}\|_{F}\\
		& \precsim_{p}\ell_{\zeta}(s)\sqrt{k_{\zeta,n}}
	\end{align*}
	
	And so, by the law of large numbers $\|\frac{1}{n}\sum_{i=1}^{n}\zeta_{n,i}\upsilon_{n,i}'\tilde{\theta}_{n}\|_{F}\precsim_{p}\sqrt{\frac{k_{\zeta,n}}{n}}$
	and $\|\frac{1}{n}\sum_{i=1}^{n}\zeta_{n,i}R_{\pi,n,i}'\tilde{\theta}_{n}\|_{F}\precsim_{p}\sqrt{\frac{\ell_{\zeta}(s)^{2}k_{\zeta,n}}{n}}$
	which is dominated by $\sqrt{\frac{k_{\zeta,n}}{n}}$. Using $\sqrt{\frac{k_{\zeta,n}}{n}}\to0$
	(which follows from Assumptions 4.5.iii, 4.2.i, and 4.2.ii), we get:
	\begin{align*}
		\big(\frac{1}{n}\sum_{i=1}^{n}(l_{\pi,n,i}'\theta_{n})^{2}\big)^{1/2}\precsim_{p} & \frac{\sqrt{\xi_{\zeta,n}^{2}ln(k_{\zeta,n})k_{\zeta,n}}}{n}+\ell_{\zeta}(s)+\lambda_{\pi}
	\end{align*}
	
	Applying the properties of the matrix norm and the triangle inequality
	we get: 
	\begin{align*}
		& \big(\frac{1}{n}\sum_{i=1}^{n}|(\hat{\pi}_{n,i}-\pi_{n,i})'\theta_{n}|^{2}\big)^{1/2}\\
		\leq & \|\hat{\Sigma}_{\kappa}\|^{1/2}\|\Sigma_{\zeta,n}^{-1}\|\|\frac{1}{n}\sum_{i=1}^{n}\zeta_{n,i}\upsilon_{n,i}'\tilde{\theta}_{n}\|_{F}+\big(\frac{1}{n}\sum_{i=1}^{n}(l_{\pi,n,i}'\theta_{n})^{2}\big)^{1/2}
	\end{align*}
	
	Combing our earlier results we see that: 
	\[
	\big(\frac{1}{n}\sum_{i=1}^{n}|(\hat{\pi}_{n,i}-\pi_{n,i})'\theta_{n}|^{2}\big)^{1/2}\precsim_{p}\sqrt{\frac{k_{\zeta,n}}{n}}+\frac{\bar{l}_{\pi,n}}{\sqrt{n}}
	\]
	
	To derive a rate for $\|\hat{\pi}_{n}-\pi_{n}\|_{L_{2}}$ first note
	that for a deterministic matrix $\beta$ of conformable dimensions
	we have: 
	\begin{align*}
		E[\|(\beta'\otimes I_{\chi})\kappa_{n,i}\|^{2}] & =\|\Sigma_{\kappa,n}^{1/2}(\beta'\otimes I_{\chi})\|_{F}^{2}\\
		& \leq\|\Sigma_{\kappa,n}\|\|\beta'\otimes I_{\chi}\|_{F}^{2}\\
		& =k_{\chi,n}\|\Sigma_{\kappa,n}\|\|\beta\|_{F}^{2}
	\end{align*}
	
	Using the above along with the triangle inequality and properties
	of the matrix norm we get: 
	\begin{align*}
		&\|\hat{\pi}_{n}-\pi_{n}\|_{L_{2}} \\ \leq&\sqrt{k_{\chi,n}}\|\Sigma_{\kappa,n}\|^{1/2}\|\Sigma_{\zeta,\lambda_{\pi}}^{-1}\|\|\frac{1}{n}\sum_{i=1}^{n}\zeta_{n,i}(\upsilon_{n,i}+R_{\pi,n,i})'\|_{F}\\
		 +&\sqrt{k_{\chi,n}}\|\Sigma_{\kappa,n}\|^{1/2}\|\hat{\Sigma}_{\zeta,\lambda_{\pi}}^{-1}-\Sigma_{\zeta,\lambda_{\pi}}^{-1}\|\|\frac{1}{n}\sum_{i=1}^{n}\zeta_{n,i}(\upsilon_{n,i}+R_{\pi,n,i})'\|_{F}\\
		 +&E[\|R_{\pi,n,i}\otimes\chi_{n,i}\|^{2}]^{1/2}+\lambda_{\pi}\sqrt{k_{\chi,n}}\|\Sigma_{\kappa,n}\|^{1/2}\|\hat{\Sigma}_{\zeta,\lambda_{\pi}}^{-1}\|\|\Sigma_{\zeta,n}^{-1}\|^{1/2}\|\Sigma_{\zeta,n}^{1/2}\beta_{\pi,n}\|_{F}
	\end{align*}
	
	By elementary properties of Kronecker products: \[E[\|R_{\pi,n,i}\otimes\chi_{n,i}\|^{2}]\leq\text{ess sup}\|R_{\pi,n,i}\|^{2}E[\|\chi_{n,i}\|^{2}]\]
	Which is bounded by a constant times $k_{\chi,n}\ell_{\zeta}(s)^{2}$.
	By Assumption 4.2.i and the properties of least squares projections
	we have: 
	\[
	\|\Sigma_{\zeta,n}^{1/2}\beta_{\pi,n}\|_{F}^{2}=E[\|\beta_{\pi,n}'\zeta_{n,i}\|^{2}]\leq E[\|\rho_{n,i}\|^{2}]\precsim k_{\rho,n}
	\]
	Next we apply a law of large numbers. Note that $E[\zeta_{n,i}(\upsilon_{n,i}+R_{\pi,n,i})']=0$
	and: 
	\begin{align*}
		&E[\|\zeta_{n,i}(\upsilon_{n,i}+R_{\pi,n,i})'\|_{F}^{2}]^{1/2}\\
		  =&E[\|\zeta_{n,i}\|^{2}\|\upsilon_{\rho,n,i}+R_{\pi,n,i}\|^{2}]^{1/2}\\
		 \leq& E[\|\zeta_{n,i}\|^{2}]^{1/2}\text{ess sup }E[\|\upsilon_{n,i}+R_{\pi,n,i}\|^{2}|X_{i},D_{i},Z_{i}]^{1/2}\\
		 \leq &E[\|\zeta_{n,i}\|^{2}]^{1/2}\text{ess sup }E[\|\upsilon_{n,i}\|^{2}|X_{i},D_{i},Z_{i}]^{1/2}\\
		 +&E[\|\zeta_{n,i}\|^{2}]^{1/2}\text{ess sup }\|R_{\pi,n,i}\|\\
		 \precsim_{p}&\sqrt{k_{\zeta,n}k_{\rho,n}}
	\end{align*}
	
	Where the final line uses Assumption 4.7.i to get that the essential
	supremum of $E[\|\upsilon_{n,i}\|^{2}|X_{i},D_{i},Z_{i}]$ grows at
	rate $k_{\rho,n}$. Now applying the law of large numbers we get:
	\[
	\|\frac{1}{n}\sum_{i=1}^{n}\zeta_{n,i}(\upsilon_{n,i}+R_{\pi,n,i})'\|_{F}\precsim_{p}\sqrt{\frac{k_{\zeta,n}k_{\rho,n}}{n}}
	\]
	
	In all we get:
	
	\[
	\|\hat{\pi}_{n}-\pi_{n}\|_{L_{2}}\precsim_{p}\sqrt{\frac{k_{\zeta,n}k_{\rho,n}k_{\chi,n}}{n}}+\sqrt{k_{\chi,n}}\ell_{\zeta}(s)+\lambda_{\pi}\sqrt{k_{\rho,n}k_{\chi,n}}
	\]
	
	Finally, using similar reasoning we get: 
	\begin{align*}
		&\frac{1}{n}\sum_{i=1}^{n}\|\hat{\pi}_{n,i}-\pi_{n,i}\| \\
		 \leq&\sqrt{k_{\chi,n}}\|\hat{\Sigma}_{\kappa,n}\|^{1/2}\|\Sigma_{\zeta,\lambda_{\pi}}^{-1}\|\|\frac{1}{n}\sum_{i=1}^{n}\zeta_{n,i}(\upsilon_{n,i}+R_{\pi,n,i})'\|_{F}\\
		 +&\sqrt{k_{\chi,n}}\|\hat{\Sigma}_{\kappa,n}\|^{1/2}\|\hat{\Sigma}_{\zeta,\lambda_{\pi}}^{-1}-\Sigma_{\zeta,\lambda_{\pi}}^{-1}\|\|\frac{1}{n}\sum_{i=1}^{n}\zeta_{n,i}(\upsilon_{n,i}+R_{\pi,n,i})'\|_{F}\\
		 +&\big(\frac{1}{n}\sum_{i=1}^{n}\|R_{\pi,n,i}\otimes\chi_{n,i}\|^{2}\big)^{1/2}+\lambda_{\pi}\sqrt{k_{\chi,n}}\|\hat{\Sigma}_{\kappa,n}\|^{1/2}\|\hat{\Sigma}_{\zeta,\lambda_{\pi}}^{-1}\|\|\Sigma_{\zeta,n}^{-1}\|^{1/2}\|\Sigma_{\zeta,n}^{1/2}\beta_{\pi,n}\|_{F}\\
		 \precsim_{p}&\sqrt{\frac{k_{\zeta,n}k_{\rho,n}k_{\chi,n}}{n}}+\sqrt{k_{\chi,n}}\ell_{\zeta}(s)+\lambda_{\pi}\sqrt{k_{\rho,n}k_{\chi,n}}
	\end{align*}
	
\end{proof}

\theoremstyle{plain} \newtheorem*{LA5}{Lemma C.5} \begin{LA5}
	
	Suppose Assumptions 4.2.i, 4.2.ii, 4.2.iv, 4.2.v, 4.4.ii, 4.5.ii,
	and 4.7.i hold. Define $l_{\alpha,n}$ by: 
	\begin{align*}
		l_{\alpha,n}(x_{1},x_{2},d) & =\hat{\alpha}_{n}(x_{1},x_{2},d)-\alpha_{n}(x_{1},x_{2},d)\\
		& -\big(\chi_{n}(x_{2},d)'\Sigma_{\chi,\lambda_{\alpha}}^{-1}\frac{1}{n}\sum_{i=1}^{n}\chi_{n,i}u_{n,i}'\big)'\otimes\chi_{n}(x_{1},d)
	\end{align*}
	
	Then for any non-random sequence $\{\theta_{n}\}_{n=1}^{\infty}$
	where $\theta_{n}$ is a length-$k_{\rho,n}k_{\chi,n}$ column vector
	and $\|\theta_{n}\|\precsim1$, we have: 
	\[
	\frac{\sqrt{n}|l_{\alpha,n}(x_{1},x_{2},d)'\theta_{n}|}{\|\chi_{n}(x_{1},d)\otimes\chi_{n}(x_{2},d)\|}\precsim_{p}\bar{l}_{\alpha,n}
	\]
	
	Where $\bar{l}_{\alpha,n}$ is defined by: 
	\[
	\bar{l}_{\alpha,n}=\bigg(\sqrt{\frac{\xi_{\chi,n}^{2}ln(k_{\chi,n})k_{\chi,n}}{n}}+\sqrt{n}(\frac{\ell_{\mathcal{\chi},n}(s)}{\xi_{\chi,n}}+\lambda_{\alpha})\bigg)
	\]
	
	Moreover $\|\hat{\alpha}_{n}(x_{1},x_{2},d)-\alpha_{n}(x_{1},x_{2},d)\|\precsim_{p}\bar{r}_{\alpha,n}$
	where $\bar{r}_{\alpha,n}$ is given by: 
	\[
	\bar{r}_{\alpha,n}=\sqrt{k_{\rho,n}}\xi_{\chi,n}^{2}\bigg(\sqrt{\frac{k_{\chi,n}}{n}}+\frac{\ell_{\mathcal{\chi},n}(s)}{\xi_{\chi,n}\sqrt{k_{\rho,n}}}+\lambda_{\alpha}\bigg)
	\]
\end{LA5}

\begin{proof} Recall that $\hat{\alpha}_{n}(x_{1},x_{2},d)=\big(\hat{\beta}_{\alpha}'\chi_{n}(x_{2},d)\big)\otimes\chi_{n}(x_{1},d)$
	where we set  $\hat{\beta}_{\alpha}=\hat{\Sigma}_{\chi,\lambda_{\alpha}}^{-1}\frac{1}{n}\sum_{i=1}^{n}\chi_{n,i}\rho_{n,i}'$.
	Let $\beta_{\alpha,n}=\Sigma_{\chi,n}^{-1}E[\chi_{n,i}\rho_{n,i}']$
	and define the function $R_{\alpha,n}$ by $R_{\alpha,n}(X_{i},D_{i})=E[\rho_{n,i}|X_{i},D_{i}]-\beta_{\alpha,n}'\chi_{n,i}$
	and let $R_{\alpha,n,i}=R_{\alpha,n}(X_{i},D_{i})$. As in the proof
	of Lemma C.4, let $\tilde{\theta}_{n}$ denote the $k_{\rho,n}$-by-$k_{\chi,n}$
	matrix obtained by partitioning the vector $\theta_{n}$ into $k_{\rho,n}$
	contiguous length-$k_{\chi,n}$ subvectors and stacking the transposes
	into a matrix. Let $vec(\cdot)$ be the vectorization operator so
	that for a $k_{1}$-by-$k_{2}$ matrix $A$, $vec(A)$ is the vector
	obtained by transposing the rows of $A$ and stacking them into a
	vector. Note that $vec(\tilde{\theta}_{n})=\theta_{n}$ and $\|\tilde{\theta}_{n}\|_{F}=\|\theta_{n}\|$
	where $\|\cdot\|_{F}$ is the Frobenius norm.
	
	Using these definitions we have: 
	\begin{align*}
		& \hat{\alpha}_{n}(x_{1},x_{2},d)-\alpha_{n}(x_{1},x_{2},d)\\
		= & \big(\chi_{n}(x_{2},d)'\Sigma_{\chi,\lambda_{\alpha}}^{-1}\frac{1}{n}\sum_{i=1}^{n}\chi_{n,i}u_{n,i}'\big)'\otimes\chi_{n}(x_{1},d)\\
		+ & l_{\alpha,n}(x_{1},x_{2},d)
	\end{align*}
	
	Where the linearization error $l_{\alpha,n}(x_{1},x_{2},d)$, is given
	by:
	
	\begin{align*}
		& l_{\alpha,n}(x_{1},x_{2},d)\\
		= & \big(\chi_{n}(x_{2},d)'(\hat{\Sigma}_{\chi,\lambda_{\alpha}}^{-1}-\Sigma_{\chi,\lambda_{\alpha}}^{-1})\frac{1}{n}\sum_{i=1}^{n}\chi_{n,i}u_{n,i}'\big)'\otimes\chi_{n}(x_{1},d)\\
		+ & \big(\chi_{n}(x_{2},d)'\Sigma_{\chi,\lambda_{\alpha}}^{-1}\frac{1}{n}\sum_{i=1}^{n}\chi_{n,i}R_{\alpha,n,i}'\big)'\otimes\chi_{n}(x_{1},d)\\
		- & R_{\alpha,n}(x_{2},d)\otimes\chi_{n}(x_{1},d)-\lambda_{\alpha}\big(\chi_{n}(x_{2},d)'\hat{\Sigma}_{\chi,\lambda_{\alpha}}^{-1}\beta_{\alpha,n}\big)'\otimes\chi_{n}(x_{1},d)
	\end{align*}
	
	Using the mixed-product property of Kronecker products we get:
	
	\begin{align*}
		& l_{\alpha,n}(x_{1},x_{2},d)'\theta_{n}\\
		= & \big(\chi_{n}(x_{1},d)\otimes\chi_{n}(x_{2},d)\big)'vec\big((\hat{\Sigma}_{\chi,\lambda_{\alpha}}^{-1}-\Sigma_{\chi,\lambda_{\alpha}}^{-1})\frac{1}{n}\sum_{i=1}^{n}\chi_{n,i}u_{n,i}'\tilde{\theta}_{n}\big)\\
		+ & \big(\chi_{n}(x_{1},d)\otimes\chi_{n}(x_{2},d)\big)'vec\big(\Sigma_{\chi,\lambda_{\alpha}}^{-1}\frac{1}{n}\sum_{i=1}^{n}\chi_{n,i}R_{\alpha,n,i}'\tilde{\theta}_{n}\big)\\
		- & R_{\alpha,n}(x_{2},d)'\tilde{\theta}_{n}\chi_{n}(x_{1},d)\\
		- & \lambda_{\alpha}\big(\chi_{n}(x_{1},d)\otimes\chi_{n}(x_{2},d)\big)'vec\big(\hat{\Sigma}_{\chi,\lambda_{\alpha}}^{-1}\beta_{\alpha,n}\tilde{\theta}_{n}\big)
	\end{align*}
	
	By the triangle inequality and properties of the matrix and Frobenius
	norms we obtain the following inequality from the above:
	
	\begin{align}
		& |l_{\alpha,n}(x_{1},x_{2},d)'\theta_{n}|\nonumber \\
		\leq & \|\chi_{n}(x_{1},d)\otimes\chi_{n}(x_{2},d)\|\|\hat{\Sigma}_{\chi,\lambda_{\alpha}}^{-1}-\Sigma_{\chi,\lambda_{\alpha}}^{-1}\|\|\frac{1}{n}\sum_{i=1}^{n}\chi_{n,i}u_{n,i}'\tilde{\theta}_{n}\|_{F}\nonumber \\
		+ & \|\chi_{n}(x_{1},d)\otimes\chi_{n}(x_{2},d)\|\|\Sigma_{\chi,\lambda_{\alpha}}^{-1}\|\|\frac{1}{n}\sum_{i=1}^{n}\chi_{n,i}R_{\alpha,n,i}'\tilde{\theta}_{n}\|_{F}\|\chi_{n}(x_{1},d)\|\nonumber \\
		+ & \|R_{\alpha,n}(x_{2},d)\|\|\tilde{\theta}_{n}\|_{F}\|\chi_{n}(x_{1},d)\|\nonumber \\
		+ & \lambda_{\alpha}\|\chi_{n}(x_{1},d)\otimes\chi_{n}(x_{2},d)\|\|\hat{\Sigma}_{\chi,\lambda_{\alpha}}^{-1}\|\|\Sigma_{\chi,n}^{-1}\|^{1/2}\|\Sigma_{\chi,n}^{1/2}\beta_{\alpha,n}\tilde{\theta}_{n}\|_{F}\label{eq:linalpherr-1}
	\end{align}
	
	By Assumptions 4.2.i, 4.2.ii, and 4.5.ii we can apply Rudelson's matrix
	law of large numbers to get $\|\Sigma_{\chi,n}-\hat{\Sigma}_{\chi}\|\precsim_{p}\sqrt{\frac{\xi_{\chi,n}^{2}ln(k_{\chi,n})}{n}}$.
	By Assumption 4.2.i $\|\Sigma_{\chi,n}^{-1}\|$ is bounded and so
	$\|\hat{\Sigma}_{\chi}^{-1}-\Sigma_{\chi,n}^{-1}\|\precsim_{p}\sqrt{\frac{\xi_{\chi,n}^{2}ln(k_{\chi,n})}{n}}$
	and $\|\hat{\Sigma}_{\chi,\lambda_{\alpha}}^{-1}-\Sigma_{\chi,\lambda_{\alpha}}^{-1}\|\precsim_{p}\sqrt{\frac{\xi_{\chi,n}^{2}ln(k_{\chi,n})}{n}}$,
	and thus $\|\hat{\Sigma}_{\chi,\lambda_{\alpha}}^{-1}\|\precsim_{p}1$.
	By Lemma C.2 $\text{ess sup}\|R_{\alpha,n}(X_{i},D_{i})\|\precsim\ell_{\mathcal{\chi},n}(s)$.
	By the properties of the Kronecker product and least squares projections:
	
	\begin{align*}
		\|\Sigma_{\chi,n}^{1/2}\beta_{\alpha,n}\tilde{\theta}_{n}\|_{F} & =\|\Sigma_{\chi,n}^{-1/2}E[\chi_{n,i}\rho_{n,i}'\tilde{\theta}_{n}]\|_{F}\\
		& \leq E[\|\rho_{n,i}'\tilde{\theta}_{n}\|_{F}^{2}]^{1/2}\\
		& =\|\Sigma_{\rho,n}^{1/2}\tilde{\theta}_{n}\|_{F}\\
		& \leq\|\Sigma_{\rho,n}\|^{1/2}\|\theta_{n}\|
	\end{align*}
	
	By Assumption 4.2.i $\|\Sigma_{\rho,n}\|^{1/2}$ is bounded and by
	supposition $\|\theta_{n}\|$ is bounded, and so $\|\Sigma_{\chi,n}^{1/2}\beta_{\alpha,n}\tilde{\theta}_{n}\|_{F}\precsim_{p}1$.
	Note that $E[\chi_{n,i}u_{n,i}'\tilde{\theta}_{n}]=0$ and $E[\chi_{n,i}R_{\alpha,n,i}'\tilde{\theta}_{n}]=0$
	and moreover:
	
	\begin{align*}
		E[\|\chi_{n,i}u_{n,i}'\tilde{\theta}_{n}\|_{F}^{2}]^{1/2} & =E[\|\chi_{n,i}\|^{2}\|u_{n,i}'\tilde{\theta}_{n}\|^{2}]^{1/2}\\
		& \leq E[\|\chi_{n,i}\|^{2}]^{1/2}\text{ess sup }E[\|u_{n,i}'\tilde{\theta}_{n}\|^{2}|X_{i},D_{i}]^{1/2}\\
		& \leq E[\|\chi_{n,i}\|^{2}]^{1/2}\text{ess sup }\|E[u_{n,i}u_{n,i}'|X_{i},D_{i}]\|^{1/2}\|\tilde{\theta}_{n}\|_{F}\\
		& \precsim_{p}\sqrt{k_{\chi,n}}
	\end{align*}
	
	Where the final inequality uses Assumption 4.7.i. Similarly we get:
	
	\begin{align*}
		E[\|\chi_{n,i}R_{\alpha,n,i}'\tilde{\theta}_{n}\|_{F}^{2}]^{1/2} & =E[\|\chi_{n,i}\|^{2}]^{1/2}\text{ess sup }\|R_{\alpha,n,i}\|\|\tilde{\theta}_{n}\|_{F}\\
		& \precsim_{p}\sqrt{\ell_{\mathcal{\chi},n}(s)^{2}k_{\chi,n}}
	\end{align*}
	
	So applying the law of large numbers $\|\frac{1}{n}\sum_{i=1}^{n}\chi_{n,i}u_{n,i}'\tilde{\theta}_{n}\|_{F}\precsim_{p}\sqrt{\frac{k_{\chi,n}}{n}}$
	and $\|\frac{1}{n}\sum_{i=1}^{n}\chi_{n,i}R_{\alpha,n,i}'\tilde{\theta}_{n}\|_{F}\precsim_{p}\sqrt{\frac{\ell_{\mathcal{\chi},n}(s)^{2}k_{\chi,n}}{n}}$,
	which is dominated by $\sqrt{\frac{k_{\chi,n}}{n}}$. Combining the
	results above with \eqref{eq:linalpherr} we get:
	
	\begin{align*}
		& \sqrt{n}|l_{\alpha,n}(x_{1},x_{2},d)'\theta_{n}|\\
		\precsim_{p} & \|\chi_{n}(x_{1},d)\otimes\chi_{n}(x_{2},d)\|\bigg(\sqrt{\frac{\xi_{\chi,n}^{2}ln(k_{\chi,n})k_{\chi,n}}{n}}+\sqrt{n}\lambda_{\alpha}\bigg)+\sqrt{n}\ell_{\mathcal{\chi},n}(s)
	\end{align*}
	
	Next we derive a rate for $\|\hat{\alpha}_{n}(x_{1},x_{2},d)-\alpha_{n}(x_{1},x_{2},d)\|$.
	By the triangle inequality and properties of the matrix norm: 
	\begin{align}
		& \|\hat{\alpha}_{n}(x_{1},x_{2},d)-\alpha_{n}(x_{1},x_{2},d)\|\nonumber \\
		\leq & \|\chi_{n}(x_{2},d)\|\|\hat{\Sigma}_{\chi,\lambda_{\alpha}}^{-1}\|\|\frac{1}{n}\sum_{i=1}^{n}\chi_{n,i}(u_{n,i}+R_{\alpha,n,i})'\|_{F}\|\chi_{n}(x_{1},d)\|\nonumber \\
		+ & \|R_{\alpha,n}(x_{2},d)\|\|\chi_{n}(x_{1},d)\|\nonumber \\
		+ & \lambda_{\alpha}\|\chi_{n}(x_{2},d)\|\|\hat{\Sigma}_{\chi,\lambda_{\alpha}}^{-1}\|\|\Sigma_{\chi,n}^{-1}\|^{1/2}\|\Sigma_{\chi,n}^{1/2}\beta_{\alpha,n}\|_{F}\|\chi_{n}(x_{1},d)\|\label{eq:alpherr2}
	\end{align}
	
	First we use the properties of least squares projections to get:
	
	\[
	\|\Sigma_{\chi,n}^{1/2}\beta_{\alpha,n}\|_{F}=\|\Sigma_{\chi,n}^{-1/2}E[\chi_{n,i}\rho_{n,i}']\|_{F}\leq E[\|\rho_{n,i}\|_{F}^{2}]^{1/2}\precsim\sqrt{k_{\rho,n}}
	\]
	
	Note that $E[\chi_{n,i}(u_{n,i}+R_{\alpha,n,i})']=0$ and moreover:
	
	\begin{align*}
		&E[\|\chi_{n,i}(u_{n,i}+R_{\alpha,n,i})'\|_{F}^{2}]^{1/2}\\
		 &= E[\|\chi_{n,i}\|^{2}\|u_{n,i}+R_{\alpha,n,i}\|^{2}]^{1/2}\\
		 \leq& E[\|\chi_{n,i}\|^{2}]^{1/2}\text{ess sup }E[\|u_{n,i}+R_{\alpha,n,i}\|^{2}|X_{i},D_{i}]^{1/2}\\
		 \leq& E[\|\chi_{n,i}\|^{2}]^{1/2}\text{ess sup }E[\|u_{n,i}\|^{2}|X_{i},D_{i}]^{1/2}\\
		 +&E[\|\chi_{n,i}\|^{2}]^{1/2}\text{ess sup }E[\|R_{\alpha,n,i}\|^{2}|X_{i},D_{i}]^{1/2}\\
		 \precsim_{p}&\sqrt{k_{\chi,n}k_{\rho,n}}
	\end{align*}
	
	Where the final line uses Assumption 4.7.i to get that the essential
	supremum of $E[\|u_{n,i}\|^{2}|X_{i},D_{i}]$ grows at rate $k_{\rho,n}$.
	So by the law of large numbers: 
	\[
	\|\frac{1}{n}\sum_{i=1}^{n}\chi_{n,i}(u_{n,i}+R_{\alpha,n,i})'\|_{F}\precsim_{p}\sqrt{\frac{k_{\chi,n}k_{\rho,n}}{n}}
	\]
	
	Combining the above with \eqref{eq:alpherr2} and using our earlier
	bounds on the remaining quantities, we get: 
	\[
	\|\hat{\alpha}_{n}(x_{1},x_{2},d)-\alpha_{n}(x_{1},x_{2},d)\|\precsim_{p}\xi_{\chi,n}^{2}\bigg(\sqrt{\frac{k_{\chi,n}k_{\rho,n}}{n}}+\frac{\ell_{\mathcal{\chi},n}(s)}{\xi_{\chi,n}}+\lambda_{\alpha}\sqrt{k_{\rho,n}}\bigg)
	\]
\end{proof} \theoremstyle{plain} \newtheorem*{LA6}{Lemma C.6}
\begin{LA6} Suppose Assumptions 4.2, 4.3.ii, 4.5.iii, 4.5.iv, 4.7,
	hold , $k_{\chi,n}\ell_{\zeta}(s)^{2}\precsim\bar{r}_{\pi,n}^{2}$ and $\hat{\pi}_{n}=\tilde{\beta}_{\pi}'\kappa_{n,i}$ for some
	estimator $\tilde{\beta}_{\pi}$(for the first-stage estimator in
	Section 3.3, $\tilde{\beta}_{\pi}=(\hat{\beta}_{\pi}\otimes I_{\chi})$
	where $I_{\chi}$ is the identity of dimension $k_{\chi,n}$). In
	addition, suppose $\|\pi_{n}-\hat{\pi}_{n}\|_{L_{2}}\precsim_{p}\bar{r}_{\pi,n}$
	and $\frac{\lambda_{0}\bar{r}_{\pi,n}^{2}}{\mu_{n}+\lambda_{0}^{2}}\to0$.
	Then we have:
	
	\[
	\|\Sigma_{\pi,\lambda_{0}}^{-1/2}(\Sigma_{\pi,n}-\hat{\Sigma}_{\hat{\pi}})\Sigma_{\pi,\lambda_{0}}^{-1/2}\|\precsim_{p}\sqrt{\frac{\xi_{\kappa,n}^{2}ln(k_{\kappa,n})}{n}}+\sqrt{\frac{\lambda_{0}\bar{r}_{\pi,n}^{2}}{\mu_{n}+\lambda_{0}^{2}}}
	\]
	
	Where $\bar{r}_{\pi,n}=\sqrt{k_{\rho,n}k_{\chi,n}}\bigg(\sqrt{\frac{k_{\zeta,n}}{n}}+\frac{\ell_{\zeta}(s)}{\sqrt{k_{\rho,n}}}+\lambda_{\pi}\bigg)$.
	In addition, $\|\hat{\Sigma}_{\hat{\pi},\lambda_{0}}^{-1/2}\Sigma_{\pi,\lambda_{0}}^{1/2}\|\precsim_{p}1$ and: \[\|\alpha_{n}(x_{1},x_{2},d)'\Sigma_{\pi,\lambda_{0}}^{-1/2}\|\precsim1+\bar{\xi}_{\kappa,n}\]
	.\end{LA6} \begin{proof} First note that by standard linear algebra,
	$\|\Sigma_{\pi,\lambda_{0}}^{-1}\|\leq\frac{\lambda_{0}}{\mu_{n}+\lambda_{0}^{2}}$
	where $\mu_{n}$ is the smallest eigenvalue of $\Sigma_{\pi,n}$ and
	thus $\|\Sigma_{\pi,\lambda_{0}}^{-1}\|\leq\lambda_{0}^{-1}$. Let
	$\|\pi_{n}-\hat{\pi}_{n}\|_{L_{2}}\precsim_{p}\bar{r}_{\pi,n}$. Using
	the properties of the matrix norm and properties of least squares
	projections we get:
	
	\begin{align*}
		& \|\Sigma_{\pi,\lambda_{0}}^{-1/2}(\Sigma_{\hat{\pi}}-\Sigma_{\pi,n})\Sigma_{\pi,\lambda_{0}}^{-1/2}\|\\
		\leq & 2\|\Sigma_{\pi,\lambda_{0}}^{-1/2}\|\|\Sigma_{\pi,\lambda_{0}}^{-1/2}E_{(Z_{i},X_{i},D_{i})}\big[\pi_{n}(Z_{i},X_{i},D_{i})\big(\pi_{n}(Z_{i},X_{i},D_{i})-\hat{\pi}_{n}(Z_{i},X_{i},D_{i})\big)'\big]\|\\
		+ & \|\Sigma_{\pi,\lambda_{0}}^{-1/2}\|^{2}\\
		\times&\|E_{(Z_{i},X_{i},D_{i})}\big[\big(\pi_{n}(Z_{i},X_{i},D_{i})-\hat{\pi}_{n}(Z_{i},X_{i},D_{i})\big)\big(\pi_{n}(Z_{i},X_{i},D_{i})-\hat{\pi}_{n}(Z_{i},X_{i},D_{i})\big)'\big]\|\\
		\leq & \|\Sigma_{\pi,\lambda_{0}}^{-1}\|^{1/2}\|\pi_{n}-\hat{\pi}_{n}\|_{L_{2}}+\|\Sigma_{\pi,\lambda_{0}}^{-1}\|\|\pi_{n}-\hat{\pi}_{n}\|_{L_{2}}^{2}\\
		& \precsim_{p}\sqrt{\frac{\lambda_{0}\bar{r}_{\pi,n}^{2}}{\mu_{n}+\lambda_{0}^{2}}}
	\end{align*}
	
	Where the expectations with $(Z_{i},X_{i},D_{i})$ subscripts treat
	the function $\hat{\pi}_{n}$ as non-random. Note that the final line
	uses that the rate $\frac{\lambda_{0}\bar{r}_{\pi,n}^{2}}{\mu_{n}+\lambda_{0}^{2}}$
	for the term corresponding to $\|\Sigma_{\pi,\lambda_{0}}^{-1}\|\|\pi_{n}-\hat{\pi}_{n}\|_{L_{2}}^{2}$
	is dominated by its square root. By the triangle inequality and properties
	of the matrix norm, we have: 
	\begin{align*}
		\|\Sigma_{\pi,\lambda_{0}}^{-1/2}\Sigma_{\hat{\pi},\lambda_{0}}^{1/2}\|^{2} & =\|\Sigma_{\pi,\lambda_{0}}^{-1/2}\Sigma_{\hat{\pi},\lambda_{0}}\Sigma_{\pi,\lambda_{0}}^{-1/2}\|\\
		& \leq1+\|\Sigma_{\pi,\lambda_{0}}^{-1/2}(\Sigma_{\hat{\pi}}-\Sigma_{\pi,n})\Sigma_{\pi,\lambda_{0}}^{-1/2}\|\\
		& \precsim_{p}1
	\end{align*}
	
	Using properties of the matrix norm, along with the fact that $\|\Sigma_{\hat{\pi},\lambda_{0}}^{-1/2}\Sigma_{\hat{\pi}}^{1/2}\|\leq1$,
	and that $\hat{\pi}_{n,i}=\tilde{\beta}_{\pi}'\kappa_{n,i}$, we have:
	\begin{align*}
		& \|\Sigma_{\hat{\pi},\lambda_{0}}^{-1/2}(\Sigma_{\hat{\pi}}-\hat{\Sigma}_{\hat{\pi}})\Sigma_{\hat{\pi},\lambda_{0}}^{-1/2}\|\\
		\leq & \|\Sigma_{\hat{\pi}}^{-1/2}(\Sigma_{\hat{\pi}}-\hat{\Sigma}_{\hat{\pi}})\Sigma_{\hat{\pi}}^{-1/2}\|\\
		= & \|\Sigma_{\hat{\pi}}^{-1/2}\tilde{\beta}_{\pi}'(\Sigma_{\kappa,n}-\hat{\Sigma}_{\kappa})\tilde{\beta}_{\pi}\Sigma_{\hat{\pi}}^{-1/2}\|\\
		\leq & \|\Sigma_{\hat{\pi}}^{-1/2}\tilde{\beta}_{\pi}'\Sigma_{\kappa,n}^{1/2}\|^{2}\|\Sigma_{\kappa,n}^{-1}\|\|\Sigma_{\kappa,n}-\hat{\Sigma}_{\kappa}\|
	\end{align*}
	
	Note that $\|\Sigma_{\hat{\pi}}^{-1/2}\tilde{\beta}_{\pi}'\Sigma_{\kappa,n}^{1/2}\|=1$.
	By Assumption 4.2.i $\|\Sigma_{\kappa,n}^{-1}\|$ is bounded, and
	by Assumptions 4.2.ii and 4.5.iv we can apply Rudelson's matrix LLN
	to get $\|\Sigma_{\kappa,n}-\hat{\Sigma}_{\kappa}\|\precsim_{p}\sqrt{\frac{\xi_{\kappa,n}^{2}ln(k_{\kappa,n})}{n}}$.
	Putting the pieces together, we get: 
	\[
	\|\Sigma_{\hat{\pi},\lambda_{0}}^{-1/2}(\Sigma_{\hat{\pi}}-\hat{\Sigma}_{\hat{\pi}})\Sigma_{\hat{\pi},\lambda_{0}}^{-1/2}\|\precsim_{p}\sqrt{\frac{\xi_{\kappa,n}^{2}ln(k_{\kappa,n})}{n}}
	\]
	
	Using the properties of the matrix norm and $\|\Sigma_{\pi,\lambda_{0}}^{-1/2}\Sigma_{\hat{\pi},\lambda_{0}}^{1/2}\|\precsim_{p}1$
	we then get: 
	\begin{align*}
		\|\Sigma_{\pi,\lambda_{0}}^{-1/2}(\Sigma_{\hat{\pi}}-\hat{\Sigma}_{\hat{\pi}})\Sigma_{\pi,\lambda_{0}}^{-1/2}\| & \leq\|\Sigma_{\pi,\lambda_{0}}^{-1/2}\Sigma_{\hat{\pi},\lambda_{0}}^{1/2}\|^{2}\|\Sigma_{\hat{\pi},\lambda_{0}}^{-1/2}(\Sigma_{\hat{\pi}}-\hat{\Sigma}_{\hat{\pi}})\Sigma_{\hat{\pi},\lambda_{0}}^{-1/2}\|\\
		& \precsim_{p}\sqrt{\frac{\xi_{\kappa,n}^{2}ln(k_{\kappa,n})}{n}}
	\end{align*}
	
	Then by the triangle inequality: 
	\begin{align*}
		\|\Sigma_{\pi,\lambda_{0}}^{-1/2}(\Sigma_{\pi,n}-\hat{\Sigma}_{\hat{\pi}})\Sigma_{\pi,\lambda_{0}}^{-1/2}\| & \leq\|\Sigma_{\pi,\lambda_{0}}^{-1/2}(\Sigma_{\pi,n}-\Sigma_{\hat{\pi}})\Sigma_{\pi,\lambda_{0}}^{-1/2}\|\\
		& +\|\Sigma_{\pi,\lambda_{0}}^{-1/2}(\Sigma_{\hat{\pi}}-\hat{\Sigma}_{\hat{\pi}})\Sigma_{\pi,\lambda_{0}}^{-1/2}\|\\
		& \precsim_{p}\sqrt{\frac{\xi_{\kappa,n}^{2}ln(k_{\kappa,n})}{n}}+\sqrt{\frac{\lambda_{0}\bar{r}_{\pi,n}^{2}}{\mu_{n}+\lambda_{0}^{2}}}
	\end{align*}
	
	Note that $\sqrt{\frac{\xi_{\kappa,n}^{2}ln(k_{\kappa,n})}{n}}+\sqrt{\frac{\lambda_{0}\bar{r}_{\pi,n}^{2}}{\mu_{n}+\lambda_{0}^{2}}}=o(1)$
	. Next note that by the triangle inequality: 
	\begin{align*}
		\|\hat{\Sigma}_{\hat{\pi},\lambda_{0}}^{1/2}\Sigma_{\pi,\lambda_{0}}^{-1/2}\|^{2} & =\|\Sigma_{\pi,\lambda_{0}}^{-1/2}\hat{\Sigma}_{\hat{\pi},\lambda_{0}}\Sigma_{\pi,\lambda_{0}}^{-1/2}\|\\
		& \leq1+\|I-\Sigma_{\pi,\lambda_{0}}^{-1/2}\hat{\Sigma}_{\hat{\pi},\lambda_{0}}\Sigma_{\pi,\lambda_{0}}^{-1/2}\|\\
		& =1+\|\Sigma_{\pi,\lambda_{0}}^{-1/2}(\Sigma_{\pi,n}-\hat{\Sigma}_{\hat{\pi}})\Sigma_{\pi,\lambda_{0}}^{-1/2}\|\\
		& \precsim_{p}1
	\end{align*}
	
	And then again using the triangle inequality: 
	\begin{align*}
		\|\hat{\Sigma}_{\pi,\lambda_{0}}^{-1/2}\Sigma_{\pi,\lambda_{0}}^{1/2}\| & \leq\|\hat{\Sigma}_{\hat{\pi},\lambda_{0}}^{1/2}\Sigma_{\pi,\lambda_{0}}^{-1/2}\|+\|\hat{\Sigma}_{\hat{\pi},\lambda_{0}}^{-1/2}\Sigma_{\pi,\lambda_{0}}^{1/2}-\hat{\Sigma}_{\hat{\pi},\lambda_{0}}^{1/2}\Sigma_{\pi,\lambda_{0}}^{-1/2}\|\\
		& =\|\hat{\Sigma}_{\hat{\pi},\lambda_{0}}^{1/2}\Sigma_{\pi,\lambda_{0}}^{-1/2}\|+\|\hat{\Sigma}_{\hat{\pi},\lambda_{0}}^{-1/2}(\Sigma_{\pi,n}-\hat{\Sigma}_{\hat{\pi}})\Sigma_{\pi,\lambda_{0}}^{-1/2}\|\\
		& \leq\|\hat{\Sigma}_{\hat{\pi},\lambda_{0}}^{1/2}\Sigma_{\pi,\lambda_{0}}^{-1/2}\|+\|\hat{\Sigma}_{\pi,\lambda_{0}}^{-1/2}\Sigma_{\pi,\lambda_{0}}^{1/2}\|\|\Sigma_{\pi,\lambda_{0}}^{-1/2}(\Sigma_{\pi,n}-\hat{\Sigma}_{\hat{\pi}})\Sigma_{\pi,\lambda_{0}}^{-1/2}\|
	\end{align*}
	
	And so, since $\|\Sigma_{\pi,\lambda_{0}}^{-1/2}(\Sigma_{\pi,n}-\hat{\Sigma}_{\hat{\pi}})\Sigma_{\pi,\lambda_{0}}^{-1/2}\|\prec_{p}1$,
	we have with probability approaching $1$ that $\|\Sigma_{\pi,\lambda_{0}}^{-1/2}(\Sigma_{\pi,n}-\hat{\Sigma}_{\hat{\pi}})\Sigma_{\pi,\lambda_{0}}^{-1/2}\|<1$
	in which case:
	
	\begin{align*}
		\|\hat{\Sigma}_{\pi,\lambda_{0}}^{-1/2}\Sigma_{\pi,\lambda_{0}}^{1/2}\| & \leq\frac{\|\hat{\Sigma}_{\hat{\pi},\lambda_{0}}^{1/2}\Sigma_{\pi,\lambda_{0}}^{-1/2}\|}{1-\|\Sigma_{\pi,\lambda_{0}}^{-1/2}(\Sigma_{\pi,n}-\hat{\Sigma}_{\hat{\pi}})\Sigma_{\pi,\lambda_{0}}^{-1/2}\|}\\
		& \precsim_{P}1
	\end{align*}
	
	It remains to bound $\|\alpha_{n}(x_{1},x_{2},d)'\Sigma_{\pi,\lambda_{0}}^{-1/2}\|$.
	As in Lemma C.4, define $\beta_{\pi,n}=\Sigma_{\zeta,n}^{-1}E[\zeta_{n,i}\rho_{n,i}']$
	and let $R_{\pi,n}(z,x,d)=E[\rho_{n,i}|Z_{i}=z,X_{i}=x,D_{i}=d]-\beta_{\pi,n}'\zeta_{n,i}$.
	By the properties of the matrix norm:
	
	\begin{align*}
		& \|\alpha_{n}(x_{1},x_{2},d)'\Sigma_{\pi,\lambda_{0}}^{-1/2}\|\\
		= & \|E[\varphi(Z_{i},x_{1},x_{2},d)\pi_{n}(Z_{i},x_{1},d)|X_{i}=x_{1},D_{i}=d]'\Sigma_{\pi,\lambda_{0}}^{-1/2}\|\\
		\leq & \|E[\varphi(Z_{i},x_{1},x_{2},d)\kappa_{n}(Z_{i},x_{1},d)|X_{i}=x_{1},D_{i}=d]'(\beta_{\pi,n}\otimes I_{\chi})\Sigma_{\pi,\lambda_{0}}^{-1/2}\|\\
		+ & \|E\big[\varphi(Z_{i},x_{1},x_{2},d)\big(\chi_{n}(x_{1},d)\otimes R_{\pi,n}(Z_{i},x_{1},d)\big)|X_{i}=x_{1},D_{i}=d\big]\|\|\Sigma_{\pi,\lambda_{0}}^{-1/2}\|
	\end{align*}
	
	By supposition $E[\varphi(Z,x_{1},x_{2},d)^{2}|X=x_{1},=d]$ is bounded
	above and by Assumption 4.2.ii $\|\chi_{n}(x_{1},d)\|\leq\xi_{\chi,n}$,
	and by Lemma C.2:
	
	\[
	\text{ess sup}\|R_{\pi,n}(Z_{i},x_{1},d)\|\precsim\ell_{\zeta,n}(s)
	\]
	
	So by Cauchy-Schwartz we get: 
	\begin{align*}
		& \|\alpha_{n}(x_{1},x_{2},d)'\Sigma_{\pi,\lambda_{0}}^{-1/2}\|\\
		\leq & \|E[\varphi(Z_{i},x_{1},x_{2},d)\kappa_{n}(Z_{i},x_{1},d)|X_{i}=x_{1},D_{i}=d]'(\beta_{\pi}\otimes I_{\chi})\Sigma_{\pi,\lambda_{0}}^{-1/2}\|\\
		+ & \xi_{\chi,n}\ell_{\zeta,n}(s)\|\Sigma_{\pi,\lambda_{0}}^{-1/2}\|
	\end{align*}
	
	Note that $\xi_{\chi,n}\ell_{\zeta,n}(s)\|\Sigma_{\pi,\lambda_{0}}^{-1/2}\|=\xi_{\chi,n}\ell_{\zeta,n}(s)\sqrt{\frac{\lambda_{0}}{\mu_{n}+\lambda_{0}^{2}}}$
	and $\sqrt{\frac{\lambda_{0}\bar{r}_{\pi,n}^{2}}{\mu_{n}+\lambda_{0}^{2}}}$
	is $o(1)$ by supposition.
	
	For the first term on the RHS above, we use the properties of the
	matrix norm to get:
	
	\begin{align*}
		& \|E[\varphi(Z_{i},x_{1},x_{2},d)\kappa_{n}(Z_{i},x_{1},d)|X_{i}=x_{1},D_{i}=d]'(\beta_{\pi}\otimes I_{\chi})\Sigma_{\pi,\lambda_{0}}^{-1/2}\|\\
		\leq & \|E[\varphi(Z_{i},x_{1},x_{2},d)\kappa_{n}(Z_{i},x_{1},d)|X_{i}=x_{1},D_{i}=d]'\Sigma_{\kappa,n}(x_{1},d)^{-1/2}\|\\
		\times & \|\Sigma_{\kappa,n}(x_{1},d)\|^{1/2}\|\Sigma_{\kappa,n}^{-1}\|^{1/2}\|\Sigma_{\kappa,n}^{1/2}(\beta_{\pi}\otimes I_{\chi})\Sigma_{\pi,\lambda_{0}}^{-1/2}\|
	\end{align*}
	
	By the properties of least squares projections: 
	\begin{align*}
		& \|E[\varphi(Z_{i},x_{1},x_{2},d)\kappa_{n}(Z_{i},x_{1},d)|X_{i}=x_{1},D_{i}=d]'\Sigma_{\kappa,n}(x_{1},d)^{-1/2}\|\\
		\leq & E[\varphi(Z,x_{1},x_{2},d)^{2}|X=x_{1},=d]^{1/2}\\
		\precsim & 1
	\end{align*}
	
	$\|\Sigma_{\kappa,n}^{-1}\|\precsim1$ and $\|\Sigma_{\kappa,n}(x_{1},d)\|\leq\bar{\xi}_{\kappa,n}^{2}$
	by Assumptions 4.2.i and 4.2.ii, and using the triangle inequality,
	properties of the matrix norm, properties of least squares projections,
	and that $\|\Sigma_{\pi,\lambda_{0}}^{-1/2}\Sigma_{\pi,n}\Sigma_{\pi,\lambda_{0}}^{-1/2}\|\leq1$,
	we get: 
	\begin{align*}
	&	\|\Sigma_{\kappa,n}^{1/2}(\beta_{\pi}\otimes I_{\chi})\Sigma_{\pi,\lambda_{0}}^{-1/2}\|\\
		 =& \|\Sigma_{\pi,\lambda_{0}}^{-1/2}E[(\beta_{\pi}\otimes I_{\chi})'\kappa_{n,i}\kappa_{n,i}'(\beta_{\pi}\otimes I_{\chi})]\Sigma_{\pi,\lambda_{0}}^{-1/2}\|\\
		& \leq\|\Sigma_{\pi,\lambda_{0}}^{-1/2}\Sigma_{\pi,n}\Sigma_{\pi,\lambda_{0}}^{-1/2}\|\\
		& +2\|\Sigma_{\pi,n}^{-1/2}E\big[\pi_{n,i}\big(\pi_{n,i}-\kappa_{n,i}'(\beta_{\pi}\otimes I_{\chi})\big)\big]\|\|\Sigma_{\pi,\lambda_{0}}^{-1/2}\|\\
		& +\|\Sigma_{\pi,\lambda_{0}}^{-1/2}\|^{2}\|E\big[\big(\pi_{n,i}-\kappa_{n,i}'(\beta_{\pi}\otimes I_{\chi})\big)\big(\pi_{n,i}-\kappa_{n,i}'(\beta_{\pi}\otimes I_{\chi})\big)\|\\
		& \leq1+2\|\Sigma_{\pi,\lambda_{0}}^{-1/2}\|E[\|\chi_{n,i}\otimes R_{\pi,n,i}\|^{2}]^{1/2}\\
		& +\|\Sigma_{\pi,\lambda_{0}}^{-1/2}\|^{2}E[\|\chi_{n,i}\otimes R_{\pi,n,i}\|^{2}]^{2}
	\end{align*}
	
	Note that $E[\|R_{\pi,n,i}\otimes\chi_{n,i}\|^{2}]\precsim k_{\chi,n}\ell_{\zeta}(s)^{2}\precsim\bar{r}_{\pi,n}^{2}$
	(see the proof of Lemma C.4), and we can use our earlier expression
	for $\|\Sigma_{\pi,\lambda_{0}}^{-1/2}\|$ to get: 
	\begin{align*}
		\|\Sigma_{\kappa,n}^{1/2}(\beta_{\pi}\otimes I_{\chi})\Sigma_{\pi,\lambda_{0}}^{-1/2}\| & \precsim_{p}1+\sqrt{\frac{\lambda_{0}\bar{r}_{\pi,n}^{2}}{\mu_{n}+\lambda_{0}^{2}}}+\frac{\lambda_{0}\bar{r}_{\pi,n}^{2}}{\mu_{n}+\lambda_{0}^{2}}\\
		& \precsim_{p}1
	\end{align*}
	
	So in all: 
	\begin{align*}
		\|\alpha_{n}(x_{1},x_{2},d)'\Sigma_{\pi,\lambda_{0}}^{-1/2}\|\precsim & 1+\bar{\xi}_{\kappa,n}+\frac{\xi_{\chi,n}}{\sqrt{k_{\chi,n}}}\sqrt{\frac{\lambda_{0}\bar{r}_{\pi,n}^{2}}{\mu_{n}+\lambda_{0}^{2}}}\\
		\precsim & 1+\bar{\xi}_{\kappa,n}
	\end{align*}
	
	where $\bar{r}_{\pi,n}=\sqrt{\frac{k_{\zeta,n}k_{\phi,n}}{n}}+\sqrt{k_{\chi,n}}\ell_{\zeta}(s)+\lambda_{\pi}\sqrt{k_{\phi,n}}$.
\end{proof}  \bibliographystyle{authordate1}
\bibliography{proxybib}

\end{document}